\documentclass[11pt]{article}
\usepackage{latexsym}
\usepackage{tabularx,booktabs,multirow,delarray,array}
\usepackage{graphicx,amssymb,amsmath,amssymb,mathrsfs}
\usepackage[linesnumbered, vlined, ruled]{algorithm2e}

\newcommand{\VD}{\mbox{$V\!D$}}
\newcommand{\CD}{\mbox{$C\!D$}}

\newcommand{\Vor}{\mbox{$V\!D$}}
\newcommand{\Tri}{\mbox{$B\!D\!D$}}

\def\calO{\mathcal{O}}
\def\calB{\mathcal{B}}
\def\calP{\mathcal{P}}
\def\calS{\mathcal{S}}
\def\calT{\mathcal{T}}

\def\spp{SPSD}
\def\algo{Algo-SPSD}

\def\calR{\mathcal{R}}
\def\calF{\mathcal{F}}

\begin{document}

\baselineskip=14.0pt

\title{
\vspace*{-0.55in} Computing Shortest Paths among Curved Obstacles in the
Plane\thanks{To appear in ACM Transactions on Algorithms, Vol.11(4), 2015. 
A preliminary version appeared in the Proceedings of the 29th Annual
Symposium on Computational Geometry (SoCG 2013).}
}

\author{
Danny Z. Chen\thanks{Department of Computer Science and Engineering,
University of Notre Dame, Notre Dame, IN 46556, USA.
E-mail: {\tt dchen@cse.nd.edu}.
D.Z.~Chen's research was supported in part by NSF
under Grant CCF-1217906.
}
\hspace*{0.3in} Haitao Wang\thanks{Corresponding
author. Department of Computer Science,
Utah State University, Logan, UT 84322, USA.
E-mail: {\tt haitao.wang@usu.edu}.
H.~Wang's research was supported in part by NSF under Grant CCF-1317143.
}}

\date{}

\maketitle

\thispagestyle{empty}

\newtheorem{lemma}{Lemma}
\newtheorem{theorem}{Theorem}
\newtheorem{corollary}{Corollary}
\newtheorem{fact}{Fact}
\newtheorem{definition}{Definition}
\newtheorem{observation}{Observation}
\newtheorem{condition}{Condition}
\newtheorem{property}{Property}
\newtheorem{claim}{Claim}
\newenvironment{proof}{\noindent {\textbf{Proof:}}\rm}{\hfill $\Box$
\rm}

\pagestyle{plain}
\pagenumbering{arabic}
\setcounter{page}{1}

\vspace*{-0.1in}
\begin{abstract}
A fundamental problem in computational geometry is to compute an obstacle-avoiding
Euclidean shortest path between two points in the plane.
The case of this problem on polygonal obstacles is well studied.
In this paper, we consider the problem version on curved obstacles,
which are commonly modeled as {\em splinegons}. A splinegon can be
viewed as replacing each edge of a polygon by a convex curved edge
(polygons are special splinegons), and the combinatorial complexity of each curved edge is assumed to be
$O(1)$. Given in the plane two points $s$ and $t$ and
a set of $h$ pairwise disjoint splinegons with a total of $n$ vertices,
we compute a shortest $s$-to-$t$ path avoiding the splinegons,
in $O(n\log n+k)$ time or $O(n+h\log^{1+\epsilon}h+k)$ time for any
arbitrarily small constant $\epsilon>0$, where $k$ is a parameter
sensitive to the geometric structures of the input and is
upper-bounded by $O(h^2)$. In particular, when all splinegons are
convex, $k$ is proportional to the number of common tangents in the
free space (called ``free common tangents") among the splinegons.
We develop techniques for solving the problem on the general
(non-convex) splinegon domain, which also improve several previous results, as follows.

\begin{enumerate}

\item We improve the previous work on the polygon case (i.e., when all splinegons are
polygons). The
polygon case was previously
solved in $O(n \log n)$ time, or
in $O(n+h^2\log n)$ time.  Thus,
our algorithm improves the $O(n+h^2\log n)$ time result, and is faster than the
$O(n\log n)$ time solution for sufficiently small $h$, e.g., $h=o(\sqrt{n\log n})$.

\item Our techniques produce an optimal output-sensitive algorithm for
a basic visibility problem of
computing all free common tangents among $h$ pairwise disjoint convex
splinegons of totally $n$ vertices. Our algorithm runs
in $O(n+h\log h+k)$ time and $O(n)$ space, where $k$ is the number
of all free common tangents. Note that $k=O(h^2)$.
Even for the special case where all splinegons are convex {\em polygons},
the previously best algorithm for this visibility problem
takes $O(n+h^2\log n)$ time.

\item We improve the previous work
for computing the shortest path between two points among
convex pseudodisks of $O(1)$ complexity each.
\end{enumerate}

In addition, a by-product of our techniques is an optimal $O(n+h\log h)$ time
and $O(n)$ space algorithm for computing the Voronoi diagram of a set of $h$ pairwise
disjoint convex splinegons with a total of $n$ vertices.
\end{abstract}
\vspace*{-0.1in}

\section{Introduction}
\label{sec:intro}

Finding Euclidean shortest paths among obstacles in the plane
is a fundamental problem in computational geometry.
The case on polygonal obstacles has been well studied (e.g.,
\cite{ref:GhoshAn91,ref:HershbergerAn99,ref:KapoorEf88,ref:KapoorAn97,ref:MitchellSh96,ref:RohnertSh86,ref:StorerSh94}).
For obstacles bounded by curves, the problem is more
difficult and only limited work is found in the literature, and we present an efficient algorithm for this curved version in this paper.

\subsection{The Geometric Setting and Our Results}

\begin{figure}[t]
\begin{minipage}[t]{0.47\linewidth}
\begin{center}
\includegraphics[totalheight=1.0in]{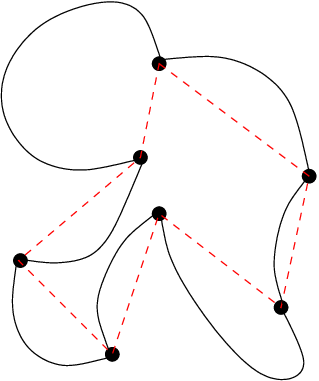}
\caption{\footnotesize A splinegon (solid
curves) defined on a polygon (red or dashed segments).}
\label{fig:splinegon}
\end{center}
\end{minipage}
\hspace*{0.03in}
\begin{minipage}[t]{0.47\linewidth}
\begin{center}
\includegraphics[totalheight=1.0in]{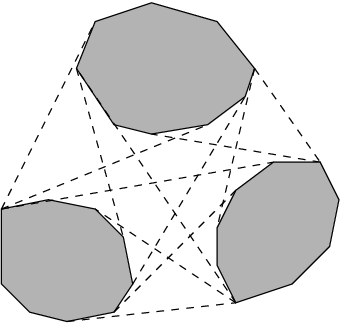}
\caption{\footnotesize The relevant visibility graph of three convex
objects.}\label{fig:relevantGraph}
\end{center}
\end{minipage}
\end{figure}

As in \cite{ref:DobkinCo90,ref:DobkinDe88,ref:MelissaratosSh92}, we
use splinegons to model planar curved objects. A (simple) {\em
splinegon} $S$ is a simple region formed by replacing each edge
$e'_i$ of a simple polygon $P$ by a curved edge $e_i$ joining the
endpoints of $e'_i$ such that the area bounded by the curve $e_i$
and the line segment $e'_i$ is convex (see Fig.~\ref{fig:splinegon}).
The vertices of $S$ are the vertices of $P$.
As in \cite{ref:DobkinCo90,ref:DobkinDe88,ref:MelissaratosSh92}, we assume that the combinatorial
complexity of each splinegon edge is $O(1)$, and {\em primitive operations} on
a splinegon edge can each be performed in $O(1)$ time, such as computing the intersections of a splinegon edge with a line, computing the tangents (if any) between two splinegon edges, finding the tangents between a point and a splinegon edge, computing the distance between two points along a splinegon edge, etc.

We study the problem of computing
{\em shortest paths in a splinegon domain}, denoted by \spp.
Given two points $s$ and $t$ and a set of $h$
pairwise disjoint splinegons, $\calS=\{S_1,\ldots,S_h\}$, with a total of $n$ vertices,
we view the splinegons as {\em obstacles} and
the plane minus the interior of obstacles is called the {\em free space}.
The \spp\ problem seeks a shortest path from $s$ to $t$ in the free space.
If the splinegons in $\calS$ are all convex, then we refer to it as the \emph{convex \spp}.
We are not aware of any previous work that solves this general \spp\ problem exactly. For the convex \spp, by generalizing the algorithm in
\cite{ref:RohnertSh86} for the convex polygonal domain, one may obtain an
$O(n+h^2\log n)$ time solution.

We develop techniques for the general \spp\ problem, and our
algorithm, denoted by \emph{\algo}, takes $O(\min\{n\log n, n+h\log^{1+\epsilon}
h\}+k)$ time, where $k$ is a parameter sensitive to the geometric structures of
the input and $k=O(h^2)$ (the exact
definition of $k$ will be given in Section \ref{sec:overview}).
Throughout this paper, we let $\epsilon>0$ be any arbitrarily
small constant.  For the convex \spp, \algo\ runs in $O(n+h\log h+k)$
time, with $k=O(h^2)$
being the number of free common tangents among the splinegons. A
{\em common tangent} of two convex splinegons is a line segment that
is tangent to both splinegons at its endpoints; the common tangent
is {\em free} if it lies entirely in the free space.

One major contribution of this paper is an optimal output-sensitive
algorithm for the following {\em relevant visibility graph problem}:
When all
splinegons in $\calS$ are convex, compute all free common tangents
of the splinegons (see Fig.~\ref{fig:relevantGraph}). Our
algorithm runs in $O(n+h\log h+k)$ time and $O(n)$ working space.
This visibility problem is a key subproblem to our
algorithm \algo. Note that similar to the argument
in \cite{ref:AsanoVi86,ref:GhoshAn91}, our algorithm is optimal.
Since computing
visibility graphs is a fundamental topic in computational geometry,
our result for this problem may be
interesting in its own right.



Another interesting subproblem that is also solved by our approach is
to compute the Voronoi diagram of $h$ pairwise
disjoint convex splinegons of $n$ vertices. By
generalizing Fortune's sweeping algorithm \cite{ref:FortuneA87}, one may obtain an
$O(n+h\log h\log n)$ time solution. Instead, we extend the
algorithm in \cite{ref:McAllisterA96} for the convex polygon case, and
show that the Voronoi diagram for our problem can be
constructed in $O(n+h\log h)$ time and $O(n)$ space, which is an
optimal solution. Further, as in
\cite{ref:McAllisterA96}, we can also construct in $O(h\log n)$ time a
``compact diagram'', which has several advantages
over the Voronoi diagram and has applications in, e.g., the
post-office problem and the retraction motion planning problem
\cite{ref:McAllisterA96}.

\subsection{Previous Work}

The polygon case of \spp\ (i.e., $\calS$ contains
polygons only) is well studied.
By constructing the visibility graph \cite{ref:GhoshAn91}, a
shortest $s$-$t$ path can be found in $O(n\log n+K)$ time, where
$K=O(n^2)$ is the size of the visibility graph.
By building a shortest path map, Storer and Reif solved this case
in $O(nh)$ time \cite{ref:StorerSh94}. Mitchell
\cite{ref:MitchellSh96} gave the first subquadratic,
$O(n^{3/2+\epsilon})$ time algorithm for it based on the continuous
Dijkstra approach.  Also using the continuous Dijkstra approach and
a conforming planar subdivision, Hershberger and Suri
\cite{ref:HershbergerAn99} presented an $O(n\log n)$ time solution.
An $O(n+h^2\log n)$ time algorithm was given in
\cite{ref:KapoorAn97} (a preliminary version is in
\cite{ref:KapoorEf88}).
Thus, our \algo\ algorithm improves the results in
\cite{ref:KapoorAn97,ref:StorerSh94} and is faster than the
$O(n\log n)$ time solution \cite{ref:HershbergerAn99} for sufficiently small values of
$h$, say $h=o(\sqrt{n\log n})$.\footnote{An unrefereed report
\cite{ref:InkuluA10} announced an algorithm for the polygon case based on the
continuous Dijkstra approach with
an $O(n+h\log h\log n)$ time. Our algorithm is superior to
it when $k=o(n+h\log h\log n)$.}

For \spp\ on curved obstacles, only limited results for
some special convex cases are known.
For the case with $n$ discs, $O(n^2\log n)$ time algorithms were
given \cite{ref:ChangSh05,ref:ChewPl85}, and a heuristic approach
\cite{ref:Kimsh04} was derived with experimental results. For disks
of the same radius, the algorithms in
\cite{ref:HershbergerAn88,ref:StorerSh94} can find a shortest
$s$-$t$ path in $O(n^2)$ time. A set of objects in the plane is
called {\em pseudodisks} if the boundaries of any two objects can
cross each other at most twice. If $\calS$ contains $n$ convex
pseudodisks of $O(1)$ complexity each,
an algorithm in \cite{ref:ChenCo11} can
find a shortest $s$-$t$ path in $O(n^2)$ time. By using our
\algo\ algorithm, the result in \cite{ref:ChenCo11} can be improved
as follows. Let $\Cup \calS$ denote the union of the convex
pseudodisks in $\calS$ and $K$ be the number of vertices on the
boundary of $\Cup\calS$. It has been shown in \cite{ref:KedemOn86} that
$K=O(n)$ and $\Cup \calS$ can be computed in $O(n\log^2 n)$ time.
Since all pseudodisks in $\calS$ are convex, $\Cup\calS$ can be
viewed as consisting of pairwise disjoint splinegons; let $H$ be the
number of splinegons in $\Cup\calS$ (obviously, $H\leq n$). By
applying \algo\ to $\Cup\calS$, a shortest $s$-$t$ path can be found
in $O(n\log^2 n+k)$ time with $k=O(H^2)$. This
improves the $O(n^2)$ time result in \cite{ref:ChenCo11} when
$H=o(n)$.  As a consequence, a robot motion planning problem
\cite{ref:ChenCo11,ref:HershbergerAn88} can also be solved
faster.

For a single splinegon $S$, a shortest $s$-$t$ path in $S$ can be
found in $O(n)$ time, and further, shortest paths from $s$ to all
vertices of $S$ can be found in $O(n)$ time
\cite{ref:MelissaratosSh92}.

There are computational difficulties in applying the continuous Dijkstra approach
\cite{ref:HershbergerAn99,ref:MitchellSh96} to our
\spp\ problem (even when all splinegons are discs) due to the curved obstacle
boundaries. For example, Mitchell's approach \cite{ref:MitchellSh96}
uses a data structure for processing wavelet dragging queries by
modeling them as high-dimensional radical-free semialgebraic range
queries. In \spp, however, such queries would involve not only
radical numbers but also inverse trigonometric operations (e.g.,
arcsine), and hence similar techniques do not seem to apply.
Using the continuous Dijkstra framework,
Hershberger {\em et al.} \cite{ref:HershbergerA13} recently proposed
an $O(n\log n)$ time algorithm for the problem \spp, based on certain
assumption on the computation of localizing the intersection of two
``bisectors''. As indicated in \cite{ref:HershbergerA13}, the bisector
computation itself is a complex problem and the complexity of
computing these bisectors is still unknown.
Without the bisector computation assumption, the approach in
\cite{ref:HershbergerA13} can compute
a $(1+\epsilon)$-factor approximate shortest path in $O(n\log
n+n\log\frac{1}{\epsilon})$ time.

Constructing the visibility graph for polygonal objects has been well studied
\cite{ref:AsanoVi86,ref:GhoshAn91,ref:GuibasLi87,ref:KapoorEf88,ref:OvermarsNe88,ref:RohnertSh86,ref:WelzlCo85}.
Ghosh and Mount finally gave an $O(n\log n+K)$ time algorithm
\cite{ref:GhoshAn91}, where
$K=O(n^2)$ is the size of the visibility graph. For the {\em relevant visibility graph}
problem \cite{ref:KapoorAn97,ref:PocchiolaTo96,ref:RohnertSh86} (or building
the {\em relevant visibility graph})
on splinegons, two special cases have been studied. When $\calS$ contains $n$
disjoint convex objects of $O(1)$ complexity each, the problem
is solvable in $O(n\log n+K)$ time \cite{ref:PocchiolaTo96},
where $K=O(n^2)$ is the number of free common tangents. If
$\calS$ contains $h$ convex {\em polygons}, as in
\cite{ref:KapoorAn97,ref:RohnertSh86}, then the problem is solvable in
$O(n+h^2\log n)$ time;
an open question was posed in \cite{ref:KapoorAn97} to solve this
case in $O(n+k\log n)$ time, where $k=O(h^2)$ is the number of
free common tangents. Note that our $O(n+h\log
h+k)$ time result is better than the solution desired by this
open question.


\section{An Overview of Our \spp\ Algorithm}
\label{sec:overview}

Our algorithm \algo\ follows the high-level scheme used in the
polygonal domain case \cite{ref:KapoorAn97}, but with the key steps
replaced by our new, generalized, and more efficient solutions for the more
difficult splinegon domain counterparts.
We present the paper in a way that each section is
highly self-contained, as discussed below.
Let $\calR$ be a rectangle containing all splinegons in $\calS$, and
$\calF$ denote the free space inside $\calR$. We view both $s$ and
$t$ as two special splinegons in $\calS$.

\begin{figure}[t]
\begin{minipage}[t]{0.53\linewidth}
\begin{center}
\includegraphics[totalheight=1.4in]{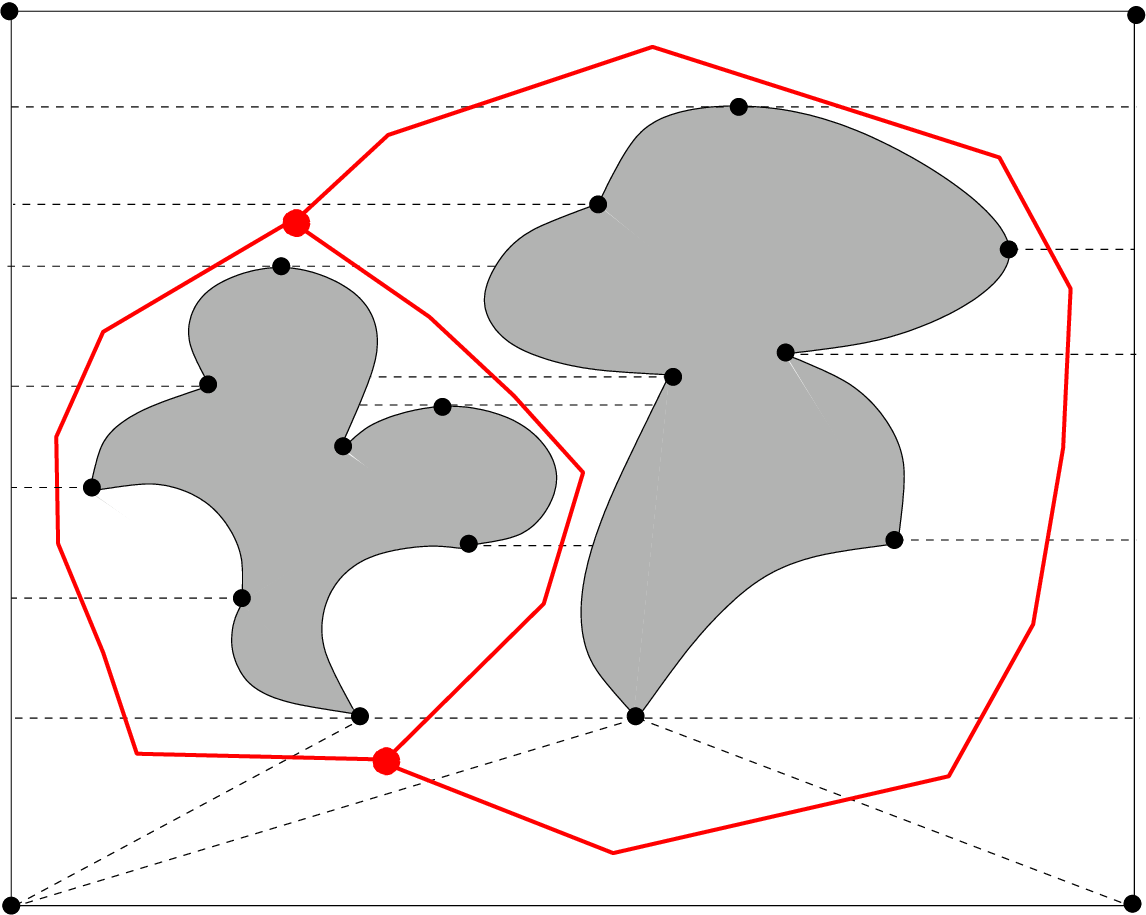}
\caption{\footnotesize Illustrating a bounded degree decomposition
of $\calF$ (with dashed segments)
and the corridors (with red solid arcs).
There are two junction regions indicated by large (red)
points inside them, connected by three solid (red)
arcs. Removal of these two junction regions results
in three corridors.} \label{fig:trimulobjnew}
\end{center}
\end{minipage}
\hspace*{0.04in}
\begin{minipage}[t]{0.45\linewidth}
\begin{center}
\includegraphics[totalheight=1.4in]{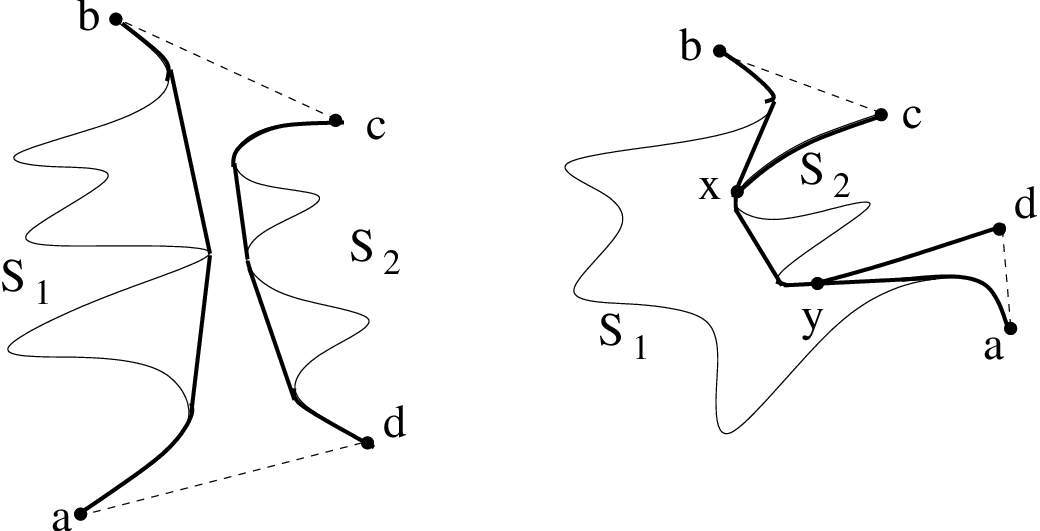}
\caption{\footnotesize Illustrating an open hourglass (left) and a
closed hourglass (right) with a corridor path linking the apices
$x$ and $y$ of the two funnels. The dashed segments are diagonals. The paths
$\pi(a,b)$ and $\pi(c,d)$ are shown with thick solid curves.}
\label{fig:corridor}
\end{center}
\end{minipage}
\vspace*{-0.15in}
\end{figure}

The first step is to decompose $\calF$ into regions each
of which has at most four sides and at most three neighbors (see
Fig.~\ref{fig:trimulobjnew}). This decomposition,
called {\em bounded degree decomposition}, serves the same purpose as a
usual triangulation in the polygonal domain case.
Melissaratos and Souvaine \cite{ref:MelissaratosSh92} computed
a bounded degree decomposition inside a simple splinegon
in linear time. By extending the triangulation algorithm for the
polygonal domain case \cite{ref:Bar-YehudaTr94} and applying the
algorithm in \cite{ref:MelissaratosSh92},
we present an $O(n+h\log^{1+\epsilon} h)$ time algorithm for
computing a bounded degree decomposition of $\calF$,
denoted by $\Tri(\calF)$ (which can also be computed in $O(n\log n)$ time by the standard sweeping techniques).
The details of this step are given in Section \ref{sec:triangulation}.

The second step, with details in Section \ref{sec:corridors}, is to
compute a {\em corridor structure} in $\Tri(\calF)$, which
consists of $O(h)$ corridors and $O(h)$ junction regions
(see Fig.~\ref{fig:trimulobjnew}). Each corridor contains an
hourglass, either open or closed (see Fig.~\ref{fig:corridor}). An
open hourglass contains two convex chains. A closed hourglass
contains two ``funnels" with a {\em corridor path} connecting the
two apices of the two funnels. Each side of a funnel is also a
convex chain. As in \cite{ref:KapoorAn97}, the above $O(h)$ convex
chains from the corridors can be used to partition the space in $\calR$ into a
set $\calS'$ of $O(h)$ convex splinegons with a total of $O(n)$ vertices
such that a shortest $s$-$t$ path
for our original \spp\ problem is also a shortest $s$-$t$ path
avoiding the convex splinegons of $\calS'$ and
possibly utilizing some corridor paths. Thus, in addition to the
presence of the $O(h)$ corridor paths, our \spp\ problem is reduced
to an instance of the convex \spp.
All the above computation can be performed in $O(n+h\log h)$ time. The key
is to solve the convex \spp\ problem on $\calS'$.

To solve the convex \spp\ on $\calS'$, we define a {\em relevant
visibility graph} $G$ (see Fig.~\ref{fig:relevantGraph}), as follows.
Let $k$ be the number of all free common tangents of the $O(h)$
convex splinegons in $\calS'$; thus $k=O(h^2)$. The node set of
$G$ consists of the endpoints of the free common tangents. Hence $G$
has $O(k)$ nodes. Each free common tangent defines an edge in $G$.
For every splinegon $S\in\calS'$, its boundary portion between any
two consecutive nodes of $G$ along the boundary of $S$ also defines an
edge. Thus $G$ has $O(k)$ edges. Clearly, a shortest $s$-$t$ path
in the free space of $\calS'$ corresponds to a shortest path from $s$ to $t$ in
$G$ (both $s$ and $t$ are nodes in $G$). Therefore, to solve the
convex \spp, we need to solve two subproblems: Constructing $G$ and
computing a shortest $s$-to-$t$ path in $G$.

The third step solves the first subproblem: Constructing $G$. Our
algorithm takes $O(n+k+h\log h)$ time.
This step, which is interesting in its own right, is
presented in Section \ref{sec:relevant}.

The fourth step solves the second subproblem: Finding a
shortest path from $s$ to $t$ in $G$. Since $G$ has $O(k)$ nodes and
$O(k)$ edges, simply running Dijkstra's algorithm on $G$ would take
$O(k\log k)$ time. To avoid the $\log k$ factor, we
extend the approach in \cite{ref:ChenCo11} for computing a shortest
path among pseudodisks, where a subproblem is to compute the
Voronoi diagram of the convex splinegons in $\calS'$. We show that
this Voronoi diagram can be computed in $O(n+h\log h)$ time, which
may be of independent interest. The details of this step are given in Section
\ref{sec:shortestpath}.

The last step, discussed in Section \ref{sec:polygoncase}, is to
incorporate the $O(h)$ corridor paths into the algorithm for the
convex \spp\ on $\calS'$ to obtain a shortest path for our original
\spp\ problem.

\section{The Bounded Degree Decomposition of the Free Space}
\label{sec:triangulation}

Recall that $\calS=\{S_1,\ldots,S_h\}$ is a set of $h$ pairwise disjoint splinegons with a total of
$n$ vertices, and $\calF$ is
the free space of $\calS$. In this
section, we compute a bounded degree decomposition $\Tri(\calF)$ of
the free space $\calF$. In the following, we first define
$\Tri(\calF)$ and then present our algorithm for it.

\subsection{Defining a Bounded Degree Decomposition}

As preprocessing, we perform a {\em monotone cut} on the edges of
the splinegons in $\calS$, as follows. For each splinegon edge $e$,
if one or both of its topmost and bottommost points lie in the
interior of $e$, then we add these points (at most two) as new
splinegon vertices, which divide the original edge $e$ into several
new edges (at most three). Since each splinegon edge is of $O(1)$
complexity, this monotone cut can be done in $O(n)$ time. After the
cut, $\calS$ contains at most $3n$ vertices. For convenience, with a
little abuse of notation, we still use $n$ to denote the number of
vertices of $\calS$ after the monotone cut.  From now on, we assume
that the monotone cut has been done on all splinegons in $\calS$.
For any object $A$ in the plane, let
$\partial A$ denote its boundary.

Recall that $\calR$  is a large rectangle containing all splinegons
in $\calS$.  Let $\partial\calF$ denote the boundary of $\calF$,
i.e., the union of the boundaries of the splinegons in $\calS$ and
$\calR$. A {\em diagonal} is an open line segment in the interior of
$\calF$ with its two endpoints on $\partial\calF$. A {\em bounded
degree decomposition} of the free space $\calF$, denoted by
$\Tri(\calF)$, is a decomposition of $\calF$ into $O(n)$ {\em bounded
degree regions} (or simply {\em regions}) each with at most four
sides and with at most three neighboring regions by adding $O(n)$
non-intersecting diagonals (see Fig.~\ref{fig:trimulobjnew}). Two
regions are {\em neighboring} if they share a diagonal on their
boundaries. Each region has at most four sides and each side is
either a diagonal or (part of) a splinegon edge. Thus the complexity
of each region is $O(1)$ (that is why we call it a ``bounded degree
region"). $\Tri(\calF)$ serves the same purpose as a triangulation in
the polygonal domain case.

For a single simple splinegon $S$, a linear time algorithm was
given in \cite{ref:MelissaratosSh92} for computing a bounded degree
decomposition of $S$.

\subsection{Computing a Bounded Degree Decomposition}

As the triangulation algorithm in \cite{ref:Bar-YehudaTr94}, our algorithm for computing $\Tri(\calF)$ consists of two main steps.
First, we find $h$ non-crossing diagonals to connect all splinegons
in $\calS$ and $\calR$ together to form a single simple splinegon
$S^*$ (so the interior of $S^*$ is $\calF$ minus the above diagonals). Second, we apply the algorithm in \cite{ref:MelissaratosSh92}
to compute a bounded degree decomposition of $S^*$, which is
$\Tri(\calF)$. The details are given below.


For the first main step, our approach generalizes the triangulation algorithm in
\cite{ref:Bar-YehudaTr94} for the polygonal domain case. We first
define a {\em visibility tree} for $\calS$, as follows. For each
splinegon $S_i\in \calS$, pick a point on its boundary (not
necessarily a vertex) and draw a ray to the right until it hits
either another splinegon in $\calS$ or $\calR$. As shown in
\cite{ref:Bar-YehudaTr94}, if we choose the origins of the rays
carefully and view each splinegon as a node, we can then ensure that
the resulting planar graph is connected and acyclic; actually, the
resulting planar graph is a visibility tree for $\calS$, denoted by
$T_{vis}(\calS)$. Clearly, $T_{vis}(\calS)$ connects all splinegons
of $\calS$ and $\calR$ into a single simple splinegon. Our task
is to compute $T_{vis}(\calS)$, which can be easily done in $O(n\log n)$ time by the standard sweeping techniques. In the following Lemma \ref{lem:new10}, we present an $O(n+h\log^{1+\epsilon}h)$ time algorithm.

Note that in the special case
where all splinegons in $\calS$ are convex, a visibility tree
$T_{vis}(\calS)$ can be computed in $O(n+h\log h)$ time by the
sweeping techniques.

\begin{lemma}\label{lem:new10}
A visibility tree $T_{vis}(\calS)$ of $\calS$ and $\calR$ can be
computed in $O(n+h\log^{1+\epsilon}h)$ time.
\end{lemma}

\begin{proof}
Our approach generalizes the algorithm in \cite{ref:Bar-YehudaTr94},
which computes a visibility tree in $O(n+h\log^{1+\epsilon} h)$ time
for a set of $h$ pairwise disjoint polygons of totally $n$ vertices.

To generalize the algorithm in \cite{ref:Bar-YehudaTr94}, we need to
make sure that each of its components can be generalized. First, the
algorithm in \cite{ref:Bar-YehudaTr94} makes use of the linear time
algorithm in \cite{ref:HoffmannSo86} for sorting the intersections
(by their $x$-coordinates) of a horizontal line and an oriented
Jordan curve. For any splinegon $S_i\in \calS$, consider the problem
of sorting the intersections of $\partial S_i$ with a horizontal
line $l$. Due to the monotone cut, each edge of $S_i$ has at most
one intersection with $l$ which can be computed in $O(1)$ time.
Further, the line $l$ breaks up the boundary of $S_i$ into disjoint
arcs either entirely below or above $l$ such that the arcs above
(resp., below) the line still form a parenthesis system, as shown in
\cite{ref:Bar-YehudaTr94,ref:HoffmannSo86}. Hence, the algorithmic
scheme in \cite{ref:HoffmannSo86} and the time analysis are still
applicable to our problem. In summary, the intersections of
$\partial S_i$ and the line $l$ can be computed in linear time (in
terms of the number of vertices of $S_i$); let $m$ be the number of
such intersections. Then these intersection points on $l$ can be
sorted in $O(m)$ time.

With the above sorting algorithm in hand, the following more general
sorting problem can be solved by using the approach in
\cite{ref:Bar-YehudaTr94}: Given a subset of $h'$ ($h'\leq h$)
splinegons in $\calS$ with a total of $n'$ ($n'\leq n$) vertices and a
horizontal line $l$, the goal is to sort the intersections of $l$
with these $h'$ splinegons. All intersections can be computed in
$O(n')$ time. Let $m$ be the number of such intersections.  Then by
following the algorithmic scheme in \cite{ref:Bar-YehudaTr94} and
using our sorting procedure for a single splinegon case, these $m$
intersections can be sorted in $O(m+h'\log h')$ time.

In addition, the algorithm in \cite{ref:Bar-YehudaTr94} needs a
point location data structure
\cite{ref:EdelsbrunnerOp86,ref:KirkpatrickOp83} on a simple polygon,
constructed in linear time, for answering each point location query
in logarithmic time. In our problem, correspondingly, we need such a
point location data structure on a simple splinegon. As shown in
\cite{ref:EdelsbrunnerOp86,ref:MelissaratosSh92}, the data structure
in \cite{ref:EdelsbrunnerOp86} can be made to work on a simple
splinegon with the same performance as for the simple polygon case.

It is also easy to verify that other parts of the algorithm in
\cite{ref:Bar-YehudaTr94} are all applicable to our problem.
Therefore, the visibility tree $T_{vis}(\calS)$ for $\calS$ can be
computed in $O(n+h\log^{1+\epsilon}h)$ time.
\end{proof}

For the second main step, we simply apply the linear time decomposition algorithm in \cite{ref:MelissaratosSh92} to $S^*$.
Below, for completeness and easy understanding of our
approach, we briefly discuss the
algorithm in \cite{ref:MelissaratosSh92} for a single splinegon.

\begin{figure}[t]
\begin{minipage}[t]{0.49\linewidth}
\begin{center}
\includegraphics[totalheight=0.9in]{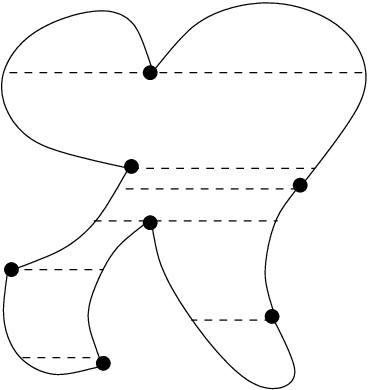}
\caption{\footnotesize Illustrating the horizontal visibility map of
a splinegon.} \label{fig:visibilitymap}
\end{center}
\end{minipage}
\hspace*{0.02in}
\begin{minipage}[t]{0.49\linewidth}
\begin{center}
\includegraphics[totalheight=0.9in]{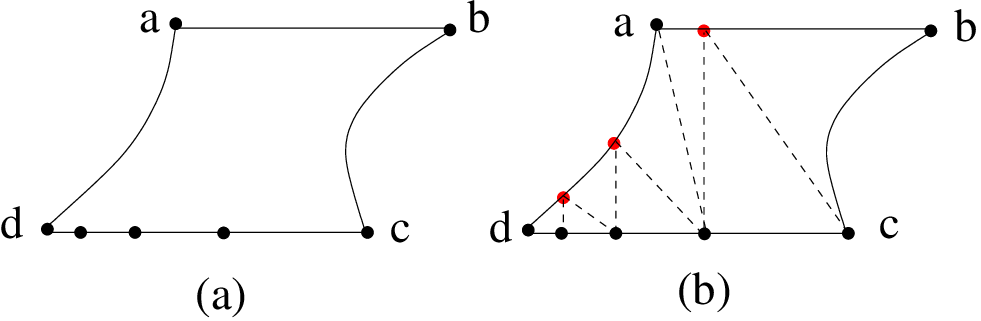}
\caption{\footnotesize Illustrating the decomposition of a
trapezoid: (a) The black points on the base $cd$ are vertices; (b) a
decomposition of the trapezoid.} \label{fig:trapezoid}
\end{center}
\end{minipage}
\vspace*{-0.15in}
\end{figure}

Let $S$ be a splinegon of $n$ vertices. The algorithm in
\cite{ref:MelissaratosSh92} first decomposes $S$ into a set of
horizontal trapezoids by computing the horizontal visibility map and
then further decomposes each trapezoid into (bounded degree) regions
 as the final decomposition of $S$.
Below are some details of it. Again, the topmost and bottommost
points of each splinegon edge are treated as vertices of $S$.
Clearly, there are $O(n)$ vertices on $\partial S$. As for the
simple polygon case, the {\em horizontal visibility map} is to draw
a horizontal line segment through each vertex of $S$, extending the
segment so long as it does not properly cross $\partial S$ (see
Fig.~\ref{fig:visibilitymap}). The visibility map of $S$ adds $O(n)$
new vertices on $\partial S$ and divides $S$ into $O(n)$ trapezoids
(each with curved {\em sides} and line segment {\em bases}). As
shown in \cite{ref:MelissaratosSh92}, Chazelle's algorithm
\cite{ref:ChazelleTr91} can be used to compute the visibility map on
the splinegon $S$ in $O(n)$ time. Since in the degenerate case there may be multiple
vertices in the interior of a trapezoid base, a trapezoid may have
many neighbors. The next step is to further decompose each trapezoid
into (bounded degree) regions such that each region has at most
three neighbors. There are many ways to decompose a trapezoid into
(bounded degree) regions. In \cite{ref:MelissaratosSh92}, one
algorithmic approach for this task was given and some of the cases
were discussed. For example, consider a trapezoid $abcd$ with $ab$
and $cd$ as bases and $ad$ and $bc$ as (curved) sides, as in
Fig.~\ref{fig:trapezoid}(a). Suppose there are multiple vertices in
the interior of the base $cd$. A further decomposition of the
trapezoid is shown in Fig.~\ref{fig:trapezoid}(b).
Refer to \cite{ref:MelissaratosSh92} for more details.
Note that although a region
may be a four-side trapezoid, it has at most three neighbors and is of $O(1)$ complexity.

By Lemma \ref{lem:new10} and the linear time decomposition algorithm
for a simple splinegon \cite{ref:MelissaratosSh92}, the following
result follows.

\begin{theorem}\label{theo:new10}
A bounded degree decomposition of the free space $\calF$ among the
splinegons of $\calS$ can be computed in $O(n\log n)$ time or $O(n+h\log^{1+\epsilon}h)$
time. If all splinegons in $\calS$ are convex, then the
decomposition can be computed in $O(n+h\log h)$ time.
\end{theorem}

\section{The Corridor Structure}
\label{sec:corridors}

In this section, based on $\Tri(\calF)$, i.e, the bounded degree
decomposition of the free space $\calF$, we compute a corridor
structure to reduce our original problem on \spp\ to an
instance of the convex \spp.
The corridor structure and its extended version
for polygonal domains have been used for solving shortest path and
visibility problems
\cite{ref:ChenTw14,ref:ChenA11ESA,ref:ChenCo12arXiv,ref:ChenCo12ICALP,ref:ChenCo13SoCG,ref:ChenL113STACS,ref:ChenVi15,ref:InkuluPl09,ref:KapoorEf88,ref:KapoorAn97}.
For our splinegonal domain, we generalize the approach in \cite{ref:KapoorAn97} for the
polygonal domain case.

Recall that both $s$ and $t$ are considered as two special
splinegons in $\calS$.
In addition to the splinegon vertices, the endpoints of the
diagonals of $\Tri(\calF)$ are also treated as the {\em vertices} of
$\Tri(\calF)$. Note that $\Tri(\calF)$ has $O(n)$ vertices. Let
$G(\calF)$ denote the planar dual graph of $\Tri(\calF)$, i.e., each
node of $G(\calF)$ corresponds to a region in $\Tri(\calF)$ and each
edge connects two nodes of $G(\calF)$ corresponding to two regions
sharing a diagonal. Because the splinegons in $\calS$ are pairwise
disjoint, the dual graph $G(\calF)$ is clearly connected, and an
$s$-$t$ path among the splinegons of $\calS$ always exists.
Since $\Tri(\calF)$ is a planar structure and each region in
$\Tri(\calF)$ has at most three neighbors, $G(\calF)$ is a
planar graph whose vertex degrees are at most three.  Since $\calF$ is connected, as in the
polygonal domain case \cite{ref:KapoorAn97}, at least one node dual to a region incident
to each of $s$ and $t$ is of degree three.

Based on $G(\calF)$, we compute a planar 3-regular graph, denoted by $G^3$ (the
degree of each node in it is three), possibly with loops and multi-edges,
as follows. First, we remove every degree-one node from $G(\calF)$
along with its incident edge; repeat this process until no
degree-one node exists. Second, remove every degree-two node from
$G(\calF)$ and replace its two incident edges by a single edge;
repeat this process until no degree-two node exists. The
resulting graph is $G^3$ (e.g., see Fig.~\ref{fig:trimulobjnew}). By
a similar argument as in \cite{ref:KapoorAn97} for the polygonal
domain case, we can show that the resulting $G^3$ has $O(h)$ faces,
nodes, and arcs. Each node of $G^3$ corresponds to a
region of $\Tri(\calF)$, which is called a {\em junction region}
(e.g., see Fig.~\ref{fig:trimulobjnew}). Removal of all junction
regions from $G^3$ results in $O(h)$ {\em corridors}, each of which
corresponds to one edge of $G^3$.

The boundary of a corridor $C$ consists of four parts (see
Fig.~\ref{fig:corridor}): (1) A boundary portion of a splinegon
$S_1\in \calS$, from a point $a$ to a point $b$; (2) a diagonal of a
junction region from $b$ to a point $c$ of a splinegon $S_2\in
\calS$ (it is possible that $S_1=S_2$); (3) a boundary portion of
the splinegon $S_2$ from $c$ to a point $d$; (4) a diagonal of a
junction region from $d$ to $a$. The two diagonals $\overline{bc}$
and $\overline{ad}$ are called the {\em doors} of $C$. Note that the
corridor $C$ itself is a simple splinegon. Let $|C|$ denote the
number of vertices of $\Tri(\calF)$ on $\partial C$. Note that a
shortest path between two points inside a simple splinegon can be
found in linear time \cite{ref:MelissaratosSh92}. Therefore, in
$O(|C|)$ time, we can compute the shortest path $\pi(a,b)$ (resp.,
$\pi(c,d)$) from $a$ to $b$ (resp., $c$ to $d$) inside $C$. The
region $H_C$ bounded by $\pi(a,b), \pi(c,d)$, and the two diagonals
$\overline{bc}$ and $\overline{da}$ is called an {\em hourglass},
which is {\em open} if $\pi(a,b)\cap \pi(c,d)=\emptyset$ and {\em
closed} otherwise (see Fig.~\ref{fig:corridor}). If $H_C$ is open,
then both $\pi(a,b)$ and $\pi(c,d)$ are convex and they are called
the {\em sides} of $H_C$; otherwise, $H_C$ consists of two
``funnels" and a path $\pi_C=\pi(a,b)\cap \pi(c,d)$ joining the two
apices of the two funnels, called the {\em corridor path} of $C$.
Each funnel side is also convex. We process all corridors as above.
The running time for processing all corridors is linear in terms of
the total number of vertices of all corridors, which is at most the
number of vertices of $\Tri(\calF)$, i.e., $O(n)$. Therefore, the
running time for processing all corridors is $O(n)$.

Let $Q$ be the union of all junction regions and hourglasses. Then
$Q$ consists of $O(h)$ junction regions, open hourglasses, funnels,
and corridor paths. Let $\pi(s,t)$ be a shortest $s$-$t$ path for
the original problem \spp. As shown in \cite{ref:KapoorAn97},
$\pi(s,t)$ must be contained in $Q$. Consider a corridor $C$. If
$\pi(s,t)$ contains an interior point of $C$ and neither $s$ nor $t$
lies on $\partial C$, then the path $\pi(s,t)$ must cross both doors
of $C$, i.e., it enters $C$ from one door and leaves $C$ from the
other. Further, if the hourglass $H_C$ for $C$ is closed, then the
corridor path of $C$ must be contained in $\pi(s,t)$. When $H_C$ is
open, since both sides of $H_C$ are convex with respect to the
interior of $H_C$, if $\pi(s,t)$ intersects both sides of $H_C$,
then it must contain a common tangent of the two sides such that
$\pi(s,t)$ goes from one side of $H_C$ to the other side via that
common tangent.

With all the properties above, let $Q'$ be $Q$ minus the corridor
paths. In other words, $Q'$ consists of $O(h)$ junction regions,
open hourglasses, and funnels. Since the sides of open hourglasses and
funnels are convex, the boundary of $Q'$ consists
of $O(h)$ reflex vertices and $O(h)$ convex chains, implying that
the complementary region $\calR\setminus Q'$ consists of a set of
splinegons of totally $O(h)$ reflex vertices and $O(h)$ convex
chains. Recall that $\calR$  is a large rectangle containing all
splinegons in $\calS$.  As in the polygonal domain case
\cite{ref:KapoorAn97}, we can partition $\calR\setminus Q'$ into a
set $\calS'$ of $O(h)$ convex splinegons with a total of
$O(n)$ vertices (e.g., by extending an angle-bisecting segment
inward from each reflex vertex) such that a shortest path $\pi(s,t)$
for the original \spp\ is also a shortest $s$-$t$ path that avoids the splinegons in $\calS'$ but possibly
contains some corridor paths. Therefore, other than the $O(h)$
corridor paths, we have reduced our original \spp\ problem to an
instance of the convex \spp. In fact, the key is to compute the free common tangents among the convex splinegons in $\calS'$.

\section{Computing the Relevant Visibility Graph of Convex Splinegons}
\label{sec:relevant}

In this section, we construct the relevant visibility graph $G$ for
the $O(h)$ convex splinegons in $\calS'$ with a total of $O(n)$ vertices.
For convenience, we slightly change the notation and consider the
following problem: Construct the relevant visibility graph $G$ for $h$
pairwise disjoint convex splinegons in a set $\calP=\{P_1,P_2,\ldots,P_h\}$
with a total of $n$
vertices. Let $\calB$ denote the set of all free common tangents of
$\calP$ and let $k=|\calB|$. Our algorithm for computing $\calB$
runs in $O(n+k+h\log h)$ time and $O(n)$ working space.
The graph $G$ can also be constructed in $O(n+k+h\log h)$ time.

Our algorithm can be viewed as a generalization of Pocchiola and
Vegter's algorithm \cite{ref:PocchiolaTo96} (and its preliminary
version \cite{ref:PocchiolaCo95}). We call it the {\em PV
algorithm}. Given a set $\calO$ of $n$ pairwise disjoint convex obstacles of $O(1)$
complexity each, the PV algorithm
computes all free common tangents of $\calO$ in
$O(n\log n+K)$ time and $O(n)$ space, where $K=O(n^2)$ is the number of all free common
tangents of the $n$ obstacles in $\calO$. It was also claimed in
\cite{ref:PocchiolaTo96} (without giving any details) that the PV
algorithm may be made to compute $\calB$ for our problem on $\calP$ in $O(n\log h+k)$ time.
It should be noted that although our improvement looks ``small'' (at most a logarithmic factor), it is theoretically quite meaningful because our algorithm is optimal.

In general, the PV algorithm relies mainly on the convexity of the
obstacles. The needed properties also hold for our setting. The
high-level scheme of our algorithm follows that of the PV algorithm,
but with certain modifications. However, the most challenging task is to
achieve an optimal time bound for the algorithm.
A similar analysis to the PV algorithm in \cite{ref:PocchiolaTo96} does not
work for our problem (or may only obtain an $O(n\log h+k)$ time
solution as claimed in \cite{ref:PocchiolaTo96}). As shown
later, our analysis needs numerous non-trivial new observations and novel
ideas. Comparing with the PV algorithm, our algorithm and
analysis explore more crucial properties and geometric structures of
the problem, which may be useful for solving other related problems
as well. Further, our modifications to the scheme of the PV
algorithm seem necessary (without them, it is not
clear to us whether the scheme in \cite{ref:PocchiolaTo96} can be generalized to
attain our optimal time bound).

\subsection{Preliminaries}
\label{subsec-termino}

We focus on showing how to compute $\calB$ since the graph $G$ can be
constructed simultaneously while $\calB$ is being computed.
In the following, we simply call each splinegon in
$\calP$ an {\em obstacle} and each (curved) splinegon edge an
{\em elementary curve}. Thus, the complexity of each elementary curve is $O(1)$.
We follow some terminology in \cite{ref:PocchiolaCo95,ref:PocchiolaTo96}.
We call a common tangent of two obstacles a {\em bitangent}.
For ease of exposition,
as in the PV algorithm \cite{ref:PocchiolaCo95,ref:PocchiolaTo96},
we assume all obstacles in $\calP$
are smooth (i.e., only one tangent line touches each boundary
point) and are in general position (i.e., no three obstacles share a
common tangent line). The algorithm can be easily generalized to the
general case. To handle the case with polygons,
for example, we can take the Minkowski sum of the polygons with an
infinitesimally small circle.

\begin{figure}[t]
\begin{minipage}[t]{0.48\linewidth}
\begin{center}
\includegraphics[totalheight=1.3in]{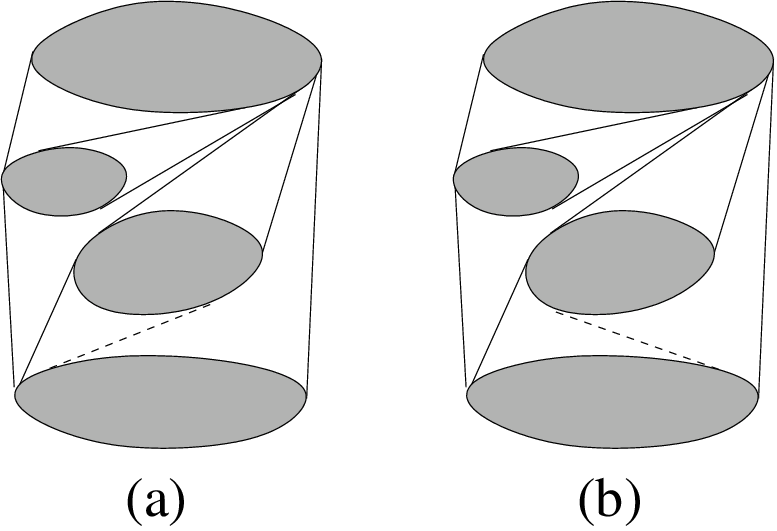}
\caption{\footnotesize A flip operation on a free bitangent $b$: (a)
The dashed bitangent is $b$ before the flip; (b) the dashed
bitangent is $\varphi(b)$ after the flip.} \label{fig:flip}
\end{center}
\end{minipage}
\hspace*{0.04in}
\begin{minipage}[t]{0.48\linewidth}
\begin{center}
\includegraphics[totalheight=1.0in]{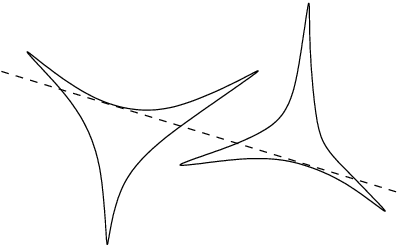}
\caption{\footnotesize The common tangent line of two
pseudo-triangles.} \label{fig:adjacenttriangle}
\end{center}
\end{minipage}
\vspace*{-0.15in}
\end{figure}

As in \cite{ref:PocchiolaCo95,ref:PocchiolaTo96}, we define
a {\em pseudo-triangulation} of the convex obstacles in $\calP$
as a subdivision of the free space induced
by a maximal number of pairwise noncrossing free bitangents (see
Fig.~\ref{fig:flip}).  The number of
free bitangents in any pseudo-triangulation of $\calP$ is $3h-3$
\cite{ref:PocchiolaCo95,ref:PocchiolaTo96}.

Let $\calT$ be a pseudo-triangulation and
$B(\calT)$ denote the set of all free bitangents that appear in $\calT$.
Any bounded free face $T$ in $\calT$ is a {\em
pseudo-triangle}, and the boundary of
$T$, denoted by $\partial T$, consists of three convex chains with
convexity towards the interior of $T$.
The three endpoints of the convex chains are called the {\em cusps} of $T$.
Denote by $B(\partial T)$ the set
of all free bitangents on $\partial T$ (i.e., each bitangent in
$B(\partial T)$ lies on $\partial T$).
For a point $p$ on $\partial T$ lying on an obstacle $P_i$, the
tangent line of $P_i$ at $p$ is called the {\em tangent line} of $T$
at $p$; for a point $p$ lying on a free bitangent of $B(\partial T)$, the line
containing the bitangent is the {\em tangent line} of $T$ at $p$.
Note that if a line is tangent to $T$ at a point $p\in \partial T$,
then the line is also tangent to the convex chain of $\partial T$ that
contains $p$. As shown in \cite{ref:PocchiolaCo95,ref:PocchiolaTo96},
any two pseudo-triangles in $\calT$
have a unique common tangent line,
i.e., a line tangent to both pseudo-triangles
(see Fig.~\ref{fig:adjacenttriangle}).
Suppose two adjacent pseudo-triangles $T$ and $T'$ in
$\calT$ share a bitangent $b\in
B(\calT)$; a {\em flip} operation on $b$ replaces $b$ by the
common tangent of $T$ and $T'$, which is a free bitangent and denoted by
$\varphi(b)$ (see Fig.~\ref{fig:flip}).
A flip operation produces another pseudo-triangulation.
If $b$ lies on the convex hull of $\calP$, then we let $\varphi(b)$ be $b$ itself.

Our algorithm for computing $\calB$, called the {\em topological flip algorithm},
performs flip operations based on a topological order, which can be
viewed as a generalization of a topological sweep
\cite{ref:EdelsbrunnerTo89}.
We define a partial order on
the free bitangents and a topological structure that is maintained by our algorithm.
To define the topological structure, below we assign directions to bitangents and tangent
lines of pseudo-triangulations, and discuss some properties.

Given a unit vector $u$, the {\em $u$-slope} of a
directed line (or segment) $l$ is defined as the
angle (in $[0,2\pi)$) of rotating $u$ counterclockwise to
the same direction as $l$.
For an undirected line (or segment) $l$, its {\em $u$-slope} is the angle
(in $[0,\pi)$) of rotating $u$ counterclockwise
to the first vector {\em parallel} to $l$; the direction of that
vector is said to be {\em consistent} with the $u$-slope of $l$.

Consider a pseudo-triangulation $\calT$.
Given a vector $u$, for each bitangent $b\in B(\calT)$, we assign
to $b$ the direction consistent with the $u$-slope of $b$.
For every pseudo-triangle $T$ of $\calT$,
let $b_T$ be the bitangent in $B(\partial T)$ with
the minimum $u$-slope.  Further, for each point $p \in \partial T$,
we assign to the tangent line $l(p)$ of $T$ at $p$ the
direction consistent with the $b_T$-slope of $l(p)$, and call $l(p)$
the {\em directed
tangent line} of $T$ at $p$. The $b_T$-slope of the directed $l(p)$ is also called
the {\em pseudo-triangle slope} (or {\em pt-slope} for short)
of $l(p)$ at the point $p \in \partial T$ \footnote{If $p$ is on a bitangent, then $p$ is also contained in another pseudo-triangle $T'$, in which case the pt-slope thus defined is with respect to $T$ only and $p$ also has another pt-slope defined with respect to $T'$. }.
For any bitangent $b\in \calB$, suppose we assign a direction to $b$
and $p$ is an endpoint of $b$ (say, $p$ is on $\partial T$ of a pseudo-triangle $T$);
then the direction
assigned to $b$ is said to be {\em compatible} with $p$ if the
directed tangent line of $T$ at $p$ has the same direction as $b$.
Note that the pt-slope of any point on $b_T$ is zero.
As moving on $\partial T$ clockwise
from $b_T$, the pt-slope of the moving point
increases continuously from $0$ to $\pi$,
until we are back to $b_T$.

Our algorithm maintains a topological structure called
{\em good pseudo-triangulation}, defined as follows.
A pseudo-triangulation $\calT$ is said to be {\em good}
(called {\em weakly greedy} in \cite{ref:PocchiolaCo95}) if
there is a way to assign every free bitangent $b\in \calB$ a direction such
that a partial order $\prec$ can be defined on the
directed bitangents of $B(\calT)$ with the following properties:
(1) For each pseudo-triangle $T$ in $\calT$, the partial order $\prec$ is a total
order, which corresponds to the pt-slope order on $B(\partial T)$
with respect to $b_T$; (2) the direction of each bitangent $b\in \calB$
is compatible with both its
endpoints; (3) for any bitangent $b\in\calB \setminus
B(\calT)$, all bitangents in
$B(\calT)$ intersecting $b$ cross the directed $b$ from left to right.

\subsection{Initialization and Outline of the Topological Flip Algorithm}
\label{subsec-intial}

Let $u_0$ be a vector with the direction of the positive $x$-axis.
We first compute an initial pseudo-triangulation $\calT_0$
of $\calP$ induced by a set
$\{b_1,b_2,\ldots,b_{3h-3}\}$ of free (undirected) bitangents such that (1) $b_1$
is the bitangent in $\calB$ with the smallest $u_0$-slope, and
(2) for any $1\leq i<3h-3$, $b_{i+1}$ is the bitangent with
the smallest $u_0$-slope in $\calB$ that does not cross any of
$b_1,b_2,\ldots,b_i$ (e.g., Fig.~\ref{fig:flip}(a) is $\calT_0$). As shown in
\cite{ref:PocchiolaCo95,ref:PocchiolaTo96}, $\calT_0$ for the obstacle set $\calO$
is a good pseudo-triangulation, and can be built in $O(n\log n)$ time.
Likewise, for our problem, $\calT_0$ of $\calP$ is also
a good pseudo-triangulation; further, we can compute $\calT_0$ even
faster, as shown in Lemma \ref{lem:15}.

\begin{lemma}\label{lem:15}
The initial good pseudo-triangulation $\calT_0$
of $\calP$ can be constructed in $O(n+h\log h)$ time.
\end{lemma}
\begin{proof}
To construct $\calT_0$ of $\calP$, we modify and generalize the
corresponding $O(n\log n)$ time
algorithm in \cite{ref:PocchiolaTo96} for the set $\calO$.
The algorithm is based on a rotational sweeping procedure, during which a {\em
visibility map}, denoted by $M(u')$ associated with the current rotational
direction $u'\in [0,\pi]$, is (implicitly) maintained. Consider a direction $u'$.
Each obstacle $P_i$ of $\calP$ contains two {\em extreme points} each
having a tangent line with
slope $u'$ such that $P_i$ is between these two tangent lines.
We denote by $V(u')$ the set of such extreme points in all obstacles of
$\calP$.

We first define $M(0)$ for $u'=0$ as follows (see Fig.~\ref{fig:startPos}).
For each extreme point in $V(u')$, we shoot one ray in the direction of $u'$
and shoot another ray in the opposite direction until hitting some
obstacles. The subdivision of the plane defined by
all these rays and the obstacles of $\calP$ is $M(0)$, which can be viewed
as similar to a trapezoidal decomposition of the free space.
In general, for any $u'>0$, $M(u')$ is defined by the rays shooting from the points
of $V(u')$ in the direction of $u'$ and in the opposite direction
until hitting some obstacles or bitangents of $\calT_0$ obtained up to
that moment
of the rotational sweep, together with the obstacles of $\calP$.

\begin{figure}[t]
\begin{minipage}[t]{0.32\linewidth}
\begin{center}
\includegraphics[totalheight=1.0in]{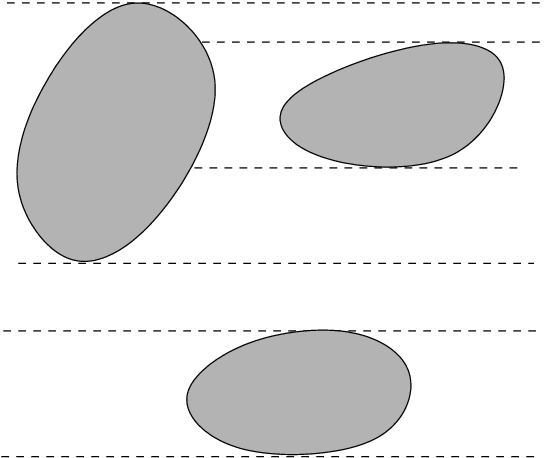}
\caption{\footnotesize
The visibility map $M(0)$.}
\label{fig:startPos}
\end{center}
\end{minipage}
\hspace*{0.02in}
\begin{minipage}[t]{0.32\linewidth}
\begin{center}
\includegraphics[totalheight=1.0in]{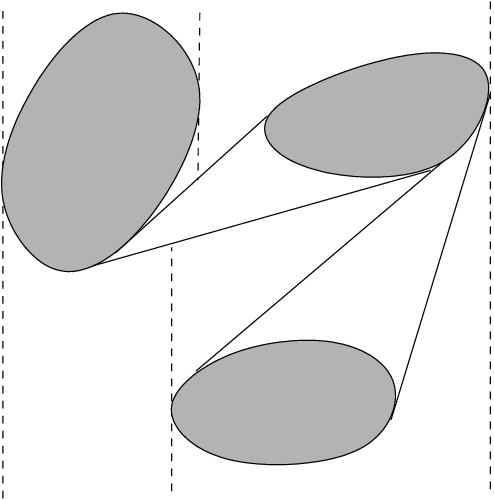}
\caption{\footnotesize
The visibility map $M(\pi/2)$.}
\label{fig:midPos}
\end{center}
\end{minipage}
\hspace*{0.02in}
\begin{minipage}[t]{0.32\linewidth}
\begin{center}
\includegraphics[totalheight=1.0in]{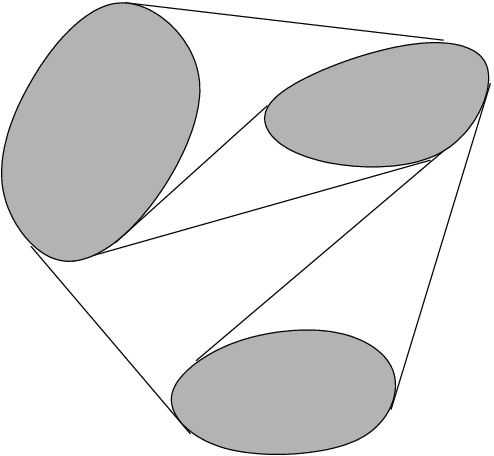}
\caption{\footnotesize
The visibility map $M(\pi)$.}
\label{fig:finalPos}
\end{center}
\end{minipage}
\vspace*{-0.15in}
\end{figure}

The algorithm first constructs $M(0)$. For this, we compute
the extreme point set $V(0)$, which takes
$O(n)$ time. Since the obstacles of $\calP$ are pairwise
disjoint, with a standard sweeping algorithm (from top to bottom),
$M(0)$ can be easily constructed in $O(n+h\log h)$ time.

Starting at $M(0)$, we rotate $u'$ from $0$ to $\pi$. During the
rotation, the topology of $M(u')$ is maintained implicitly. Specifically,
the topology of $M(u')$ does not change until $u'$ becomes equal to
the slope of a free bitangent $b$ of $\calT_0$.
When a new free bitangent $b$ is detected,
a ``quadrangular" region (which contains two points of $V(u')$)
in $M(u')$ will disappear, and some rays shooting
from $V(u')$ will first hit $b$ instead of some obstacles or other
free bitangents of $\calT_0$ already found (we
then view these rays as hitting $b$ without going further). At the
same time, a ``triangular" region (which contains only one point of
$V(u')$) will emerge (refer to \cite{ref:PocchiolaTo96} for the details). If two triangular regions contain a same point
of $V(u')$ at their boundaries, then they are incident along a ray
shooting from this point, and we merge these two regions by removing
this ray. The resulting new visibility map is $M(u')$ with
the newly detected free bitangent $b$ of $\calT_0$. We keep rotating $u'$ in this manner.
Fig.~\ref{fig:midPos} illustrates $M(u')$ for $u'=\pi/2$ and Fig.~\ref{fig:finalPos} illustrates $M(u')$ for $u'=\pi$.
Note that every new free bitangent is detected from two
obstacles along the boundary of a quadrangular region.
In this way, after the rotation of $u'$ is over, $\calT_0$
is obtained.

The key to this procedure is to determine the rotation
events, i.e., which free bitangent will be encountered
next in the rotation. For this, the same strategy of the original ($O(n\log n)$
time) algorithm for $\calO$ in
\cite{ref:PocchiolaTo96} is applied. The only difference is that
a bitangent of two $O(1)$ complexity obstacles in \cite{ref:PocchiolaTo96} is found
in $O(1)$ time, while in our problem we compute each free
bitangent of $\calT_0$ in $O(\log n)$ time. Since $\calT_0$ has $O(h)$ free
bitangents, computing all of them during the rotational sweep takes
$O(h\log n)$ time. Note that $h\log n=O(n+h\log h)$.

In summary, constructing $\calT_0$ takes $O(n+h\log h)$ time.
The lemma thus follows.
\end{proof}

After computing $\calT_0$, we assign to every bitangent $b\in B(\calT_0)$
the direction consistent with its $u_0$-slope (in $[0,\pi)$).
For each bitangent $b\in \calB\setminus B(\calT_0)$, as shown
in \cite{ref:PocchiolaCo95,ref:PocchiolaTo96}, we can always assign a
direction to $b$ that is compatible with both $b$'s
endpoints; for the purpose of discussion,
we assume that this direction has been assigned to $b$ (the algorithm
does not explicitly perform this assignment).  Let $\calT=\calT_0$.
A (directed) bitangent $b\in B(\calT)$ is {\em minimal} if it has the
smallest $u_0$-slope among all free bitangents on the boundaries of
both the left and right adjacent pseudo-triangles of $b$ in $\calT$.
A minimal bitangent always exists in a good pseudo-triangulation
$\calT$ \cite{ref:PocchiolaCo95,ref:PocchiolaTo96}.
To compute the bitangents in $\calB\setminus B(\calT_0)$,
the topological flip algorithm keeps flipping a minimal
bitangent in $B(\calT)$ and generating another
good pseudo-triangulation of $\calP$.
The next lemma (proved in
\cite{ref:PocchiolaCo95} and applicable to our problem)
shows that any minimal bitangent in $B(\calT)$ can be flipped.

\begin{lemma}\label{lem:40}{\em \cite{ref:PocchiolaCo95}}
For any good pseudo-triangulation $\calT$ of $\calP$
(initially, $\calT=\calT_0$), let $b$ be a minimal
free bitangent in $B(\calT)$ with a $u_0$-slope less than $\pi$. Then
the pseudo-triangulation $\calT'$ of $\calP$ obtained by flipping $b$ is also
good, and the assigned direction of any free bitangent $t\in \calB \setminus \{b\}$
does not change after the flip of $b$, whereas the direction of $b$
is reversed.
\end{lemma}

As shown in \cite{ref:PocchiolaCo95,ref:PocchiolaTo96},
we only need to flip a minimal
bitangent in $B(\calT)$ with a $u_0$-slope less than $\pi$,
and this ensures that
the algorithm will terminate and all free bitangents in $\calB$ will be generated.
Note that once a bitangent is flipped, since its direction is
reversed, its $u_0$-slope becomes no smaller than $\pi$, and thus
it will never be flipped again.

The effectiveness of our algorithm hinges on its
ability to perform the $k$ flips in $O(n+k)$ time, and the key is
to determine a minimal bitangent $b^*$ of $\calT$ efficiently.
To this end, for each pseudo-triangle $T$ of $\calF$, we will
choose and store a critical portion of its boundary,
as $Awake[T]$, which is used to find $b^*$.
After obtaining $b^*$ and a new good pseudo-triangulation, we
need to update $Awake$ for some (two) new pseudo-triangles induced by the flip.
To update
$Awake$ efficiently, we also choose and store a boundary portion of each
pseudo-triangle $T$, as $Asleep[T]$. In other words, $Awake$ is used to find
$b^*$ and $Asleep$ is used to update $Awake$ ($Asleep$
itself also needs to be updated). A key difference between the PV
algorithm and ours is that $Asleep[T]$ refers to different portions of
$T$'s boundary.

Both $Awake$ and $Asleep$ are implemented as ``splittable queues''
\cite{ref:PocchiolaTo96} that
support three operations: enqueue, dequeue, and split.
Our algorithm uses two phases for handling each flip: Phase I computes
$b^*$; Phase II updates $Awake$ and $Asleep$. To bound the
running time, it suffices to prove the following {\bf key claim}:
The total number of enqueue operations for all $k$ flips is $O(n+k)$.
Actually, for each flip, only $O(1)$
sequences of enqueue operations are needed and each
sequence involves either a free common tangent or
a boundary portion of a single obstacle.

A main difference between the PV algorithm and ours is on proving
the key claim.  In the PV algorithm, it is fairly easy: Since
every obstacle is  of $O(1)$ complexity, each enqueue sequence
needs only $O(1)$ enqueue operations. Our problem is more challenging as
the complexity of the boundary of an obstacle (i.e., a splinegon)
can be $\Omega(n)$ and thus an enqueue
sequence may take as many as $\Omega(n)$ enqueue operations.
To prove the key claim, we must
conduct a global analysis that requires many new observations and
analysis techniques, which is the most challenging part.
Section \ref{sec:keylemma} is devoted entirely to this task.

\subsection{Conducting the Flips}
\label{sec:conducting}

Given $\calT_0$, we determine the set of minimal bitangents in $B(\calT_0)$,
denoted by $C$.  Then, we take an arbitrary bitangent $b$
from $C$, flip $b$, and update $C$. We repeat this process until
$C = \emptyset$ (by then $\calB$ is obtained and all bitangents in the resulting pseudo-triangulation have $u_0$-slope at least $\pi$). The key is to perform all
$k$ ($=|\calB|$) flips in $O(n+k)$ time.

Let $\calT$ be a good pseudo-triangulation. For any bitangent $t$
in $B(\calT)$, denote by $Ltri(t)$ (resp., $Rtri(t)$)
the pseudo-triangle of $\calT$ (if any) that is bounded by the directed $t$ and is on the
left (resp., right) of $t$.
Suppose we are about to flip a minimal bitangent $b$ in $\calT$. Let $R=Rtri(b)$ and
$L=Ltri(b)$. To compute $\varphi(b)$, an easy way is to walk clockwise
along $\partial R$ and
$\partial L$ synchronously, starting from $b$,
until finding $\varphi(b)$. But, this is too
expensive. A more efficient approach is to first ``jump" to a certain location on
$\partial R$ and $\partial L$ and then do the synchronous walking.
To implement this idea, we need some ``crucial points" on $\partial T$ for
each pseudo-triangle $T\in \calT$, as defined below.

For any directed free bitangent $b$, we denote its two endpoints
by $Tail(b)$ and $Head(b)$, respectively, such that $b$'s direction
is from $Tail(b)$ to $Head(b)$, and call them {\em tail} and {\em head} of $b$.

\begin{figure}[t]
\begin{minipage}[t]{\linewidth}
\begin{center}
\includegraphics[totalheight=1.5in]{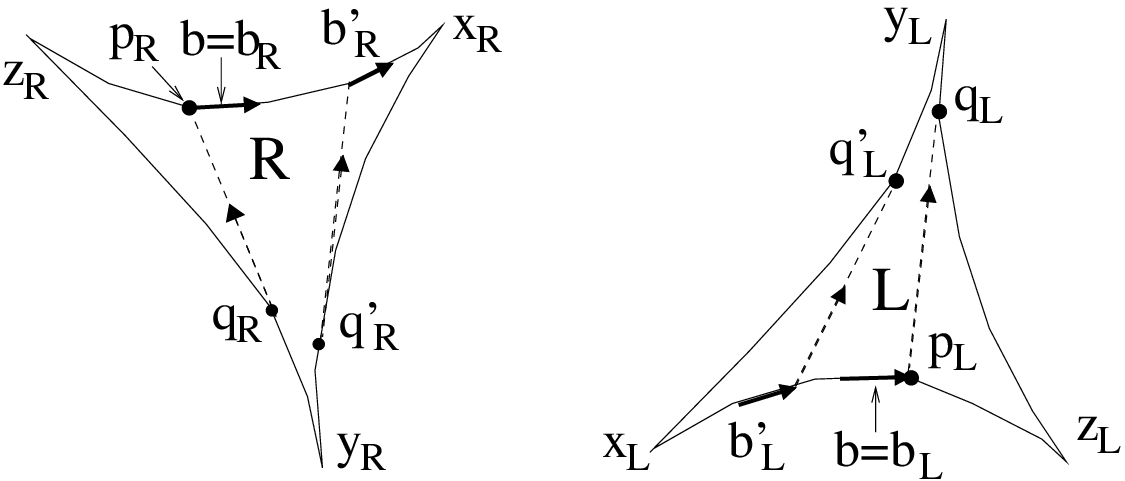}
\caption{\footnotesize Left: $R=Rtri(b)$ and $b=b_R$; the {\em awake}
points are those on $\widehat{x_Rq_R}$; to flip $b$, the walk on
$\partial R$ starts from $q'_R$ for Case 1 and from $x_R$ for Cases 2
and 3. Right: $L=Ltri(b)$ and $b=b_L$; the {\em awake}
points are those on $\widehat{x_Lq_L}$; to flip $b$, the walk on
$\partial L$ starts from $q'_L$ for Case 1 and from $x_L$ for Cases 2
and 3.} \label{fig:awake}
\end{center}
\end{minipage}
\vspace*{-0.15in}
\end{figure}

Consider a pseudo-triangle $T\in \calT$.
We define the {\em basepoint} of $T$, denoted by $p_T$,
to be the {\em tail} of $b_T$ (i.e., the smallest $u_0$-slope bitangent in
$B(\partial T)$) if $T=Rtri(b_T)$, and the {\em head} of $b_T$ if $T=Ltri(b_T)$ (e.g., see Fig.~\ref{fig:awake}).
Starting at $p_T$, if we move along $\partial T$ clockwise,
the successive cusps of $T$ encountered are denoted by $x_T$, $y_T$, and
$z_T$ (if $p_T$ is a cusp, we let it be $z_T$). The {\em forward}
(resp., {\em backward}) {\em $T$-view} of any point $p$ on $\partial T$ is
the intersection point of $\partial T$ with the directed tangent line $l(p)$
of $T$ at $p$, i.e., lying ahead of $p$ (resp., behind $p$).
Let $q_T$ denote the special point on $\partial T$
whose forward (resp., backward) $T$-view is $p_T$ if $T=Rtri(b_T)$
(resp., $T=Ltri(b_T)$) (e.g., see Fig.~\ref{fig:awake}). For any two points
$p_1$ and $p_2$ on $\partial T$, let $\widehat{p_1p_2}$ denote the
portion of $\partial T$ from $p_1$ {\em clockwise} to $p_2$.

For a pseudo-triangle $T$, a point $p\in \partial T$ is said to be {\em
awake} if and only if $p\in \widehat{x_Tq_T}$ (see Fig.~\ref{fig:awake}). We let $Awake[T]$
represent the awake portion $\widehat{x_Tq_T}$ of $\partial T$.
Also, we let $Asleep[T]$ represent the portion $\widehat{w_T p_T}$ of $\partial T$,
where the point $w_T= q_T$ if $q_T\in \widehat{y_Tz_T}$ and
$w_T= y_T$ if $q_T\not\in \widehat{y_Tz_T}$.
Note that our definition of $Asleep[T]$ is different from that in
\cite{ref:PocchiolaCo95,ref:PocchiolaTo96}, in which $Asleep[T]$ represents
$\widehat{w_Tz_T}$. Comparing with the algorithm in \cite{ref:PocchiolaCo95,ref:PocchiolaTo96}, our modification on defining $Asleep[T]$ seems necessary for obtaining the optimal algorithm, and without the modification, it is not clear to us whether the algorithmic scheme in \cite{ref:PocchiolaCo95,ref:PocchiolaTo96} can be generalized to attain our optimal time bound.

To perform a flip on $b$, we use $Awake[T]$ to find $\varphi(b)$
and use $Asleep[T]$ to help update $Awake[T]$.
After every flip, $Awake[T]$ and $Asleep[T]$
are updated accordingly. Both $Awake[T]$ and
$Asleep[T]$ are stored as {\em splittable queues},
which support three types of operations on a list: (1) {\em enqueue} an atom, either at the
head or the tail of the list; (2) {\em dequeue} the head or the tail of the list;
(3) {\em split} the list at an atom $x$, which is preceded by a
{\em search} for $x$ in the list.
An {\em atom} can be a bitangent or an elementary
curve. Further, an elementary curve may be divided into multiple pieces by
the endpoints of some free bitangents in a good pseudo-triangulation, in which
case an atom may be only a portion of an elementary curve. In any case,
the complexity of each atom is $O(1)$.
A data structure for splittable queues was given in
\cite{ref:PocchiolaCo95,ref:PocchiolaTo96}, whose performance is shown below.

\begin{lemma}\label{lem:60}{\em \cite{ref:PocchiolaCo95,ref:PocchiolaTo96}}
A sequence of $O(n+k)$
enqueue, dequeue, and split operations on a collection of $n$
initially empty splittable queues can be performed in $O(n+k)$ time.
\end{lemma}

For a pseudo-triangle $T$, a
portion of $\partial T$ is called an {\em obstacle arc} if that
portion lies entirely on the boundary of a certain obstacle.

Initially, we compute $Awake[T]$ and $Asleep[T]$ for each
pseudo-triangle $T$ in $\calT_0$,
using $O(n)$ enqueue operations.
Consider a flip on a minimal bitangent $b$
in the current good pseudo-triangulation $\calT$. Let $R=Rtri(b)$ and
$L=Ltri(b)$. Let $b^*=\varphi(b)$, $p^*=Tail(b^*)$, and $q^*=Head(b^*)$.
Let $\calT'$ be the resulting pseudo-triangulation after the flip.
Let $R'=Rtri(b^*)$ and $L'=Ltri(b^*)$.
To maintain the minimal bitangents in
$\calT'$, we need to find the bitangent with the smallest $u_0$-slope in $B(\partial
T)$ (i.e., $b_{T}$), for each $T\in \{R',L'\}$.
Below, we only discuss the case for $R'$
(the case for $L'$ is similar).

\begin{figure}[t]
\begin{minipage}[t]{\linewidth}
\begin{center}
\includegraphics[totalheight=1.4in]{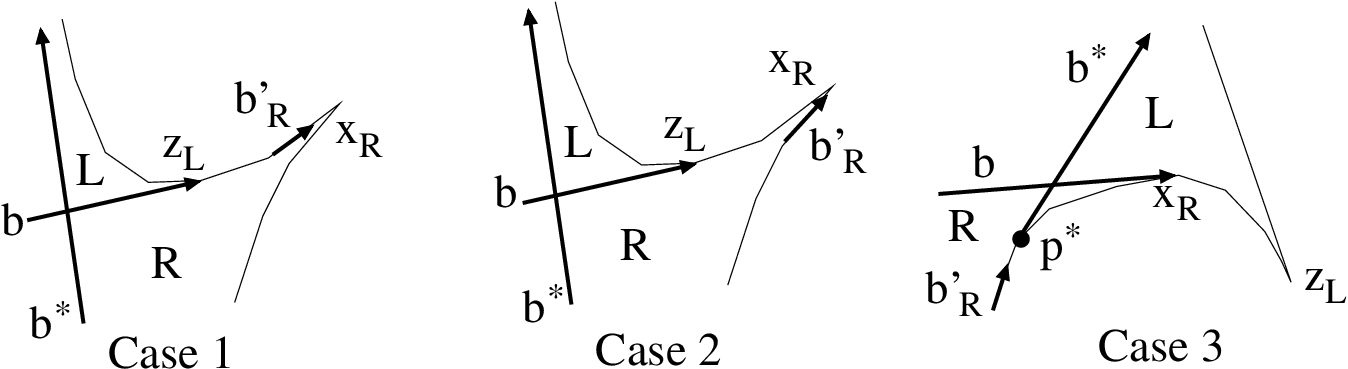}
\caption{\footnotesize The three possible cases when flipping $b$;
$b^*=\varphi(b)$ and $b'_R$ is the next bitangent of $b$ along
$\partial R$ clockwise.}
\label{fig:maincases}
\end{center}
\end{minipage}
\vspace*{-0.15in}
\end{figure}

Let $b'_R$ be the next bitangent of $b$ in $B(\partial R)$
clockwise along $\partial R$. Notice that $b'_R$ is the bitangent in
$B(\partial R)\setminus\{b\}$ with the minimum $u_0$-slope.
Let $\Gamma_R$ be the portion of $\partial R$
from $Head(b)$ clockwise to the first encountered
point of $b'_R$. Clearly, $\Gamma_R$ is an obstacle arc, and
$b_{R'}$ is one of the two bitangents $b'_R$ and $b^*$
(the one with a smaller $u_0$-slope).
As in \cite{ref:PocchiolaCo95,ref:PocchiolaTo96}, there are three
main cases (see Fig.~\ref{fig:maincases}).

{\em Case 1:} $b$ and $b'_R$ are not separated by the cusp $x_R$ of $R$ (i.e.,
$\Gamma_R$ does not contain $x_R$). Then
$R'$ ($=Rtri(b^*)$) is also $Rtri(b'_R)$ and $p^*$ does not lie on $\Gamma_R$.
Thus, $b_{R'}$ is $b'_R$.

{\em Case 2:} $b$ and $b'_R$ are separated by $x_R$
and $p^*$ does not lie on $\Gamma_R$.
Then $b_{R'}$ is $b'_R$. In this case, $x_{R}$ is either $Head(b'_R)$ (see
Fig.~\ref{fig:maincases}) or $Head(b)$. As in
\cite{ref:PocchiolaCo95,ref:PocchiolaTo96} and shown later, $R'$ is
also $Ltri(b'_R)$.

{\em Case 3:} $b$ and $b'_R$ are separated by $x_R$ and $p^*$
lies on $\Gamma_R$. Then, $b_{R'}$ is $b^*$ (e.g., see
Fig.~\ref{fig:maincases}).

In the following, we divide the processing of a flip on $b$ into two
phases as in \cite{ref:PocchiolaCo95,ref:PocchiolaTo96}:
Phase I finds $b^*$; Phase II updates
$Awake$ and $Asleep$ accordingly (i.e., computing $Awake[R']$,
$Awake[L']$, $Asleep[R']$, and $Asleep[L']$).
We only discuss the case for $R'$ (the case for $L'$ is similar).

\subsection{Phase I of a Flip Operation}
\label{sec:PhaseI}

Let $q_R'$ be the point on $\partial R$ whose forward $R$-view is
$Tail(b_R')$ (see Fig.~\ref{fig:awake}); similarly, let $q_L'$ be
the point on $\partial L$ whose backward $L$-view is $Head(b_L')$
(similar to $b_R'$, $b_L'$ is the next bitangent of $b$ in
$B(\partial L)$ clockwise along $\partial L$).  To compute
$b^*=\varphi(b)$, Cases 2 and 3 are handled in the same way while
Case 1 is different. For now, we assume that we already know whether
it is Case 1, and further, when Case 1 occurs, we know $b'_R$. We
will show later how to detect the cases and find $b'_R$ for Case 1.
An easy but useful observation is that $p^*=Tail(b^*)$ must be on
$\widehat{x_Rq_R}$ ($=Awake[R]$), and in Case 1, $p^*$ is even on
$\widehat{q'_Rq_R}$. Similarly,
$q^*=Head(b^*)$ lies on $\widehat{x_Lq_L}$ ($=Awake[L]$), and even
on $\widehat{q'_Lq_L}$ in Case 1. To compute $b^*$, we walk
clockwise along $\partial R$ and $\partial L$ synchronously, as
follows. In Cases 2 and 3, the walk on $\partial R$ starts at $x_R$;
in Case 1, we first {\em split} $Awake[R]$ at $q'_R$ and then start
walking from $q'_R$. The split operation on $Awake[R]$ at $q'_R$ in
Case 1 is preceded by a search for $q'_R$ in $Awake[R]$, which is
guided by the position of $Tail(b_R')$ with respect to the directed
tangent lines of $R$ at the endpoints of the atoms in $Awake[R]$.
Similar things are done on $\partial L$.

We perform the following three main steps for computing the
endpoints $p^*$ and $q^*$ of $b^*$.

Step (1): For Case 1, we {\em split} $Awake[R]$ ($=\widehat{x_Rq_R}$) at $q'_R$ into
$AwakeMin[R]$ and $AwakeMax[R]$ such that the atoms in the former
queue have the smaller pt-slopes; for Cases 2 and 3, we simply let
$AwakeMin[R]=\emptyset$ and $AwakeMax[R]=
Awake[R]$. Thus, in any case, $p^*$ always lies on an atom in
$AwakeMax[R]$. We compute $AwakeMax[L]$ and $AwakeMin[L]$ similarly.

Step (2): Compute $b^*$. To do so, the synchronous walks on $\partial
R$ and $\partial L$ can be implemented by {\em dequeuing} atoms from
$AwakeMax[R]$ and $AwakeMax[L]$, until $p^*$ and $q^*$ are found.

Step (3): Note that the atom at one end of $AwakeMax[R]$ now contains
$p^*$. We cut that atom at $p^*$ by setting $p^*$ as an {\em
endpoint} such that $AwakeMax[R]$ now represents the portion
$\widehat{p^*q_R}$. We do similar things for $AwakeMax[L]$ and $q^*$
($AwakeMax[L]$ thus represents the portion $\widehat{q^*q_L}$).

These steps can be done at the cost of at most two split operations
(at $Awake[R]$ and $Awake[L]$), followed by multiple successive
dequeue operations, but no enqueue operation is needed at all.

To determine whether Case 1 occurs and find $b_R'$ in Case 1, we use
the following {\em preparing procedure}.
We walk from $Head(b)$ along $\partial R$ clockwise until first
encountering either $b'_R$ or $x_R$. If we encounter $b'_R$ first,
then it is Case 1 and we have $b'_R$ as well. Otherwise, it is Case 2 or
Case 3. In addition, once $b^*$ is computed, whether it is Case 2 or Case 3 can be immediately determined.

This finishes the description of Phase I.

\subsection{Phase II of a Flip Operation}
\label{subsubsec-Phase-II}

In Phase II, our task is to compute $Awake[T]$ and $Asleep[T]$
for each $T\in \{R',L'\}$. Specifically, after Phase II, $Awake[T]$ should
represent $\widehat{x_Tq_T}$ of $\partial T$ and $Asleep[T]$ should
represent $\widehat{w_Tp_T}$ of $\partial T$, where $w_T= q_T$ if
$q_T\in\widehat{y_Tz_T}$ and $w_T=y_T$ otherwise.
Note that since our definition of $Asleep$ is
different from that for the PV algorithm \cite{ref:PocchiolaCo95,ref:PocchiolaTo96},
our algorithmic procedures in Phase II also differ from those in
the PV algorithm.

Recall that after Phase I, $AwakeMax[R]$ (resp.,
$AwakeMax[L]$) represents $\widehat{p^*q_R}$ (resp.,
$\widehat{q^*q_L}$) in all three cases. In Case 1, $AwakeMin[R]$ (resp.,
$AwakeMin[L]$) represents $\widehat{x_Rq'_R}$ (resp.,
$\widehat{x_Lq'_L}$), and in Cases 2 and 3, $AwakeMin[R]=AwakeMin[L]=\emptyset$.
Recall that $p^*\in \widehat{x_Rq_R}$ and $q^*\in \widehat{x_Lq_L}$.

We only show how to compute $Awake[R']$ and $Asleep[R']$.
The case for $L'$ can be handled similarly. We discuss for Cases 1, 2, and 3
individually. No split operation is needed in Phase II.
Note that after computing $b^*=\varphi(b)$ in Phase I, as in the PV
algorithm, the six new cusps, i.e.,
$x_T$, $y_T$, and $z_T$ for $T\in \{R',L'\}$, can be determined in
$O(1)$ time.  Hence, we assume all of them are already known.

Recall that a splittable queue, e.g., $Awake[T]$ for a
pseudo-triangle $T$, represents
a list of consecutive atoms on $\partial T$. We define the {\em
head} and {\em tail} of the queue such that if moving from the
head to the tail along the list, it is clockwise on $\partial T$.
Note that the splittable queue allows to
enqueue an atom at either the head or the tail of the queue.
Recall that $p_T$ is the basepoint of $T$.
For two points $p_1$ and $p_2$
on $\partial T$, if $p_1\in \widehat{p_Tp_2}$, we say $p_1$ is {\em
before} $p_2$; otherwise, $p_1$ is {\em after} $p_2$ (when $p_1=p_2$,
        $p_1$ is both before and after $p_2$).
In the following discussion, when a point $p$ is on both $\partial R'$
and $\partial L$ (resp., $\partial R$), we sometimes do not differentiate
whether $p$ is on $\partial R'$ or $\partial L$ (resp., $\partial R$).

\begin{algorithm}[h]
\caption{Construction of $Awake$ and $Asleep$ of $R'$ for Case 1}
\label{algo:case1}
\SetAlgoNoLine
\KwOut{$Awake[R']=\widehat{x_{R'}q_{R'}}$ and $Asleep[R']=\widehat{w_{R'}p_{R'}}$}
\BlankLine
$Awake[R']\leftarrow AwakeMin[R]$ \tcc*[l]{done for $Awake[R']$}
$Asleep[R']\leftarrow AwakeMax[L]$ \;
Enqueue $\widehat{z_Lp_{R'}}$ to the tail of $Asleep[R']$
\tcc*[l]{charge the enqueue to $Q$}\label{line:case110}
Enqueue $b^*$ to the head of $Asleep[R']$
\tcc*[l]{charge the enqueue to $\calB$}\label{line:case120}
\eIf{$y_{R'}=p^*$}{Do nothing \;}
(\tcc*[h]{$y_{R'}=y_R$}){Starting at $p^*$, enqueue atoms
along $\partial R'$ counterclockwise to the head of $Asleep[R']$
until either $q'_R$ or $y_R$ is encountered \tcc*[l]{charge the enqueue to $D$}
\label{line:case130}
}
\end{algorithm}

\subsubsection{Case 1}

In Case 1, recall that $b_{R'}=b'_R$, $R'=Rtri(b'_R)$,
$p_{R'}=Tail(b'_R)$, and $q_{R'}=q'_R$. Note that $x_{R'}=x_R$,
$y_{R'}$ is $y_R$ or $p^*$, and $z_{R'}$ is $y_L$ or $q^*$. The
pseudocode is in Algorithm \ref{algo:case1}. (Note that every
``enqueue" in the pseudocode is commented by ``charge the enqueue to
...". The comments are used for analysis in Section
\ref{sec:PhaseII} and can be ignored at the moment.) The details
are discussed below.

The goal of computing $Awake[R']$ is to let it represent
$\widehat{x_{R'}q_{R'}}$. Since in Case 1, $AwakeMin[R]=
\widehat{x_Rq'_R}$, and $x_{R'}=x_R$ and $q_{R'}=q'_R$, by
setting $Awake[R']= AwakeMin[R]$, we are done.

The goal of computing $Asleep[R']$ is to let it represent
$\widehat{w_{R'}p_{R'}}$. Note that in Case 1, $AwakeMax[L]$ $=
\widehat{q^*z_{L}}$ (since $q_L=z_L$) and $p^*\in \widehat{q'_Rq_R}$.
Thus $p^*$ is after $q'_R$ ($=q_{R'}$). Regardless of whether
$y_{R'}$ is $y_R$ or $p^*$, $p^*$ is after $y_{R'}$. Thus, $p^*$
is after $w_{R'}$, which is either $q_{R'}$ or $y_{R'}$.
Since $z_L$ is $Head(b)$, $z_L\in \widehat{z_{R'}p_{R'}}$ holds. Hence, $b^*$ is asleep on $\partial R'$,
and $\widehat{q^*z_{L}}$, which is stored in $AwakeMax[L]$, is
part of $Asleep[R']$. Thus, we first set $Asleep[R']= AwakeMax[L]$.
Since $p_{R'}=Tail(b'_R)$, we
enqueue $\widehat{z_Lp_{R'}}$ to the tail of $Asleep[R']$.
Now $Asleep[R']$ represents $\widehat{q^*p_{R'}}$.
Since $b^*$ is asleep, we enqueue $b^*$ to the head of $Asleep[R']$.
Now $Asleep[R']$ represents $\widehat{p^*p_{R'}}$.
Note that $y_{R'}$ is $p^*$ or $y_R$.

\begin{itemize}
\item
If $y_{R'}= p^*$, then since $p^*$ is after $q'_R=q_{R'}$,
$w_{R'}=y_{R'}=p^*$. Thus we are done with computing $Asleep[R']$.

\item
If $y_{R'}= y_R$, then starting at $p^*$, we enqueue atoms
along $\partial R'$ {\em counterclockwise} to the head of $Asleep[R']$
until we first encounter $q'_R$ or $y_R$. Note that the first point of
$q'_R$ ($=q_{R'}$) or $y_{R}$ ($=y_{R'}$) thus encountered is $w_{R'}$.
\end{itemize}

This completes the construction of $Asleep[R']$.

\begin{algorithm}[h]
\caption{Construction of $Awake$ and $Asleep$ of $R'$ for Case 2.1}
\label{algo:case21}
\SetAlgoNoLine
\KwOut{$Awake[R']=\widehat{x_{R'}q_{R'}}$ and $Asleep[R']=\widehat{w_{R'}p_{R'}}$}
\BlankLine
$Asleep[R']\leftarrow\emptyset$ \tcc*[l]{done for $Asleep[R']$}
$Awake[R']\leftarrow AwakeMax[L]$ \;
Enqueue $\widehat{z_Lx_R}$ of $\partial R$ to the tail of
$Awake[R']$ \tcc*[l]{charge the enqueue to $Q$}\label{line:case210}
Enqueue $b^*$ to the head of $Awake[R']$
\tcc*[l]{charge the enqueue to $\calB$}\label{line:case220}
\eIf{$x_{R'}=p^*$}{Do nothing \;}
(\tcc*[h]{$x_{R'}=y_R$}){Enqueue $\widehat{y_Rp^*}$ to the head of
$Awake[R']$ \tcc*[l]{charge enqueue to $D$}\label{line:case230}}
\end{algorithm}

\subsubsection{Case 2}

\begin{figure}[h]
\begin{minipage}[t]{0.49\linewidth}
\begin{center}
\includegraphics[totalheight=1.5in]{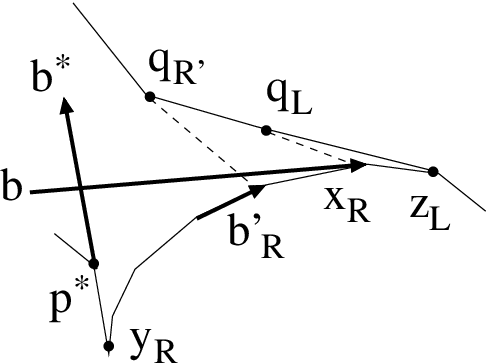}
\caption{\footnotesize Illustrating Case 2.2 in Phase II: $x_R=Head(b)$.}
\label{fig:case21}
\end{center}
\end{minipage}
\hspace*{-0.02in}
\begin{minipage}[t]{0.49\linewidth}
\begin{center}
\includegraphics[totalheight=1.5in]{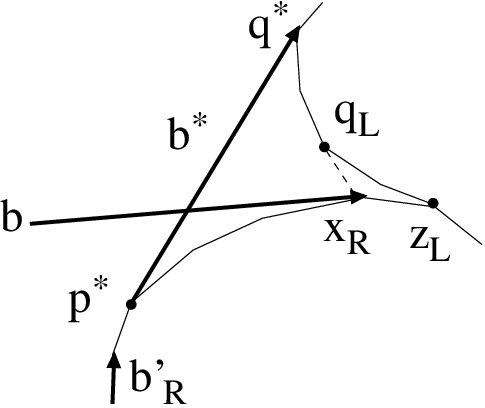}
\caption{\footnotesize Illustrating Case 3 in Phase II.}
\label{fig:case3}
\end{center}
\end{minipage}
\vspace*{-0.15in}
\end{figure}

In Case 2, recall that $b_{R'}=b'_R$ and $R'=Ltri(b'_R)$. Note that
$x_{R'}$ is $y_R$ or $p^*$, $y_{R'}$ is $y_L$ or $q^*$, and $z_{R'}$
is $z_L$ or $x_R$. Depending on whether $x_R$ is $Head(b'_R)$ or
$Head(b)$, there are two subcases. {\em Case 2.1:} $x_R=Head(b'_R)$
(see Fig.~\ref{fig:maincases}) and {\em Case 2.2:} $x_R=Head(b)$
(see Fig.~\ref{fig:case21}). In Case 2.1, $z_{R'}=x_R$ and in Case
2.2, $z_{R'}=z_L$.

{\em Case 2.1:} $x_R=Head(b'_R)$ (see Fig.~\ref{fig:maincases}). In
this subcase, $Head(b'_R)=p_{R'}=q_{R'}=x_R=z_{R'}$. The pseudocode
is in Algorithm \ref{algo:case21}. The details are discussed below.

The goal of computing $Awake[R']$ is to let it represent
$\widehat{x_{R'}q_{R'}}$.
Recall that $AwakeMax[L]$ represents $\widehat{q^*z_L}$ on $\partial L$.
Regardless of whether $x_{R'}$ is $y_R$ or $p^*$, both $b$ and the
portion $\widehat{q^*z_L}$ are awake on $\partial R'$, i.e., part of
$\widehat{x_{R'}q_{R'}}$. Thus, we first set $Awake[R']= AwakeMax[L]$ and
then enqueue $\widehat{z_Lx_R}$ of $\partial R$ to the tail of $Awake[R']$.
We also enqueue $b^*$ to the head of $Awake[R']$. 
Now $Awake[R']$ represents $\widehat{p^*q_{R'}}$.
Recall that $x_{R'}$ is either $y_R$ or $p^*$. If $x_{R'}$ is $p^*$, then we are
done with computing $Awake[R']$. Otherwise ($x_{R'}=
y_R$), we enqueue $\widehat{y_Rp^*}$ to the head of $Awake[R']$.

This completes the construction of $Awake[R']$ for Case 2.1.

Now we compute $Asleep[R']$ which represents
$\widehat{w_{R'}p_{R'}}$. Due to $q_{R'}=p_{R'}=z_{R'}$, $w_{R'}$ is
$q_{R'}$. Hence $\widehat{w_{R'}p_{R'}}$ is only a point, which is not
essential to our algorithm. We simply set $Asleep[R']=\emptyset$.

This completes the construction of $Asleep[R']$ for Case 2.1.

\begin{algorithm}[h]
\caption{Construction of $Awake$ and $Asleep$ of $R'$ for Case 2.2}
\label{algo:case22}
\SetAlgoNoLine
\KwOut{$Awake[R']=\widehat{x_{R'}q_{R'}}$ and $Asleep[R']=\widehat{w_{R'}p_{R'}}$}
\BlankLine
Determine whether $q_{R'} \in \widehat{p_{R'}p^*}$ or $p^*\in
\widehat{p_{R'}q_{R'}}$ on $\partial R'$ by checking the position of $p_{R'}$ with
respect to the line containing $b^*$ \;
\eIf{$q_{R'}\in \widehat{p_{R'}p^*}$}{Starting at $x_{R'}$,
enqueue the atoms to the tail of $Awake[R']$
along $\partial R'$ clockwise until $q_{R'}$ is found
\tcc*[l]{charge enqueue to $D$}\label{line:case2210}}
(\tcc*[h]{$p^*\in \widehat{p_{R'}q_{R'}}$})
{$Awake[R']\leftarrow AwakeMax[L]$ \;
 Dequeue atoms from the tail of $Awake[R']$ until $q_{R'}$ is
 found \;
 Enqueue $b^*$ to the head of $Awake[R']$ \tcc*[l]{charge the enqueue to
 $\calB$}\label{line:case2220}
 \eIf{$x_{R'}=p^*$}{ Do nothing \;}
(\tcc*[h]{$x_{R'}=y_R$})
 { Enqueue $\widehat{y_Rp^*}$ to the head of $Awake[R']$
 \tcc*[l]{charge the enqueue to $D$}\label{line:case2230}}
}
\tcc*[h]{done with $Awake[R']$, the following is for $Asleep[R']$}\\
$Asleep[R']\leftarrow Asleep[L]$ \;
Enqueue $\widehat{x_Rp_{R'}}$ to the tail of $Asleep[R']$
\tcc*[l]{charge the enqueue to $Q$}\label{line:case2240}
 \eIf{$w_L=y_L$}{ Do nothing \;}
(\tcc*[h]{$w_L=q_L$})
 { Starting at $q_L$, enqueue atoms along $\partial R'$ {\em counterclockwise}
 to the head of $Asleep[R']$ until either $q_{R'}$ or $y_{R'}$ is found
 \tcc*[l]{charge the enqueue to $S$}\label{line:case2250}}
\end{algorithm}

{\em Case 2.2:} $x_R=Head(b)$ (see Fig.~\ref{fig:case21}). Recall
that $z_{R'}=z_L$. The pseudocode is in Algorithm \ref{algo:case22}.
The details are discussed below.

We first compute $Awake[R']$ which represents
$\widehat{x_{R'}q_{R'}}$. Clearly,
$x_{R'}$ is before $q_{R'}$ on $\partial R'$.
Since we know $b^*$ and the basepoint
$p_{R'}$ of $R'$, by checking the position of $p_{R'}$ with
respect to the line containing $b^*$, we can determine whether
$q_{R'}$ is before $p^*$ on $\partial R'$ in $O(1)$ time.
Depending on whether $q_{R'}$ is before
$p^*$ on $\partial R'$, there are two subcases.

\begin{itemize}
\item
$q_{R'}$ is before $p^*$ on $\partial R'$, i.e.,
$q_{R'}\in \widehat{p_{R'}p^*}$. Recall that $x_{R'}$ is
either $y_R$ or $p^*$ in Case 2, and
$x_{R'}$ is before $q_{R'}$ on $\partial R'$.
Since $q_{R'}$ is before $p^*$, $x_{R'}$ is before $p^*$,
and thus $x_{R'}=y_R$.

We first set $Awake[R']=\emptyset$, and then starting at $x_{R'}$,
we enqueue the atoms to the tail of $Awake[R']$
along $\partial R'$ clockwise until we
find $q_{R'}$. Now $Awake[R']= \widehat{x_{R'}q_{R'}}$.

\item
$q_{R'}$ is after $p^*$ on $\partial R'$, i.e.,
$p^*\in \widehat{p_{R'}q_{R'}}$. Then clearly, $q^*\in
\widehat{p_{R'}q_{R'}}$. As can be easily seen, $q_{R'}$ is before $q_L$ on
$\partial R'$ (see Fig.~\ref{fig:case21}).
Hence $q_{R'}$ lies on $\widehat{q^*q_L}$
of $\partial L$, which is stored in $AwakeMax[L]$.

We set $Awake[R']= AwakeMax[L]$, and then
dequeue atoms from the tail of $Awake[R']$ until we find $q_{R'}$.
Now $Awake[R']$ represents $\widehat{q^*q_{R'}}$.
We then enqueue $b^*$ to the head of $Awake[R']$.
Now $Awake[R']$ represents $\widehat{p^*q_{R'}}$.

Recall that $x_{R'}$ is either $y_R$ or $p^*$. If $x_{R'}= p^*$, then we are
done with constructing $Awake[R']$. Otherwise ($x_{R'}=
y_R$), we enqueue $\widehat{y_Rp^*}$ to the head of $Awake[R']$.
\end{itemize}

This completes the construction of $Awake[R']$ for Case 2.2.

Next, we compute $Asleep[R']$ (for Case 2.2), which represents
$\widehat{w_{R'}p_{R'}}$. Recall that $w_{R'}$ is $q_{R'}$ or
$y_{R'}$, and $y_{R'}$ is either $q^*$ or $y_L$. Below, we will show
(implicitly) in a constructive manner
that $\widehat{w_Lp_L}$, which is stored in $Asleep[L]$, is part of
$\widehat{w_{R'}p_{R'}}$. The point $w_L$ is either $q_L$ or $y_L$.
We need to discuss the relations among the positions of
$q^*,y_L,q_{R'}$, and $q_L$.
Recall $q_{R'}$ is before $q_L$ on $\partial R'$ (see
Fig.~\ref{fig:case21}) and $q^*$ is before $q_L$ on $\partial L$.

We first set $Asleep[R']=
Asleep[L]$, which represents $\widehat{w_Lp_L}$ on $\partial L$. Recall that
$p_L=Head(b)=x_R$ and $p_{R'}=Head(b_R')$. We then enqueue
$\widehat{x_Rp_{R'}}$ to the tail of $Asleep[R']$.
Now $Asleep[R']$ represents $\widehat{w_Lp_{R'}}$.
Depending on whether $w_L$ is $y_L$
or $q_L$, there are two subcases.

\begin{itemize}
\item

If $w_L=y_L$, then $y_L$ is after $q_{L}$ on $\partial L$. Since $q^*$ is
before $q_L$ on $\partial L$, $q^*$ is before $y_L$ on $\partial L$, implying
$y_{R'}= y_L$.
Because $y_L$ is after $q_{L}$, $q_L$ is before
$y_{R'}$ ($=y_L$).
Since $q_{R'}$ is before $q_L$ on $\partial R'$ and $q_{L}$
is before $y_{R'}$ on $\partial R'$, $q_{R'}$ is before $y_{R'}$ on $\partial R'$. Thus,
$w_{R'}$ is $y_{R'}$, which is $y_L$ ($=w_L$). Therefore, we are
done with the construction of $Asleep[R']$.

\item
If $w_L=q_L$, then $y_L$ is before $q_{L}$ on $\partial L$ and thus $Asleep[R']$
currently represents $\widehat{q_Lp_{R'}}$. Next, starting at $q_L$, we
enqueue atoms along $\partial R'$ {\em counterclockwise} to the head
of $Asleep[R']$ until we first find $q_{R'}$ or encounter $y_{R'}$
(regardless of whether $y_{R'}$ is $y_L$ or $q^*$). Note that the
first point of $q_{R'}$ and $y_{R'}$ thus encountered is $w_{R'}$.
Now $Asleep[R']$ represents $\widehat{w_{R'}p_{R'}}$ and we are done with the
construction of $Asleep[R']$.
\end{itemize}

This finishes the construction of $Asleep[R']$ for Case 2.2.

We thus complete the construction of both $Awake[R']$ and $Asleep[R']$ for Case 2.

\begin{algorithm}[h]
\caption{Construction of $Awake$ and $Asleep$ of $R'$ for Case 3}
\label{algo:case3} \SetAlgoNoLine
\KwOut{$Awake[R']=\widehat{x_{R'}q_{R'}}$ and
$Asleep[R']=\widehat{w_{R'}p_{R'}}$} \BlankLine
$Asleep[R']\leftarrow\emptyset$ \tcc*[l]{done for $Asleep[R']$}
$Awake[R']\leftarrow Asleep[L]$ \; Enqueue $\widehat{x_Rp^*}$ to the
tail of $Awake[R']$ \tcc*[l]{charge the enqueue to
$Q$}\label{line:case310} Enqueue $b^*$ to the head of $Awake[R']$
\tcc*[l]{charge the enqueue to $\calB$}\label{line:case320}
\eIf{$w_L=y_L$}{Do nothing \;} (\tcc*[h]{$w_L=q_L$}){Enqueue
$\widehat{x_{R'}q_L}$ to the head of $Awake[R']$ \tcc*[l]{charge
enqueue to $S$}\label{line:case330}}
\end{algorithm}

\subsubsection{Case 3}

In Case 3 (see Fig.~\ref{fig:case3}), $b_{R'}=b^*$,
$q_{R'}=p_{R'}=p^*$, and $p_L=Head(b)=x_R$.
Note that $x_{R'}$ is $y_L$ or $q^*$, $y_{R'}=z_L$, and $z_{R'}=p^*$.
 The pseudocode is in Algorithm \ref{algo:case3}.
The details are discussed below.

We first compute $Awake[R']$, which represents
$\widehat{x_{R'}q_{R'}}$. Below, we will show (implicitly) in a
constructive manner that $\widehat{w_Lp_L}$, which is stored in
$Asleep[L]$, is part of $\widehat{x_{R'}q_{R'}}$.
We set $Awake[R']=Asleep[L]$ and then enqueue
$\widehat{x_Rp^*}$ ($=\widehat{p_Lq_{R'}}$) to the tail of $Awake[R']$.
Now $Awake[R']$ represents $\widehat{w_Lq_{R'}}$. Recall that $w_L$ is
either $y_L$ or $q_L$, and $q_L$ is after $q^*$ on $\partial L$.

\begin{itemize}
\item

If $w_L=y_L$, then $q_L$ is before $y_L$ on $\partial L$. Thus $y_L$ is after
$q^*$ on $\partial L$ and
$x_{R'}$ is $y_L$ ($=w_L$). We are then done with computing $Awake[R']$.

\item

If $w_L=q_L$, then $q_L$ is after $y_L$ on $\partial L$,
i.e., $y_L\in \widehat{x_Lq_L}$.
If $x_{R'}=q^*$, then $q^*$ is on $\widehat{y_Lz_L}$;
otherwise, $x_{R'}=y_L$. In either case, $\widehat{x_{R'}q_L}$ lies on
$\widehat{y_Lq_L}$.
We then enqueue $\widehat{x_{R'}q_L}$ to the head of $Awake[R']$.
Now $Awake[R']$ represents $\widehat{x_{R'}q_{R'}}$ and we are done
with computing $Awake[R']$.
\end{itemize}

This completes the construction of $Awake[R']$ for Case 3.

We next compute $Asleep[R']$ ($=
\widehat{w_{R'}p_{R'}}$). Recall that $p_{R'}=q_{R'}=p^*=z_{R'}$.
Thus $w_{R'}=q_{R'}$, and
$\widehat{w_{R'}p_{R'}}$ is only a point, which is not
essential to our algorithm. We simply set $Asleep[R']=\emptyset$.

This completes the construction of $Asleep[R']$ for Case 3.


\subsection{The Time Complexity of the Topological Flip Algorithm}
\label{sec:complexity}

In this section, we give the outline of the time analysis for our
algorithm, with details given in an independent Section \ref{sec:keylemma}.
A {\bf key lemma} that we need to prove is the following.

\begin{lemma}\label{lem:keylemma}
The total number of {\em enqueue} operations in Phase II of the entire
algorithm is $O(n+k)$.
\end{lemma}

Recall that the initialization procedure performs $O(n)$ enqueue
operations and no enqueue
occurs in Phase I at all. By Lemma
\ref{lem:keylemma}, the total number of enqueues
in the entire algorithm is $O(n+k)$. Thus, the total number of dequeues
in the entire algorithm is also $O(n+k)$ since it cannot be
bigger than the total number of enqueues.
Further, at most two splits
are needed for each flip in Phase I, and no split is used in Phase
II, implying that the total number of splits
in the algorithm is
$O(k)$. Thus, there are totally $O(n+k)$ operations on the
splittable queues in the entire algorithm. By Lemma \ref{lem:60},
the total time for performing all $k$ flip operations is $O(n+k)$.

In addition, as shown by Lemma \ref{lem:preparing} (given in Section
\ref{sec:reverse}), the total time of the preparing procedure (used only in
Phase I) over the entire algorithm is $O(n+k)$.

We conclude that all $k$ flips can be performed in $O(n+k)$ time.
Therefore, the overall running time for computing all free bitangents in $\calB$
is $O(n+h\log h+k)$.
At any moment of the algorithm, the space needed is for
storing the current good pseudo-triangulation and all splittable
queues, which is $O(n)$.
If we incorporate the needed graph information into the above
algorithm, then the relevant visibility graph $G$ can be built in
the same amount of time.

%
%

It remains to prove the key lemma (Lemma \ref{lem:keylemma}), which is
quite challenging.

As in the PV algorithm \cite{ref:PocchiolaCo95,ref:PocchiolaTo96},
only a constant number of enqueue sequences are involved in
Phase II for each flip and each enqueue sequence is on either a
free bitangent or a boundary
portion of one single obstacle (see Phase II for more
details of this). For the PV algorithm, since every obstacle is
of $O(1)$ complexity, each enqueue sequence can be implemented as
$O(1)$ enqueue operations. Consequently, Lemma \ref{lem:keylemma} easily
follows in \cite{ref:PocchiolaCo95,ref:PocchiolaTo96}.
For our problem, however, since the complexity of an
obstacle can be $\Omega(n)$, each enqueue sequence may take as many
as $\Omega(n)$ enqueue operations. Thus, the simple proof for
the PV algorithm does not appear to work for our problem. To prove the key lemma,
we instead conduct the analysis in a global fashion, as follows.

As one of our key proof ideas, we introduce a new concept
``reverse", which was not used in the previous analysis
\cite{ref:PocchiolaCo95,ref:PocchiolaTo96}. Consider
a flip on a free bitangent $b$ in Phase I
(everything here, such as $b^*$, $b'_R$, $R$, $L$, $\Gamma_R$, etc.,
is defined in the
same way as in Sections \ref{sec:conducting} and \ref{sec:PhaseI}).
Let $p$ be a point on
$\partial T$ of a pseudo-triangle $T$ in the current good
pseudo-triangulation $\calT$ before the flip such that $p$ lies on an obstacle
$P$. Let $l_1(p)$ be the directed tangent line of $T$ at $p$.
Suppose after the flip of $b$, $p$ lies on $\partial T'$ of a new
pseudo-triangle $T'$ ($\neq T$); let $l_2(p)$ be the directed
tangent line of $T'$ at $p$. By the definition of tangent lines
of a pseudo-triangle, both $l_1(p)$ and $l_2(p)$ are tangent to the
obstacle $P$ at $p$. Since by our assumption, the boundary of each
obstacle is smooth, $l_1(p)$ and $l_2(p)$ lie on the same undirected
line. But, it is possible that $l_1(p)$ and $l_2(p)$ have opposite
directions (e.g., if $p$ is on $\Gamma_R\setminus \{Tail(b'_R)\}$ in
Case 1; see Fig.~\ref{fig:maincases} and recall that $\Gamma_R$ refers to the portion of $\partial R$
from $Head(b)$ clockwise to the first encountered
point of $b'_R$.). When this occurs, we say that the point $p$ is {\em
reversed} due to the flip of $b$. Note that $p$ can be reversed only if the
pseudo-triangle $T$ is either $R$ or $L$.

For example, in Case 1 (resp., Case 2), all points on $\Gamma_R$
except the endpoint $Tail(b'_R)$ (resp., $Head(b'_R)$) are reversed
(see Fig.~\ref{fig:maincases}). In Case 3, all points on
$\widehat{x_Rp^*}\setminus \{p^*\}$ (here, $x_R=$ $Head(b)$) are
reversed (note $\widehat{x_Rp^*}$ is part of $\Gamma_R$).
Of course, the algorithm does not do the ``reversal" explicitly.

For any atom, 
if it is (part of) an elementary curve, then we say that it
is {\em reversed} if all its interior points are reversed; if it is
a bitangent $t$, then it is {\em reversed} if the direction of $t$ is
reversed.


Our overall proof strategy is to associate the enqueue operations in Phase
II of the entire algorithm with different ``classes'' of operations and
prove a bound for each such class.
For this, we denote by $n_E$ the number of enqueue operations in Phase II of the
entire algorithm, by $n_Q$ the number of all reversed atoms in the
entire algorithm, by $n_D$ the number of dequeue operations in Phase I
of the entire algorithm, and by $n_S$ the number of certain {\em special
enqueue operations} in Phase II of the entire algorithm (which are
defined in Section \ref{sec:PhaseII}). Recall that
$k=|\calB|$.

Then, to prove Lemma \ref{lem:keylemma} is to show $n_E=O(n+k)$. To this
end, we prove that $n_E\leq n_Q+n_D+n_S+k$ and $n_Q=O(n+k)$,
$n_D=O(n+k)$, and $n_S=O(n+k)$.

The detailed proof is given in Section \ref{sec:keylemma}, which is
organized as follows. In Section \ref{sec:reverse}, we prove
$n_Q=O(n+k)$. To this end, we prove that any point on any obstacle
boundary can be reversed at most once in the entire algorithm.
We show (in Observation \ref{obser:15})
that the total number of all atoms involved in the algorithm is $O(n+k)$.
We also show (in Lemma \ref{lem:preparing})
in Section \ref{sec:reverse} that the total running time of the
preparing procedure in Phase I of the entire algorithm is $O(n+k)$.
In Section \ref{sec:dequeue}, we prove $n_D=O(n+k)$. To this end, we
prove that every atom can be dequeued at most $O(1)$ times in Phase I
of the entire algorithm.
In Section \ref{sec:PhaseII}, we prove $n_E\leq n_Q+n_D+n_S+k$ and
$n_S=O(n+k)$. We show that for any enqueue operation in Phase II,
say, on an atom $A$, $A$ must belong to one of the following cases:
a reversed atom, an atom dequeued in Phase I,
the current enqueue on $A$ being a special enqueue operation, or a
free bitangent in $\calB$.

The proof in Section \ref{sec:keylemma}, which uses many new
observations and analysis ideas, is long and technically difficult and
complicated. Nevertheless, it does provide lots of insights into the problem
and explores many essential properties, which may help
deal with other related problems as well.

\section{The Proof of Lemma \ref{lem:keylemma}}
\label{sec:keylemma}

This section is devoted to proving the {\bf key lemma} (Lemma \ref{lem:keylemma}).

Recall that to prove Lemma \ref{lem:keylemma} is to show
$n_E=O(n+k)$. To this end, our strategy is to prove that $n_E\leq
n_Q+n_D+n_S+k$ and $n_Q=O(n+k)$, $n_D=O(n+k)$, and $n_S=O(n+k)$.

Note that the above goals that we want to prove do not appear to be
obtained easily from the results in
\cite{ref:PocchiolaCo95,ref:PocchiolaTo96} although many properties
in the PV algorithm \cite{ref:PocchiolaCo95,ref:PocchiolaTo96} on
the obstacle set $\calO$ hold in our problem on the splinegon set
$\calP$. As a simple example, in the PV algorithm, since each
obstacle in $\calO$ is of constant complexity, each enqueue sequence
only needs $O(1)$ enqueue operations, and thus $n_E=O(n+k)$ simply
follows as there are $O(k)$ enqueue sequences in the entire
algorithm. By contrast, in our problem since a splinegon may be of
$\Omega(n)$ complexity, an enqueue sequence may need $\Omega(n)$
enqueue operations, and thus if we use similar analysis to the PV
algorithm, we would only obtain $n_E=O(nk)$ as there are $O(k)$
enqueue sequences. As another example, in the PV algorithm on
$\calO$, each enqueue sequence involves a portion of the boundary of
an obstacle, which can be treated as a single atom since each
obstacle in $\calO$ is of constant complexity. Thus, it does not
matter if a boundary portion of an obstacle is involved in multiple
enqueue sequences since there are totally $O(k)$ enqueue sequences.
In our problem, however, it does matter if a boundary portion of an
obstacle is involved in multiple enqueue sequences because that
boundary portion may contain many atoms (e.g., elementary curves),
potentially causing the number of enqueue operations (i.e., $n_E$)
not bounded by $O(n+k)$. Hence, to show $n_E=O(n+k)$ we have to give
more careful argument that requires exploring more observations
on the problem.

The rest of this section is organized as follows.
In Section \ref{sec:reverse}, we prove
$n_Q=O(n+k)$. In Section \ref{sec:dequeue}, we prove $n_D=O(n+k)$.
In Section \ref{sec:PhaseII}, we show $n_E\leq n_Q+n_D+n_S+k$
and $n_S=O(n+k)$. In addition, we show in Section
\ref{sec:reverse} that the total running time of the preparing
procedure in Phase I over the entire algorithm is $O(n_Q)$, which is $O(n+k)$.

\subsection{Bounding the Number of Reversed Atoms (i.e., $n_Q=O(n+k)$)}
\label{sec:reverse}

Consider a flip operation
on a free bitangent $b$ in Phase I. Everything here is defined in the same way as in Sections
\ref{sec:conducting} and \ref{sec:PhaseI}, e.g., $b^*$, $b'_R$, $R$,
$L$, $R'$, $L'$, $\Gamma_R$, etc.

Recall that $\Gamma_R$ is the portion of $\partial R$
from $Head(b)$ clockwise to the first encountered
point of $b'_R$.
In Case 1 (resp., Case 2), all points on $\Gamma_R$ except the
endpoint $Tail(b'_R)$ (resp., $Head(b'_R)$) are reversed (see
Fig.~\ref{fig:maincases}). In Case 3, all points on
$\widehat{x_Rp^*}\setminus \{p^*\}$ (here, $x_R=$ $Head(b)$) are
reversed (note that $\widehat{x_Rp^*}$ is part of $\Gamma_R$). Observe
that $\Gamma_R$ is an obstacle arc. The following observation is
self-evident.

\begin{observation}\label{obser:10}
After the flip of $b$, let $\alpha$ be the reversed portion on
$\partial T$, with $T\in\{R,L\}$.
Let $T'$ be $R'$ (resp., $L'$) if $\alpha$ lies on $\partial R'$ (resp.,
$\partial L'$).
Then the following properties hold.
\begin{enumerate}

\item
$\alpha$ is an obstacle arc.
One (resp., the other) endpoint of $\alpha$ is an endpoint of $b$ (resp., $b_{T'}$).

\item
Let $\alpha_b$ (resp., $\alpha_{b_{T'}}$) be the endpoint of $b$ (resp.,
$b_{T'}$) on $\alpha$. From $\alpha_b$ to $\alpha_{b_{T'}}$ along $\alpha$,
it is counterclockwise with respect to the obstacle on which $\alpha$ lies.
We call $\alpha_b$ the {\em obstacle-ccw-start} endpoint of $\alpha$.
\item
Every point on $\alpha$ except the endpoint $\alpha_{b_{T'}}$ is reversed due to the
flip of $b$. For any point $p\in \alpha\setminus \{\alpha_{b_{T'}}\}$,
$\alpha$ is called the {\em hosting arc} of $p$.
\end{enumerate}
\end{observation}

Note that only points on $\partial R$ or $\partial L$ can be reversed
due to the flip of $b$.
An important property of the reversed portions after every flip is
given in the lemma below.

\begin{lemma}\label{lem:170}
After a flip operation on $b$, suppose $\alpha$ is the reversed
portion on $\partial T$ of a pseudo-triangle $T$, $T\in \{R, L\}$.
Then no interior point of $\alpha$ can be an endpoint of any
bitangent in $\calB$.
\end{lemma}

\begin{proof}
First, by Lemma \ref{lem:40}, the direction of any free bitangent $t\in
\calB$ can be reversed only by a flip operation on $t$. Thus, after the flip
of $b$, the direction of $b$ is reversed, whereas
the direction of any other free bitangent in $\calB$ does not
change.
By Observation \ref{obser:10}, $\alpha$ is an obstacle arc,
say, on an obstacle $P$.

Assume to the contrary that there is an interior point $q$ of $\alpha$
that is an endpoint of a free bitangent $t\in \calB$. Note that $t\neq
b$ since an endpoint of $b$ is an endpoint of $\alpha$ by Observation \ref{obser:10}.
By the definition of good pseudo-triangulation, the direction of $t$
is compatible with its endpoint $q$ before the flip of $b$.
Recall that the direction of $t$
is compatible with its endpoint $q$ if the directed tangent line of $T$ at
$q$, denoted by $l_1(q)$, has the same direction as $t$.

Suppose $\alpha$ lies on a pseudo-triangle $T'$ right after the
flip of $b$.
Let $l_2(q)$ be $l_1(q)$ but with the reversed direction.
Then both $l_1(q)$ and $l_2(q)$ are tangent to the obstacle $P$ at $q$. Since $q$ is
reversed due to the flip of $b$, $l_2(q)$ is the directed tangent line
of $T'$ at $q$ right after the flip of $b$.
As a free bitangent, $t$'s direction does not change after the flip of
$b$ ($\neq t$),
the direction of $l_2(q)$ is opposite to that of $t$, making
$t$ not compatible with its endpoint $q$ after the flip of $b$.
But this contradicts with
the definition of good pseudo-triangulation. Thus $q$ cannot
be an endpoint of any free bitangent in $\calB$, and
the lemma follows.
\end{proof}

The next lemma is critical.

\begin{lemma}\label{lem:70}
During the topological flip algorithm, any point on the boundary of
any obstacle in $\calP$ can be reversed at most once.
\end{lemma}

\begin{proof}
By Lemma \ref{lem:40}, the direction of any free bitangent $t\in
\calB$ can be reversed only by a flip operation on $t$.
Since the algorithm does not flip any
bitangent more than once, the direction of each free bitangent is
reversed only once throughout the algorithm.

Consider a flip of a free bitangent $b$ of a good pseudo-triangulation $\calT$.
Suppose a point $a$ on $\partial T$ of a pseudo-triangle $T$
is reversed for the first time due to the flip of $b$.
Below we prove that $a$ cannot be reversed again. We
only discuss the case when $a$ lies on $\partial R$ (the case on
$\partial L$ is similar).

Let $p=Head(b)$; let $q=Tail(b'_R)$ in Case 1, $q=Head(b'_R)$ in
Case 2, and $q=Tail(b^*)$ in Case 3, respectively (see Fig.~\ref{fig:maincases}).
Due to the flip of $b$, in each case, the reversed portion is
$\widehat{pq}\setminus\{q\}$.
Also, the direction of $b$ is reversed. By Observation
\ref{obser:10}, $\widehat{pq}$ is an obstacle arc, say, on obstacle $P$.
Assume to the contrary that later a point $a\in \widehat{pq}\setminus\{q\}$
is reversed for the second time.
Note that $a$ cannot be $p$ since otherwise $b$ would not be
compatible with $p$ unless $b$ is reversed twice.

At the second reversal of $a$,
let $\widehat{p'q'}$ be the hosting arc of $a$ with $p'$ being the
obstacle-ccw-start endpoint (see Fig.~\ref{fig:lemmareverse}).
By Observation \ref{obser:10}, $\widehat{p'q'}$ is
an obstacle arc, say, on obstacle $P'$.
Since the point $a$ is on both $\widehat{pq}\in P$ and
$\widehat{p'q'}\in P'$, $a$ lies on both $P$ and $P'$. Therefore,
$P=P'$ since our obstacles in $\calP$ are pairwise disjoint.
Also by Observation \ref{obser:10}, both $p'$ and $q'$ are
endpoints of some free bitangents in $\calB$.

\begin{figure}[t]
\begin{minipage}[t]{\linewidth}
\begin{center}
\includegraphics[totalheight=1.5in]{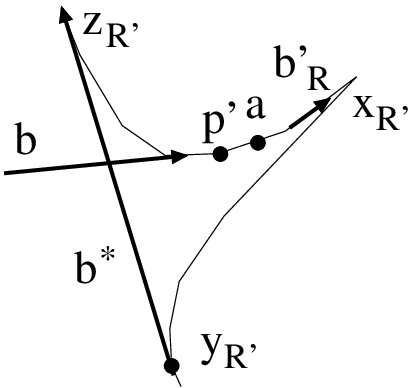}
\caption{\footnotesize Illustrating the proof of Lemma \ref{lem:70}.}
\label{fig:lemmareverse}
\end{center}
\end{minipage}
\vspace*{-0.15in}
\end{figure}

Since $p'$ is the obstacle-ccw-start endpoint of $\widehat{p'q'}$, by
Observation \ref{obser:10}, when moving from $p'$ to $q'$ on
$\widehat{p'q'}$, it is counterclockwise with respect to $P$.
Similarly, when moving from $p$ to $q$ on
$\widehat{pq}$, it is counterclockwise with respect to $P$.
Since $a$ is on both $\widehat{pq}$ and $\widehat{p'q'}$, there are three
possible cases: (i) $p$ is an interior point of $\widehat{p'q'}$, (ii) $p'$
is an interior point of $\widehat{pq}$, and (iii) $p= p'$. We argue
below that any case can not occur. Consequently, the point $a$
cannot be reversed again.

As the point $p$ is an endpoint of $b$, by Lemma \ref{lem:170},
$p$ cannot be an interior point of $\widehat{p'q'}$.
Similarly, $p'$ cannot be an interior point of $\widehat{pq}$.
For case (iii), since the point $p'$
is the obstacle-ccw-start endpoint of $\widehat{p'q'}$,
$p'$ is reversed. If case (iii) occurs, $p'$ ($=p$) is the
endpoint of $b$, and thus the direction of $b$ has to be reversed again since
otherwise $b$ would not be compatible with the reversed $p'$. But this
contradicts with the fact that $b$ cannot be flipped twice in the
algorithm.

The lemma thus follows.
\end{proof}

The following lemma bounds the number of all different atoms
involved in the algorithm.

\begin{observation}\label{obser:15}
The total number of different atoms involved in the algorithm is $O(n+k)$.
\end{observation}

\begin{proof}
In our problem, an atom is of one of the following three types:
(1) A free bitangent, (2) an elementary curve, and (3) a portion of an
elementary curve. It is easy to see that the number of type (1) atoms
is $O(k)$ and the number of type (2) atoms is $O(n)$. To prove this observation,
it suffices to show that the number of type (3) atoms is $O(n+k)$.

After initialization, in the initial good pseudo-triangulation $\calT_0$,
there are $O(n)$ type (3) atoms.
Subsequently in the algorithm, each new type (3) atom is produced only due to a flip
operation. Note that after every flip operation, at most $O(1)$
new type (3) atoms can be produced. Since there are $k$ flips, plus those in $\calT_0$, the
number of type (3) atoms is bounded by $O(n+k)$.
\end{proof}

We then have the following result.

\begin{lemma}\label{lem:nq}
The number of reversed atoms in the entire algorithm (i.e., $n_Q$)
is $O(n+k)$.
\end{lemma}

\begin{proof}
By Lemma \ref{lem:70}, each atom can be reversed at most once.
Consequently, Observation \ref{obser:15} leads to the lemma.
\end{proof}


\begin{lemma}\label{lem:preparing}
The overall running time of the preparing procedure in Phase I of
the entire algorithm is $O(n+k)$.
\end{lemma}

\begin{proof}
Recall that the preparing procedure is to determine whether Case 1
occurs and find $b_R'$ if Case 1 holds. Since the three cusps of $R$
are maintained by the algorithm, $x_R$ is known. We walk from
$Head(b)$ along $\partial R$ clockwise until first encountering
either $b'_R$ or $x_R$. If we encounter $b'_R$ first, then it is
Case 1 and we have $b'_R$ as well. Otherwise, it is either Case 2 or
Case 3.

It is easy to see that the running time of the above walking is
proportional to the number of atoms of the reversed portion due to
the flip of $b$. Therefore, the total time of the preparing procedure for
the entire algorithm is at most proportional to the total number of reversed atoms
in the entire algorithm, i.e., $O(n_Q)$, which is $O(n+k)$ by
Lemma \ref{lem:nq}. The lemma thus follows.
\end{proof}

\subsection{Bounding the Number of Dequeue Operations in Phase I (i.e., $n_D=O(n+k)$)}
\label{sec:dequeue}

Our goal in this section is to prove Lemma \ref{lem:90} below.

\begin{lemma}\label{lem:90}
In Phase I of the entire algorithm, the total number of
dequeue operations (i.e., $n_D$) is $O(n+k)$.
\end{lemma}

To prove Lemma \ref{lem:90}, we will first prove the following lemma (Lemma \ref{lem:80}),
which states that
any atom can be dequeued in Phase I of the entire algorithm at most twice
before its (only) reversal and can be dequeued in Phase I
at most twice after its reversal (if any).
Note that Lemma \ref{lem:80} does not include the dequeue operations in Phase II.

\begin{lemma}\label{lem:80}
For any atom $A$ in the algorithm, regardless of whether it
has been reversed previously, $A$ can be dequeued
in Phase I at most twice before its next reversal (if any) in the algorithm.
\end{lemma}

By Lemma \ref{lem:70}, every point on $\partial\calP$
can be reversed at most once. This implies that, by Lemma \ref{lem:80}, every atom
can be dequeued in Phase I at most four times in the entire algorithm.
Since the total number of atoms in the algorithm is $O(n+k)$, Lemma
\ref{lem:90} follows.

The rest of this section gives the proof for Lemma \ref{lem:80}. The
proof, which is quite long and technically complicated, is based on
many new geometric observations and analysis techniques.

The following lemma (proved in
\cite{ref:PocchiolaCo95,ref:PocchiolaTo96}) will be repeatedly
referred to by our analysis later.

\begin{lemma}\label{lem:50}{\em \cite{ref:PocchiolaCo95,ref:PocchiolaTo96}}
For any pseudo-triangle $T$ of a good pseudo-triangulation,
(1) if $y_T$ lies on $\widehat{x_Tq_T}$ of $\partial T$, then
$\widehat{y_Tq_T}$ is an obstacle arc, and
(2) if $z_T\neq p_T$, then
$\widehat{z_Tp_T}$ is an obstacle arc.
\end{lemma}

For example, in the left figure of Fig.~\ref{fig:awake}, since $y_R$ lies on $\widehat{x_Rq_R}$ and $z_R\neq p_R$, according to Lemma \ref{lem:50}, both $\widehat{y_Rq_R}$ and $\widehat{z_Rp_R}$ are obstacle arcs; similarly, in the right figure of Fig.~\ref{fig:awake}, both $\widehat{y_Lq_L}$ and $\widehat{z_Lp_L}$ are obstacle arcs.

For any bitangent $t \in B(\calT)$ of a good pseudo-triangulation
$\calT$ in the algorithm, we view $t$ as defining two atoms, one for
$Rtri(t)$ and the other for $Ltri(t)$, and we say that these two
atoms are {\em defined} by $t$. Similarly, we let every point on $t$
have two copies that belong to $Rtri(t)$ and $Ltri(t)$,
respectively. Thus, for any point $p\in \partial T$ for a
pseudo-triangle $T$ in $\calT$, if $p$ is on a bitangent $t\in
B(\partial T)$, then $p$ refers to the copy of the point on $t$ that
belongs to $T$. In this way, every point on $\partial T$ belongs to
exactly one pseudo-triangle in $\calT$, i.e.,the pseudo-triangle
$T$. Since there are $k$ bitangents in $\calB$, by Observation
\ref{obser:15}, the total number of atoms is still $O(n+k)$. When an
atom $A$ is (part of) an elementary curve, we also say $A$ is an
obstacle arc.

Before presenting the main proof, we give
some observations, which will be useful later.

\subsubsection{Some Observations}

\begin{observation}\label{obser:tangentline}
Consider a pseudo-triangle $T$ in a good pseudo-triangulation
at a (time) moment $\xi_1$ of the algorithm.
For any point $p\in \partial T$, let $l_1(p)$ be the directed tangent line of $T$
at $p$. Suppose later at a moment $\xi_2$ of the algorithm, the point
$p$ lies on $\partial T'$ for a pseudo-triangle $T'$, and
let $l_2(p)$ be the directed tangent line of $T'$ at $p$. Then, $l_1(p)$ and
$l_2(p)$ lie on the same undirected line. Further, if $p$ is not
reversed during the time period from $\xi_1$ to $\xi_2$, then $l_1(p)$
has the same direction as $l_2(p)$; otherwise, $l_1(p)$ and $l_2(p)$
have opposite directions
\end{observation}
\begin{proof}
The observation can be easily proved by the definition of a directed
tangent line of a pseudo-triangle.
If $A$ is defined by a free bitangent, then
$l_1(p)$ and $l_2(p)$ both lie on the same undirected line that contains $A$.
If $A$ is an obstacle arc, then
$l_1(p)$ and $l_2(p)$ also both lie on the same undirected line
that is tangent to the obstacle where $p$ lies.
Thus, in any case, $l_1(p)$ and $l_2(p)$ lie on the same undirected
line.

If the atom $A$ has not been reversed during the time period
from $\xi_1$ to $\xi_2$, then clearly $l_1(p)$ and $l_2(p)$ are of
the same direction. Otherwise, by Lemma \ref{lem:70}, $p$ is reversed
only once during the time from $\xi_1$ to $\xi_2$, and therefore,
$l_1(p)$ and $l_2(p)$ have opposite directions.
\end{proof}


\begin{observation}\label{obser:20}
Consider a pseudo-triangle $T$ in a good pseudo-triangulation
at a (time)
moment $\xi_1$ of the algorithm. For any two points $p$ and $q$ on $\partial T$, let
$l(p)$ and $l(q)$ be their corresponding directed tangent lines of $T$.
Then the following properties hold.
\begin{enumerate}
\item
If the $l(q)$-slope of $l(p)$ is less than $\pi$, then the $l(q)$-slope of
$l(p)$ is no bigger than the pt-slope of $l(p)$ in $T$.
\item
If at a later moment $\xi_2$ of the algorithm, the point $p$ lies on $\partial T'$ of
a pseudo-triangle $T'$ ($T'=T$ is possible) such
that the same directed $l(p)$ is the directed tangent line of $T'$ at $p$
then the pt-slope of $l(p)$ at the moment $\xi_1$ is
no smaller
than the pt-slope of $l(p)$ at the moment
$\xi_2$.
\end{enumerate}
\end{observation}
\begin{proof}
For the first part, observe that the pt-slopes of all points on $\partial T$
($l(p)$ and $l(q)$ included)
are no more than $\pi$. Suppose the $l(q)$-slope of $l(p)$ is less than
$\pi$. If we rotate $b_{T}$ around its tail counterclockwise, we
will encounter first the direction of $l(q)$ and then the direction of
$l(p)$ (otherwise, the pt-slope of $l(q)$ would be larger than
$\pi$). This means that the pt-slope of $l(p)$ is the sum of the pt-slope of
$l(q)$ and the $l(q)$-slope of $l(p)$. The first part thus follows.

The second part can be proved by a simple induction.
Recall that if the point $p$ lies on a bitangent in $B(\partial T)$,
then $p$ refers to the copy of the original point that belongs to $T$.
If $T'=T$, then it is easy to see that the property holds.
Now consider the first flip operation after which
the point $p$ lies on $\partial T'$ of a pseudo-triangle $T'\not=T$
such that $l(p)$ is still the directed tangent line of $T'$ at $p$.
Since $T'\not=T$, the above flip must be on the free bitangent $b_T$.

Let $\xi'_1$ (resp., $\xi'_2$) be the moment right before (resp.,
after) the flip of $b_{T}$. At the moment $\xi'_1$, one endpoint of
$b_{T'}$ must be on $\partial T$ although $b_{T'}$ may be $\varphi(b_T)$.
Recall that due to the flip of $b_T$, every free bitangent in
$\calB\setminus\{b_T\}$ does not change direction. So $b_{T'}$ does
not change direction due to the flip of $b_T$.
Consequently, the $b_{T}$-slope of $b_{T'}$ on $\partial T$
at the moment $\xi_1'$ is the pt-slope of the endpoint of $b_{T'}$
that is on $\partial T$, which must be less than $\pi$.
Note that the $b_{T}$-slope of $l(p)$ is the pt-slope of $l(p)$ on $\partial T$ at
the moment $\xi_1'$, which must be less than $\pi$.
Similarly, the $b_{T'}$-slope of $l(p)$ is the pt-slope of $l(p)$ on $\partial T'$ at
the moment $\xi_2'$, which must be less than $\pi$.
Therefore, if we rotate $b_T$ around its tail counterclockwise, we
will encounter first the direction of $b_{T'}$ and then the direction
of $l(p)$ (otherwise, the $b_{T}$-slope of $b_{T'}$ would be larger than
$\pi$ at the moment $\xi_1'$). This implies that the $b_T$-slope of
$l(p)$ at the moment $\xi_1'$ is no smaller than the $b_{T'}$-slope of
$l(p)$ at the moment $\xi_2'$, i.e., the pt-slope of
$l(p)$ at the moment $\xi'_1$ is no smaller than the pt-slope of
$l(p)$ at the moment $\xi'_2$.

Consider the first flip after which the point $p$ lies on $\partial T''$ of a
pseudo-triangle $T''\not= T'$
such that $l(p)$ is still the directed tangent
line of $T''$ at $p$.
Let $\xi'_3$ be the moment right after this flip. By
a similar argument, the pt-slope of
$l(p)$ at the moment $\xi'_2$ is no smaller than the pt-slope of
$l(p)$ at the moment $\xi'_3$.
Inductively, the second part holds.
\end{proof}

Note that Observation \ref{obser:20}(2) above actually tells us that for any
point $p\in \partial T$ of a pseudo-triangle $T$ in the algorithm,
regardless of whether $p$ has been reversed previously, the pt-slope
of $p$ is monotonically decreasing during the rest of the algorithm until it
is possibly reversed.

\subsubsection{The Main Proof of Lemma \ref{lem:80}}

In the following proof of Lemma \ref{lem:80},
all dequeue operations refer to those in Phase I.

Consider the dequeued atoms on
$\partial R$ due to the flip of $b$ for the pseudo-triangle $R=Rtri(b)$. In
Case 1, the dequeued atoms are on $\widehat{q'_Rp^*}$; in Cases 2 and
3, the dequeued atoms are on $\widehat{x_Rp^*}$. In all three cases,
the dequeued atoms of $\partial R$ are also on $\partial R'$. Recall $R'=Rtri(b^*)$
and $L'=Ltri(b^*)$.  The situation on the
pseudo-triangle $L=Ltri(b)$ is similar.

Let $A$ be an arbitrary atom of the good pseudo-triangulation $\calT$ right
before the flip of $b$.
Note that $A$ may have been reversed before
(and thus cannot be reversed again),
but this is not important to our proof.
Observe that due to the flip of $b$, $A$ can be dequeued only if
(a) $A$ is awake before the flip of $b$, and (b) $A$ lies on
either $\partial R$ or $\partial L$. Note that
$b=b_R=b_L$ since $b$ is minimal.
In other words, suppose $A$ lies on $\partial T$ of a pseudo-triangle
$T$;
if $A$ is dequeued due to the flip of $b$, then (a) $T$ is either $R$
or $L$, (b) $A$ must be awake on $\partial T$
(before the flip of $b$), and (c) $b = b_T$.

One key observation used in our proof is: For any
pseudo-triangle $T$ with $T=Rtri(b_T)$ (resp., $T=Ltri(b_T)$),
if a point $p$ is awake on $\partial T$,
then the forward (resp., backward)
$T$-view of $p$ has a smaller pt-slope
than $p$ in $T$ (see Fig.~\ref{fig:awake}).

To prove Lemma \ref{lem:80}, we first prove the following statement,
which we call Sub-lemma \ref{lem:80}(a).

\vspace*{0.08in}
\noindent
{\bf Sub-lemma \ref{lem:80}(a)}. Suppose $A$ is an atom on $\partial T$ of a pseudo-triangle $T$ with
$T=Rtri(b_T)$ in a good pseudo-triangulation $\calT$ and $A$ is dequeued due to
a flip operation on $b_T$. Also, suppose at any later moment before the reversal
of $A$, $A$ lies on $\partial T'$ of a
pseudo-triangle $T'$ with $T'=Rtri(b_{T'})$ in another good pseudo-triangulation
$\calT'$. Then $A$ cannot be dequeued due to the flip of $b_{T'}$.
\vspace*{0.08in}

To prove Sub-lemma \ref{lem:80}(a), our main idea is to show that $A$ cannot be awake when $b_{T'}$ is flipped (and thus cannot be dequeued again).
The proof, however, consists of a lengthy and complicated case
analysis with considerable details.
We analyze the three main cases (Cases 1, 2, and 3).

\vspace*{0.05in}
\noindent
{\bf The proof of Case 1}.
In Case 1, all dequeued atoms lie on $\widehat{q'_Rp^*}$.
Let $A$ be an arbitrary atom on $\widehat{q'_Rp^*}$.
Let $\xi_1$ be the moment {\em right after} the flip of $b$.
Hence the atom $A$ is on $\partial R'$ at the moment $\xi_1$.
Suppose at a later moment $\xi_2$ of the algorithm,
the atom $A$ lies on $\partial T'$ of a pseudo-triangle $T'$ of a good
pseudo-triangulation
$\calT'$ with $T'=Rtri(b_{T'})$ and $A$ has not been reversed since the moment
$\xi_1$.
As discussed above, the atom $A$ on $\partial T'$ can be dequeued only due to the flip
of $b_{T'}$. Thus, during the time between the moment $\xi_2$ and the
flip of $b_{T'}$,
$A$ cannot be dequeued on $\partial T'$.
Note that once it is formed, the pseudo-triangle $T'$ remains unchanged in the
algorithm (and thus, $A$ is not reversed) until $b_{T'}$ is flipped.
Without loss of generality,
we let $\xi_2$ be the moment {\em right before} the flip of $b_{T'}$.
We prove below that $A$ cannot be awake at the moment $\xi_2$
and thus cannot be dequeued due to the flip of $b_{T'}$.
In the following discussion, for simplicity, when we mention $R'$
(resp., $T'$), we always refer to the moment $\xi_1$ (resp., $\xi_2$)
unless otherwise stated.

Let $p$ be an arbitrary interior point on $A$,
i.e., $p$ is not an endpoint of $A$.

Let $l_2(p)$ be the directed tangent line of $T'$ at $p$
(at the moment $\xi_2$),
and $l_1(p)$ be the directed tangent line of
$R'$ at $p$ (at the moment $\xi_1$). Since $p$ is not
reversed during the time period from $\xi_1$ to $\xi_2$, by
Observation \ref{obser:tangentline}, $l_1(p)$ and $l_2(p)$ are the
same. Below, we simply use $l(p)$ to refer to both $l_1(p)$ and $l_2(p)$.
Since $p$ is not reversed due to the flip of $b$,
$l(p)$ is also the directed tangent line of $R$ at $p$ before the flip
of $b$.

Below, we prove that at the moment $\xi_2$, the point $p$ is not awake
on $\partial T'$ (i.e., $p$ does not lie on
$Awake[T']=\widehat{x_{T'}q_{T'}}$) and thus the atom $A$ cannot be
awake.
Let $q \in \partial T'$ be $p$'s forward $T'$-view point along $l(p)$
(at the moment $\xi_2$).
Let $l(q)$ be the directed tangent line of $T'$ at $q$.

Assume to the contrary that the point $p$ is awake on $\partial T'$ (at the moment
$\xi_2$). Then it immediately implies that $p$ lies on
$Awake[T']=\widehat{x_{T'}q_{T'}}$
and the pt-slope of $l(p)$ is larger than that of $l(q)$ in $T'$ (see
Fig.~\ref{fig:awake}). Thus, $l(q)$ must cross $l(p)$ from left to right.

Note that right before the flip of $b$, we have
$p_L=q_L=z_L=Head(b)$, i.e., $p_L=z_L$ is the basepoint of
$L=Ltri(b)$ and $q_L$ is the point on $\partial L$ whose backward
$L$-view (on $\partial L$) is $p_L$ and $q_L=z_L$ in this case (see
Fig.~\ref{fig:maincases}). Let $c=Tail(b'_R)$.

\begin{figure}[t]
\begin{minipage}[t]{0.49\linewidth}
\begin{center}
\includegraphics[totalheight=1.7in]{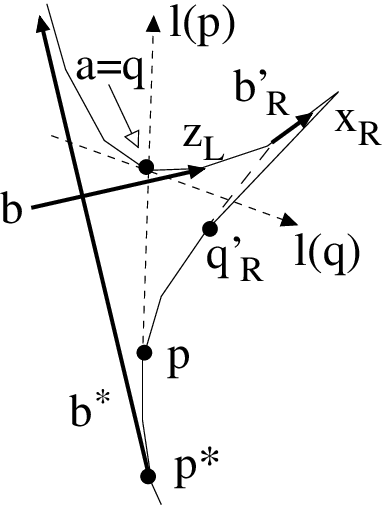}
\caption{\footnotesize Illustrating the case that the point $q$ is $a$
(here $a\in \widehat{y_Lz_L}$).}
\label{fig:forwardviewcase}
\end{center}
\end{minipage}
\hspace*{0.02in}
\begin{minipage}[t]{0.49\linewidth}
\begin{center}
\includegraphics[totalheight=1.7in]{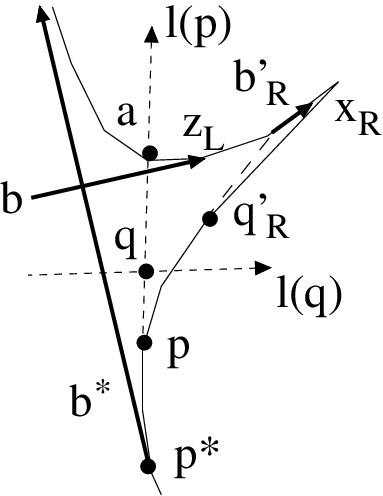}
\caption{\footnotesize Illustrating the case that $q$ lies on $l(p)$
before $a$ (here $a\in \widehat{y_Lz_L}$).} \label{fig:forwardview}
\end{center}
\end{minipage}
\vspace*{-0.15in}
\end{figure}

Recall that the point $p$ is an interior point of an atom $A$ on
$\widehat{q'_Rp^*}$. Recall that $q'_R$ is the point on $\partial R$
whose forward $R$-view is $Tail(b'_R)$. If $q'_R$ is on a bitangent
$t\in B(\partial R)$, since $p$ is the interior point of $t$,
every point on $t$ can be viewed as
$q'_R$; in this case, we let $q'_R$ be the endpoint of $t$ such that
$t$ lies entirely on $\widehat{x_Rq'_R}$. This step can be done when
we conduct the split operation on
$Awake[R]$ at $q'_R$ (i.e., change the criterion when searching $q'_R$
in $Awake[R]$). Note that the above requirement for
$q'_R$ does not change the running time of Lemma \ref{lem:60}. In this
way, $l(p)$
must intersect either $\widehat{y_Lz_L}$ on $\partial L$ or
$\widehat{z_Lc}$ on $\partial R$ (see Fig.~\ref{fig:forwardviewcase}
or Fig.~\ref{fig:forwardview}).

Let $a$ be the intersection point between $l(p)$ and
$\widehat{y_Lz_L}$ or $\widehat{z_Lc}$ (see
Fig.~\ref{fig:forwardviewcase} or Fig.~\ref{fig:forwardview}). Since
$\widehat{z_Lc}$ is the reversed portion on $\partial R$ due to the
flip of $b$, by Observation \ref{obser:10}, $\widehat{z_Lc}$ is an
obstacle arc. Since $q_L=z_L$, by Lemma \ref{lem:50}, the part
$\widehat{y_Lq_L}$ ($=\widehat{y_Lz_L}$) is an obstacle arc. Hence,
the point $a$ must be on an obstacle, say $P$. Consider the position
of $q$, which is $p$'s forward $T'$-view point on $\partial T'$
along $l(p)$ (at the moment $\xi_2$). Since $a$ lies on an obstacle
$P$, $q$ can be either at $a$ or on $l(p)$ between $p$ and $a$ (but
before $a$). In the following, we show that either case cannot occur,
and consequently our assumption that the point $p$ is awake on
$\partial T'$ is not correct.

If $q$ is at the point $a$ (see Fig.~\ref{fig:forwardviewcase}),
then $l(q)$ is the directed tangent line of $T'$ at $a$ ($=q$).
Since $a$ lies on the obstacle $P$, $l(q)$ is tangent to $P$ at $a$.
Note that the point $a$ lies on $\partial R'$ at the moment $\xi_1$
(right after the flip of $b$). Let $l_1(q)$ be the directed tangent
line of $R'$ at $a$ (at the moment $\xi_1$). Since $a$ lies on $P$,
$l_1(q)$ is also tangent to $P$ at $a$, and therefore, $l_1(q)$ and
$l(q)$ both lie on the same undirected line. There are two subcases
to consider: $a$ lies on $\widehat{z_Lc}$ or on
$\widehat{y_Lz_L}\setminus \{z_L\}$. Below, we show that neither
case can occur.

\begin{enumerate}
\item[(i)]
If $a$ ($=q$) lies on $\widehat{z_Lc}$ ($a$ can be $z_L$), since
$q'_R$ is an endpoint of an atom on $\widehat{q'_Rp^*}$ and $p$ is
an interior point of the atom $A$, we have $p\not= q'_R$ and $a\not=c$.
Since $\widehat{z_Lc}$ is
reversed due to the flip of $b$, by Lemma \ref{lem:70},
$\widehat{z_Lc}$ will not be reversed again after the moment $\xi_1$.
Thus, $l(q)$ has the same direction as $l_1(q)$. Recall that $l(q)$
crosses $l(p)$ from left to right. Thus, $l_1(q)$ crosses $l(p)$
from left to right. However, at the moment
$\xi_1$, both $p$ and $q$ ($=a$) are on $\partial R'$ and it is easy
to see that the pt-slope of $l(p)$ is less than
the pt-slope of $l_1(q)$ in $R'$. Thus, $l(p)$ must cross $l_1(q)$ from left
to right, or equivalently, $l_1(q)$ must cross $l(p)$ from right to
left. Hence, we obtain that $l_1(q)$ crosses $l(p)$ both from left
to right and from right to left, which is a contradiction. Thus, $a$
cannot lie on $\widehat{z_Lc}$.

\item[(ii)]
For the subcase when $a$ ($=q$) lies on $\widehat{y_Lz_L}\setminus
\{z_L\}$ (see Fig.~\ref{fig:forwardviewcase}),
if $a$ has not been reversed since $\xi_1$, then at the
moment $\xi_2$, the direction of $l(q)$ is the same as $l_1(q)$'s.
By a similar argument as for the former subcase (i), we can show that a
contradiction also occurs.

In the following, we assume that $a$ is reversed (only once) during
the time from $\xi_1$ to $\xi_2$. Thus,
$l_1(q)$ and $l(q)$ have opposite directions. Since $l(q)$ crosses
$l(p)$ from left to right, $l_1(q)$ crosses $l(p)$ from right to
left. Since $l(q)$ crosses
$l(p)$ from left to right,
the $l(q)$-slope of $l(p)$ must be less than $\pi$. Since both
$l(p)$ and $l(q)$ are directed tangent lines of $T'$,
by Observation \ref{obser:20}(1), the pt-slope of $l(p)$ in
$T'$ is no smaller than the $l(q)$-slope of $l(p)$ (at the moment
$\xi_2$).

Below, we consider $l(q)$ as a physically directed line that is not associate
with any time moment. We claim the $l(q)$-slope of $l(p)$ is
larger than the $b_{R'}$-slope of $l(p)$ at the moment $\xi_1$.
Indeed, the $b_{R'}$-slope of $l_1(q)$ is the
pt-slope of the point $q$ ($=a$) on $\partial R'$, which is larger
than zero and less than $\pi$. Since $l_1(q)$ and $l(q)$ have
opposite directions, the $l(q)$-slope of $b_{R'}$ is less than $\pi$
and larger than zero. Note that $l(p)$ is the directed tangent line
of $R'$. So the $b_{R'}$-slope of $l(p)$ is the pt-slope of $l(p)$
in $R'$ (at the moment $\xi_1$), which is less than $\pi$. To
summarize, we have: (1) the $l(q)$-slope of $b_{R'}$ is less than
$\pi$ and larger than zero, (2) the $b_{R'}$-slope of $l(p)$ is less
than $\pi$, and (3) the $l(q)$-slope of $l(p)$ is less than $\pi$.
Therefore, the $l(q)$-slope of $l(p)$ is the sum of the $l(q)$-slope
of $b_{R'}$ and the $b_{R'}$-slope of $l(p)$. Since the $l(q)$-slope
of $b_{R'}$ is larger than zero, the
claim is true. Since the
$b_{R'}$-slope of $l(p)$ is the pt-slope of $l(p)$ in $R'$,
we obtain that the $l(q)$-slope of $l(p)$ is larger
than the pt-slope of $l(p)$ in $R'$ at the moment $\xi_1$.


To summarize what have been deduced above, we have: (i) the
$l(q)$-slope of $l(p)$ is larger than the pt-slope of $l(p)$ in $R'$
at the moment $\xi_1$, and (ii) at the moment $\xi_2$, the pt-slope
of $l(p)$ in $T'$ is no smaller than the $l(q)$-slope of $l(p)$.
These imply that the pt-slope of $l(p)$ in $R'$ at the moment
$\xi_1$ is smaller than the pt-slope of $l(p)$ in $T'$ at the moment
$\xi_2$. However, $l(p)$ is the directed tangent line of both the
pseudo-triangles $R'$ and $T'$; by Observation \ref{obser:20}(2), the
pt-slope of $l(p)$ in $R'$ at the moment $\xi_1$ must be no smaller
than the pt-slope of $l(p)$ in $T'$ at the moment $\xi_2$, which
incurs a contradiction.

Hence, we conclude that $a$ cannot lie on $\widehat{y_Lz_L}\setminus
\{z_L\}$.
\end{enumerate}

The above analysis shows that $q$ cannot be at the point $a$.

We then discuss the case when $q$ lies on $l(p)$
between $p$ and $a$ (but before $a$). In other words,
$l(p)$ intersects $l(q)$ (at $q$) before $a$ at the moment $\xi_2$,
as shown in Fig.~\ref{fig:forwardview}.
Since the interior of the line segment $\overline{pa}$ connecting $a$ and $p$ (lying on
$l(p)$) does not intersect any
obstacle, it follows that $q$ lies on a directed free bitangent $t_2$
in $B(\partial T')$ at the moment $\xi_2$.
Clearly, $t_2$ has the same direction as $l(q)$ and thus $t_2$
crosses $l(p)$ from left to right.
We let $\overline{t_2}$ be a physical copy of $t_2$ (i.e., they are at
the same location with the same direction) but
$\overline{t_2}$ is not associated with any time moment.
Then $\overline{t_2}$ crosses $l(p)$ from left to right as well.
Note that the interior of $R'$ is free of
obstacles and the segment $\overline{pa}$ is contained in $R'$.
Because $\overline{pa}$ intersects $\overline{t_2}$ before $a$,
we claim that there must be
a (directed) bitangent $t'_1 \in B(\partial R')$ on $\widehat{x_{R'}p}$
or $\widehat{pz_{R'}}$ such that $\overline{t_2}$ crosses $t_1'$ from left to right.
This claim is proved in the next paragraph.

Recall that $\partial R'$ consists of three convex chains, i.e.,
$\widehat{x_{R'}y_{R'}}$, $\widehat{y_{R'}z_{R'}}$,
$\widehat{z_{R'}x_{R'}}$. Note that $\widehat{x_{R'}p} \cup
\widehat{pz_{R'}}=\widehat{x_{R'}y_{R'}}\cup \widehat{y_{R'}z_{R'}}$.
Since the interior of $R'$ is free of
obstacles and $\overline{pa}$ (which is contained in $R'$)
intersects the directed free bitangent $\overline{t_2}$ (at $q$) before $a$, $\overline{t_2}$ must cross
$\partial R'$ somewhere, at $\widehat{x_{R'}p}$ or $\widehat{pz_{R'}}$
(and possibly at other locations of $\partial R'$).
We discuss below the subcase when $\overline{t_2}$ crosses
$\widehat{x_{R'}p}$ (the other subcase can be analyzed similarly).
Let $w$ be the first point of $\widehat{x_{R'}p}$ encountered as walking
on $\overline{t_2}$ from $q$ in the direction of $\overline{t_2}$. Then
since $\overline{t_2}$ is a free bitangent, the point $w$
must lie on another free bitangent $t_w$ on $\widehat{x_{R'}p}$.
Below, we show that $\overline{t_2}$ crosses $t_w$ from left to right.
Note that in Case 1, the
portion $\widehat{wp}$ of $\partial R'$ does not contain the basepoint $p_{R'}$
($=Tail(b_R')$) of $R'$. Further, $\widehat{wp}$ is to the right of both
$\overline{t_2}$ and $l(p)$.
Let $w'$ be the endpoint of the bitangent $t_w$ that lies on $\widehat{wp}$.
As a portion of $\widehat{wp}$, $\widehat{w'p}$ does not contain the
basepoint $p_{R'}$ of $R'$ and $\widehat{w'p}$ is to the right of both
$\overline{t_2}$ and $l(p)$.
Suppose we move a point $w''$ from $p$ along $\widehat{w'p}$ to
$w'$; let $\rho(w'')$ be the ray originating at $w''$ and
shooting in the direction of the directed tangent line of $R'$ at
$w''$ (i.e., $\rho(w'')$ is the directed half-line of the directed tangent line
of $R'$ at $w''$). Then since $\widehat{w'p}$ does not contain the
basepoint $p_{R'}$ of $R'$, the direction of $\rho(w'')$ changes continuously
as we walk along $\widehat{w'p}$.  In particular, when $w''$ is at $p$, $\rho(w'')$ lies on
$l(p)$ and has the same direction as $l(p)$; when $w''$ arrives
at $w'$, $\rho(w'')$ contains $t_w$ and has the same direction as
$t_w$. Note that $l(p)$ crosses $\overline{t_2}$ from right to left.
Since the direction of $\rho(w'')$ changes continuously for $w''\in \widehat{w'p}$
and $\widehat{w'p}$ is to the right of both $\overline{t_2}$ and $l(p)$,
during the movement of $w''$ from $p$ to $w'$ on $\widehat{w'p}$, the
ray $\rho(w'')$ always crosses $\overline{t_2}$ from right to left. In
particular, when $w''$ arrives at $w'$, $\rho(w')$ crosses $\overline{t_2}$ from
right to left. Since $\rho(w')$
has the same direction as $t_w$ and $\overline{t_2}$ crosses $t_w$ (at
$w$), the directed bitangent $t_w$ crosses $\overline{t_2}$ from right to left, or
equivalently, $\overline{t_2}$ crosses $t_w$ from left to right. Letting
$t'_1= t_w$, the claim holds.


Let $t'_2$ be the version of the bitangent $t'_1$ at the moment
$\xi_2$ (i.e., $t'_1$ and $t'_2$ are defined by the same undirected
free bitangent but may have different directions).
So $t'_2$ has opposite direction to $t'_1$ if and only if
$t'_1$ is flipped during the time period from $\xi_1$ to $\xi_2$.
Since $\overline{t_2}$ crosses $t'_1$, $t_2$ crosses $t'_2$ at the
moment $\xi_2$.
Recall that at the moment $\xi_2$, $t_2$ is in $B(\partial T')$.
Thus, since $t_2$ crosses $t'_2$, $t'_2$ cannot be in $B(\calT')$ for the
good pseudo-triangulation $\calT'$ at the moment $\xi_2$.
Because $t'_1\in B(\partial R')$ at the moment $\xi_1$, there must be one
and only one flip operation on $t_1'$ during the time from
$\xi_1$ to $\xi_2$, which reverses the direction of $t_1'$.
Hence, $t'_1$ and $t'_2$ have opposite directions. Since
$\overline{t_2}$ crosses $t'_1$ from left to right (at the moment $\xi_1$)
and $\overline{t_2}$ has the same direction as $t_2$, $t_2$ crosses
$t'_2$ from right to left at the moment $\xi_2$.
However, at the moment $\xi_2$, we have $t_2\in B(\calT')$ and
$t'_2\not\in B(\calT')$ for the current good pseudo-triangulation
$\calT'$; by the third property of the definition of good
pseudo-triangulation, $t_2$ should cross $t'_2$ from left
to right. This incurs a contradiction.


Consequently, the case when $q$ lies on $l(p)$ between $p$ and $a$
(but before $a$) cannot occur.

Therefore, our assumption that the point $p$ is awake on $\partial T'$ at the moment
$\xi_2$ is not correct. In other words,
the point $p$ cannot be awake at the moment $\xi_2$.
Consequently, the atom $A$ cannot be awake at the moment $\xi_2$ (i.e.,
right before the flip of $b_{T'}$).


As a summary for Case 1, we conclude that when $T'=Rtri(b_{T'})$, the atom $A$
cannot be dequeued due to the flip of $b_{T'}$.
This finishes the proof of Case 1 for Sub-lemma \ref{lem:80}(a).

\begin{figure}[t]
\begin{minipage}[t]{\linewidth}
\begin{center}
\includegraphics[totalheight=1.5in]{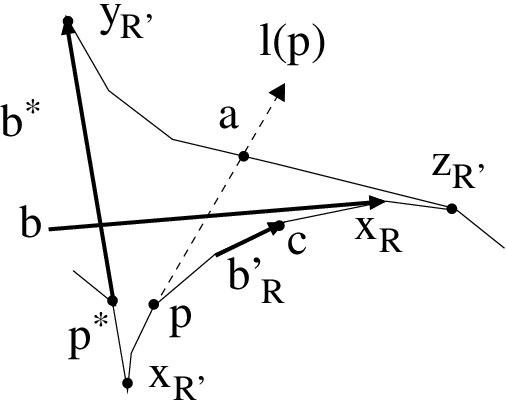}
\caption{\footnotesize Illustrating Case 2.2 in Phase I: $x_R=Head(b)$.}
\label{fig:case21new}
\end{center}
\end{minipage}
\vspace*{-0.15in}
\end{figure}

In Case 2, all dequeued atoms lie on $\widehat{x_Rp^*}$.
Depending on whether $x_R$ is $Head(b'_R)$ or $Head(b)$, there are two
subcases. {\em Case 2.1:} $x_R=Head(b'_R)$ (see
Fig.~\ref{fig:maincases}) and
{\em Case 2.2:} $Head(b)=x_R$ (see Fig.~\ref{fig:case21new}).
For convenience, Case 2.2 will be analyzed after Case 3.

\vspace*{0.05in}
\noindent
{\bf The proof of Case 2.1}. In Case 2.1,
let $A$ be an arbitrary atom on $\widehat{x_Rp^*}$ (dequeued due to the flip of $b$),
and $\xi_1$ be the moment right after the flip of $b$.
Suppose at a later moment $\xi_2$ of the algorithm,
the atom $A$ lies on $\partial T'$ of a pseudo-triangle $T'$
of a good pseudo-triangulation $\calT'$ with $T'=Rtri(b_{T'})$ and $A$ has not been reversed
since the moment $\xi_1$. Without loss of generality,
let $\xi_2$ be the moment right before the flip of $b_{T'}$.
By a similar analysis as for Case 1, we can prove that
$A$ cannot be awake at the moment $\xi_2$
and thus cannot be dequeued due to the flip of $b_{T'}$.
Below we sketch the similarity and (minor) difference between the analysis for Case 2.1 and Case 1.

Let $p$ be an arbitrary interior
point on $A$, and $l(p)$ be the directed tangent line
of $T'$ at $p$ at the moment $\xi_2$. Note that in Case 2.1,
$Head(b)=z_L=q_L=p_L$ still holds (see Fig.~\ref{fig:maincases}).
Further, although Case 2.1 does not involve with $q'_R$,
the same critical structure for this case as for Case 1 is that
$l(p)$ must intersect either $\widehat{y_Lz_L}$ on $\partial L$ or
$\widehat{z_Lx_R}$ on $\partial R$, both lying on the same obstacle, say $P$.
To see this, first, since $\widehat{z_Lx_R}$ is the reversed
portion on $\partial R$ due to the flip of
$b$, by Observation \ref{obser:10}, $\widehat{z_Lx_R}$ is an obstacle arc.
Second, due to $q_L=z_L$,
by Lemma \ref{lem:50}, the portion $\widehat{y_Lq_L}$
($=\widehat{y_Lz_L}$) is an obstacle arc. Let
$a$ be the intersection of $l(p)$ with $\widehat{y_Lz_L}$ or $\widehat{z_Lx_R}$ on $\partial P$.
As in Case 1, the fact still holds
that $p$'s forward $T'$-view point $q$ on $\partial T'$ at the moment
$\xi_2$ is either at the point $a$ or on $l(p)$ between $p$ and $a$ (but before $a$).
Hence, the rest of the analysis simply follows as in Case 1.

We conclude that in Case 2.1, when $T'=Rtri(b_{T'})$, the atom $A$
cannot be awake at the moment $\xi_2$ and consequently
cannot be dequeued due to the flip of $b_{T'}$.
This finishes the proof of Case 2.1.

\vspace*{0.05in}
\noindent
{\bf The proof of Case 3}. For Case 3, all dequeued atoms lie on $\widehat{x_Rp^*}$ of
$\partial R$, which are
immediately reversed after the flip of $b$ (see Fig.~\ref{fig:maincases}). That is, for
each atom $A$ on $\widehat{x_Rp^*}$, after it is dequeued, it is reversed
immediately as well. Hence in this case, for each dequeued atom $A$, it is
obviously true that $A$ is not dequeued again
before its forthcoming reversal (due to the flip of $b$).

\vspace*{0.05in}
\noindent
{\bf The proof of Case 2.2}. For Case 2.2 (see Fig.~\ref{fig:case21new}),
all dequeued atoms lie on $\widehat{x_Rp^*}$. Let
$c=$ $Head(b'_R)$. Note that
the portion $\widehat{x_Rc}$ is reversed due to the flip of $b$.
Similarly to the analysis for Case 3, it is clearly true that
any dequeued atom on $\widehat{x_Rc}$ is not dequeued again
before its reversal.

For the portion $\widehat{cp^*}$ of $\widehat{x_Rp^*}$,
let $A$ be an arbitrary atom on $\widehat{cp^*}$,
and $\xi_1$ be the moment right after the flip of $b$.
Suppose at a later moment $\xi_2$ of the algorithm,
the atom $A$ lies on $\partial T'$ of a pseudo-triangle $T'$
of a good pseudo-triangulation $\calT'$ with $T'=Rtri(b_{T'})$ and $A$ has not been reversed
since the moment $\xi_1$.
Further, let $\xi_2$ be the moment right before the flip of $b_{T'}$.
Next, we prove that $A$ cannot be awake at the moment $\xi_2$
and thus cannot be dequeued due to the flip of $b_{T'}$.
Parts of the proof use similar analysis techniques as for Case
1, which we will only sketch.

Let $p$ be an arbitrary interior point on $A$,
and $l(p)$ be the directed tangent line
of $T'$ at $p$ at the moment $\xi_2$.
For simplicity, we always associate $R'$ (resp,
$T'$) with the moment $\xi_1$ (resp., $\xi_2$).
As in Case 1, $l(p)$ is also the
directed tangent line of $R'$ at $p$.

Let $q\in \partial T'$ be $p$'s forward $T'$-view point on $l(p)$,
and $l(q)$ be the directed tangent line of $T'$ at $q$ (at the moment $\xi_2$).
Recall that
a point $w$ on $\partial T'$ is awake if and only if $w
\in\widehat{x_{T'}q_{T'}}$ ($=Awake[T']$).

Assume to the contrary that the atom $A$ on $\partial T'$ is awake at the moment
$\xi_2$. Then it immediately implies that the point $p\in A$ lies on
$Awake[T']=\widehat{x_{T'}q_{T'}}$
and the pt-slope of $l(p)$ is larger than that of $l(q)$ in $T'$ (see
Fig.~\ref{fig:awake}). Thus, $l(q)$ must cross $l(p)$ from left to right.

We first briefly explain why the argument in Case 1 does not work in
this case. Note that $l(p)$ must intersect $\widehat{y_{R'}z_{R'}}$
on $\partial R'$ (see Fig.~\ref{fig:case21new}); let $a$ be the
intersection point of $l(p)$ and $\widehat{y_{R'}z_{R'}}$. In Case
1, the corresponding point $a$ must lie on an obstacle and thus
$p$'s forward $T'$-view point $q$ on $\partial T'$ at the moment
$\xi_2$ is either at the point $a$ or on $l(p)$ between $p$ and $a$
(but before $a$). However, in Case 2.2, the point $a$ may not lie on
an obstacle. Consequently, the point $q$ may lie on $l(p)$ beyond
the point $a$, i.e., ``behind" the chain $\widehat{y_{R'}z_{R'}}$.
This makes the argument in Case 1 not applicable to Case 2.2. A new
analysis approach is given below.

Clearly, the point $q$ can be on either an
obstacle or a bitangent in
$B(\partial T')$. In the following, we show that neither of these two
cases can occur (the analysis techniques are somewhat similar),
and consequently our assumption that the atom $A$ is awake on
$\partial T'$ is not correct.

We first consider the case when $q$ lies on a bitangent in
$B(\partial T')$, denote by $t_q$ (at the moment $\xi_2$). Since
$t_q$ has the same direction as $l(q)$, $t_q$ crosses $l(p)$ from
left to right. Let $t_q'(\xi_1)$ be the version of the bitangent
$t_q$ at the moment $\xi_1$ (i.e., $t_q'(\xi_1)$ and $t_q$ are
defined by the same undirected free bitangent but may have different
directions). So $t_q$ has opposite direction to $t_q'(\xi_1)$ if and
only if $t_q'(\xi_1)$ is flipped during the time period from $\xi_1$
and $\xi_2$. Recall that by our general position assumption, no
three obstacles have a common tangent line. Note that $l(p)$ must
intersect $\widehat{y_{R'}z_{R'}}$ on $\partial R'$ (see
Fig.~\ref{fig:case21new}); let $a$ be the intersection point of
$l(p)$ and $\widehat{y_{R'}z_{R'}}$. Depending on the relations
between the free bitangent $t_q'(\xi_1)$ and the chain
$\widehat{y_{R'}z_{R'}}$, there are further four subcases to
consider: (i) $t'_q(\xi_1)$ is part of $\widehat{y_{R'}z_{R'}}$;
(ii) $t'_q(\xi_1)$ crosses $\widehat{y_{R'}z_{R'}}$ (i.e.,
$t'_q(\xi_1)$ crosses a bitangent in $\widehat{y_{R'}z_{R'}}$ since
$t'_q(\xi_1)$ is a free bitangent); (iii) $t'_q(\xi_1)$ does not
intersect $\widehat{y_{R'}z_{R'}}$ and $q$ lies on $l(p)$ between
$p$ and $a$; (iv) $t'_q(\xi_1)$ does not intersect
$\widehat{y_{R'}z_{R'}}$ and $q$ lies on $l(p)$ beyond $a$. Note
that in subcase (ii) above, $q$ can be either between $p$ and $a$ or
beyond $a$ on $l(p)$. We prove below that none of these four
subcases can occur.

\begin{enumerate}
\item[(i)] $t'_q(\xi_1)$ is part of $\widehat{y_{R'}z_{R'}}$.
So $q$ lies on $\widehat{y_{R'}z_{R'}}$ of $\partial R'$ in this subcase.
Clearly, the point $p$ lies on $\widehat{cy_{R'}}$ of $\partial
R'$ (recall $c=Head(b'_R)$). Recall that $b_{R'}=b'_R$ (in Case 2).
Thus, the pt-slope of $l(p)$ in $R'$ is less than that of $t'_q(\xi_1)$, and
thus $l(p)$ must cross $t'_q(\xi_1)$ from left to right.
Since $t'_q(\xi_1)\in
B(\partial R')$ and $t_q\in B(\partial T')\subseteq B(\calT')$ for the good
pseudo-triangulation $\calT'$ at the moment $\xi_2$, the bitangent $t'_q(\xi_1)$ is
not flipped during the time period from $\xi_1$ to $\xi_2$ since
otherwise $t_q$ would not be in $B(\calT')$. This implies
that $t_q'(\xi_1)$ and $t_q$ have the same direction. Thus $l(p)$ must cross
$t_q$ from left to right, or equivalently, $t_q$ must cross $l(p)$
from right to left, contradicting with that $t_q$ crosses
$l(p)$ from left to right. Hence, this subcase cannot occur.

\item[(ii)] $t'_q(\xi_1)$ crosses $\widehat{y_{R'}z_{R'}}$
(i.e., $t'_q(\xi_1)$ crosses a bitangent on $\widehat{y_{R'}z_{R'}}$).
As in Case 1, since the interior of $R'$ is free of obstacles,
it is easy to see that $t'_q(\xi_1)$ must also cross $\widehat{z_{R'}x_{R'}}$
or $\widehat{x_{R'}y_{R'}}$ of $\partial R'$.
Note that due to the flip of $b$, the portion $\widehat{x_Rc}$ on $\partial R$ is
reversed. By Observation \ref{obser:10}, $\widehat{x_Rc}$ is an obstacle arc.
Note that $z_L=z_{R'}$ in Case 2.2 (see Fig.~\ref{fig:case21new}).
Since $p_L=Head(b)=x_R$, by Lemma \ref{lem:50},
$\widehat{z_Lp_L}=\widehat{z_{R'}x_R}$ is an obstacle arc.
Thus, $t'_q(\xi_1)$ cannot cross the subchain $\widehat{z_{R'}c}$ of
$\widehat{z_{R'}x_{R'}}$ on $\partial R'$.
We let $\overline{t_q}$ be a physical copy of $t_q$ (i.e., they are at
the same location with the same direction)
but $\overline{t_q}$ is not associated with any time
moment. Then $\overline{t_q}$ crosses $\widehat{cx_{R'}}$
or $\widehat{x_{R'}y_{R'}}$ of $\partial R'$.
By a similar analysis as Case 1, we can show that there must be a
directed free bitangent $t'_1\in B(\partial R')$ on $\widehat{cp}$ or
$\widehat{py_{R'}}$ such that $\overline{t_q}$ crosses $t'_1$ from left to right.
Again, as the analysis for Case 1, this will incur a contradiction
with the third property of good pseudo-triangulation (at the moment
$\xi_2$). Therefore, this subcase cannot occur.

\item[(iii)] $t'_q(\xi_1)$ does not intersect $\widehat{y_{R'}z_{R'}}$ and $q$ lies on
$l(p)$ between $p$ and $a$.
In this subcase,
since the interior of $R'$ is free of obstacles,
$t'_q(\xi_1)$ must cross $\widehat{z_{R'}x_{R'}}$ or $\widehat{x_{R'}y_{R'}}$
of $\partial R'$. The rest of the analysis follows that of
subcase (ii) above. We conclude that this subcase cannot occur.

\item[(iv)] $t'_q(\xi_1)$ does not intersect $\widehat{y_{R'}z_{R'}}$ and $q$ lies on
$l(p)$ beyond $a$. Note that $q\neq a$ since $t'_q(\xi_1)$ does not intersect
$\widehat{y_{R'}z_{R'}}$.
Clearly, $a$ lies on a free bitangent in
$\widehat{y_{R'}z_{R'}}$; let $t'_a$ denote this bitangent (at
the moment $\xi_1$).
Note that $l(p)$ crosses $t'_a$ from left to right since the
pt-slope of $l(p)$ is less than that of $t'_a$ in $R'$.
Let $t_a(\xi_2)$ be the version of
$t'_a$ at the moment $\xi_2$ (i.e., $t'_a$ and $t_a(\xi_2)$ are
defined by the same undirected free bitangent but may have different directions).
Since $q\neq a$, $l(p)$ intersects $t_a(\xi_2)$ (at $a$) in the interior
of the pseudo-triangle $T'$, and thus $t_a(\xi_2)$ cannot be in
$B(\calT')$ for the good
pseudo-triangulation $\calT'$ at the moment $\xi_2$.
Since $t'_a\in B(\partial R')$, there is one and only one flip on $t'_a$ during the
time from $\xi_1$ to $\xi_2$. Thus, $t'_a$ and $t_a(\xi_2)$ have opposite
directions. This implies that $l(p)$ crosses $t_a(\xi_2)$ from right to left, or
equivalently, $t_a(\xi_2)$ crosses $l(p)$ from left to right.
Consider the two endpoints of $t_a(\xi_2)$ with respect to $\partial
T'$. There are three possibilities: (a) Both endpoints of $t_a(\xi_2)$ are on $\partial
T'$; (b) only one endpoint of $t_a(\xi_2)$ is on $\partial T'$; (c)
neither endpoint of $t_a(\xi_2)$
is on $\partial T'$. We show below that none of the above
possibilities can occur,
and thus this subcase cannot occur.

\begin{enumerate}
\item
Both endpoints of $t_a(\xi_2)$ are on $\partial T'$. Note that for
any free bitangent $t^*$ with $t^*\not\in B(\partial T')$, the two
endpoints of $t^*$ cannot be both on $\partial T'$. Since
$t_a(\xi_2)\not\in B(\calT')$, we have $t_a(\xi_2)\not\in B(\partial
T')$ and thus this possibility cannot occur.

\item Only one endpoint of $t_a(\xi_2)$ is on $\partial T'$. The analysis here
utilizes some analysis techniques for Case 1
(specifically for the subcase of $a=q\in \widehat{y_Lz_L}\setminus\{z_L\}$).
As in Case 1, we will show that the pt-slope of
$l(p)$ in $R'$ at the moment $\xi_1$
is smaller than the pt-slope of $l(p)$ in $T'$ at the moment
$\xi_2$. Since
$l(p)$ is the directed tangent line of both the pseudo-triangles
$R'$ and $T'$ at $p$, by Observation \ref{obser:20}, the pt-slope of $l(p)$ in
$R'$ (at the
moment $\xi_1$) must be no smaller than the pt-slope of $l(p)$ in $T'$
(at the moment $\xi_2$), which incurs a contradiction.

Let $l(t_a(\xi_2))$ be the directed line that contains $t_a(\xi_2)$ with the same
direction as $t_a(\xi_2)$.
We consider $l(t_a(\xi_2))$ as a physically directed line that
is not associated with any time moment.
Recall that $t_a(\xi_2)$ and $t'_a$ have opposite directions,
$l(t_a(\xi_2))$ and $t'_a$ have opposite directions.
We claim that the $l(t_a(\xi_2))$-slope of $l(p)$ is
larger than the $b_{R'}$-slope of $l(p)$ at the moment $\xi_1$.
Indeed, at the moment $\xi_1$, the $b_{R'}$-slope of $t'_a$ is the
pt-slope of the point $a$ on $\partial R'$, which is larger
than zero and less than $\pi$. Since $t'_a$ and $l(t_a(\xi_2))$ have
opposite directions, the $l(t_a(\xi_2))$-slope of $b_{R'}$ is less than $\pi$
and larger than zero. Note that $l(p)$ is the directed tangent line
of $R'$. So the $b_{R'}$-slope of $l(p)$ is the pt-slope of $l(p)$
in $R'$ (at the moment $\xi_1$), which is less than $\pi$.
Recall that $t_a(\xi_2)$ crosses $l(p)$ from left to right,
implying the $l(t_a(\xi_2))$-slope of $l(p)$ is less than $\pi$.
Therefore, the $l(t_a(\xi_2))$-slope of $l(p)$ is the sum of
the $l(t_a(\xi_2))$-slope
of $b_{R'}$ and the $b_{R'}$-slope of $l(p)$. Since the $l(t_a(\xi_2))$-slope
of $b_{R'}$ is larger than zero, the claim is true. Since the
$b_{R'}$-slope of $l(p)$ is the pt-slope of $l(p)$ in $R'$,
we obtain that the $l(t_a(\xi_2))$-slope of $l(p)$ is larger
than the pt-slope of $l(p)$ in $R'$ at the moment $\xi_1$.


Let $a'$ be the endpoint of $t_a(\xi_2)$ on $\partial T'$. Since the
direction of $l(t_a(\xi_2))$ is the same as $t_a(\xi_2)$,
$l(t_a(\xi_2))$ is the directed tangent line of $T'$ at
$a'$. Recall that $t_a(\xi_2)$ crosses $l(p)$ from left to right,
implying the
$l(t_a(\xi_2))$-slope of $l(p)$ is less than $\pi$.
By Observation \ref{obser:20}, the $l(t_a(\xi_2))$-slope of $l(p)$ is no
bigger than the pt-slope of $l(p)$ in $T'$ (at the moment $\xi_2$).

We thus obtain that the pt-slope of $l(p)$ in $R'$ at the
moment $\xi_1$ is smaller than the pt-slope of $l(p)$ in $T'$
at the moment $\xi_2$.
Consequently, this possibility cannot occur.

\item Neither endpoint of $t_a(\xi_2)$ is on $\partial T'$. In this situation, since
$t_a(\xi_2)$ intersects $l(p)$ (at $a$) in the interior of $T'$,
$t_a(\xi_2)$ must cross $\partial T'$ somewhere. Note
that $l(p)$ intersects $t_a(\xi_2)$ (at $a$)
before $q$ and $t_a(\xi_2)$ crosses $l(p)$ from left to right.
Recall in our proof by contradiction we assume the atom $A$ is awake on
$\partial T'$ and
the point $p\in A$ lies on $Awake[T']=\widehat{x_{T'}q_{T'}}$ (see
Fig.~\ref{fig:awakeagain}). Note that $q_{T'}$ always lies on
$\widehat{x_{T'}z_{T'}}$ and so does the point $p$.
Since the interior of $T'$ is free of obstacles and $t_a(\xi_2)$ crosses
$l(p)$ from left to right, similar to the analysis for Case 1,
there must be a directed free bitangent $t_1'\in B(\partial T')$ lying on
$\widehat{x_{T'}p}$ or $\widehat{pz_{T'}}$ such that $t_a(\xi_2)$ crosses
$t_1'$ from left to right ($t_a(\xi_2)$ may also cross
$\widehat{z_{T'}x_{T'}}$, but we are not interested in that).
Thus $t_1'$ crosses $t_a(\xi_2)$ from right to left.
However, since $t_1'\in B(\calT')$ and $t_a(\xi_2)\not \in B(\calT')$ for the
good pseudo-triangulation $\calT'$ at the
moment $\xi_2$, by the third property of good
pseudo-triangulation, $t_1'$ should cross $t_a(\xi_2)$ from left to right. But this is a
contradiction. Hence, this possibility cannot occur.

\begin{figure}[t]
\begin{minipage}[t]{\linewidth}
\begin{center}
\includegraphics[totalheight=1.5in]{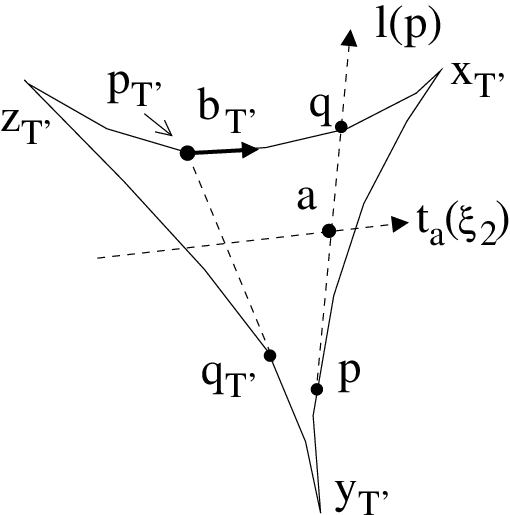}
\caption{\footnotesize Illustrating the situation when no
endpoint of $t_a(\xi_2)$ is on $\partial T'$: $t_a(\xi_2)$ crosses $l(p)$ from left
to right.}
\label{fig:awakeagain}
\end{center}
\end{minipage}
\vspace*{-0.15in}
\end{figure}

\end{enumerate}
\end{enumerate}

We conclude that the case when $q$ lies on a bitangent in
$B(\partial T')$ cannot occur.

Below we discuss the case when $q$ lies on an
obstacle. For simplicity, here we view $q$ as a
physical point not associated with
any time moment. Consider the position of $q$ with respect to the chain
$\widehat{y_{R'}z_{R'}}$ on $\partial R'$. There are two subcases to consider: (i)
$q$ lies on $\widehat{y_{R'}z_{R'}}$; (ii) $q$ does not lie on
$\widehat{y_{R'}z_{R'}}$ (i.e, $l(p)$ crosses a free bitangent in
$\widehat{y_{R'}z_{R'}}$). We show below that neither subcase can occur.

\begin{enumerate}
\item[(i)] $q$ lies on $\widehat{y_{R'}z_{R'}}$. Let $l'(q)$ be the
directed tangent line of $R'$ at $q$ (at the moment $\xi_1$).
Recall that $l(q)$ is the
directed tangent line of $T'$ at $q$ (at the moment $\xi_2$) and
$l(q)$ crosses $l(p)$ from left to right.
Since $q$ lies on an obstacle,
$l(q)$ and $l'(q)$ both lie on the same undirected line but may have
opposite directions. The following analysis is very similar to that for Case 1
(specifically for the subcase of $a=q\in \widehat{y_Lz_L}\setminus\{z_L\}$).

If the point $q$ has not been reversed since the moment $\xi_1$, then
$l(q)$ has the same direction as $l'(q)$.
As analyzed before, the pt-slope of $l(p)$ is
smaller than that of $l'(q)$ in $R'$, and thus $l(p)$ crosses $l'(q)$
from left to right, or equivalently, $l'(q)$ crosses $l(p)$
from right to left, which contradicts with that
$l(q)$ ($=l'(q)$) crosses $l(p)$ from left to right.

If the point $q$ is reversed (at most once by Lemma \ref{lem:70})
during the time from $\xi_1$ to $\xi_2$, then $l'(q)$ and $l(q)$
have opposite directions. By the same analysis as for subcase (iv
(b)) of the former case (i.e., the case when $q$ lies on a bitangent
in $B(\partial T')$), we can show that the pt-slope of $l(p)$ in
$R'$ at the moment $\xi_1$ is smaller than the pt-slope of $l(p)$ in
$T'$ at the moment $\xi_2$, which contradicts with Observation
\ref{obser:20}. We omit the details. Thus, this subcase cannot
occur.

\item[(ii)] $q$ does not lie on $\widehat{y_{R'}z_{R'}}$
(i.e, $l(p)$ crosses a free bitangent on $\widehat{y_{R'}z_{R'}}$).
Clearly, $l(p)$ must cross a free bitangent on
$\widehat{y_{R'}z_{R'}}$ before arriving at $q$. Then, the analysis
follows in exactly the same way as that for subcase (iv) of the
former case, and we can conclude that this subcase cannot occur
either.
\end{enumerate}

In summary, we prove that the atom $A$ cannot be awake at the moment
$\xi_2$ of the algorithm.

For Case 2.2, we conclude that when $T'=Rtri(b_{T'})$, the atom $A$ on $\partial T'$
cannot be dequeued due to the flip of $b_{T'}$.

Based on the above detailed case analysis, Sub-lemma \ref{lem:80}(a) is proved.

By an analogous analysis, we can also prove
the following statement, called Sub-lemma \ref{lem:80}(b).

\vspace*{0.08in}
\noindent
{\bf Sub-lemma \ref{lem:80}(b)}. Suppose $A$ is an atom on $\partial T$ of a pseudo-triangle $T$ with
$T=Ltri(b_T)$ in a good pseudo-triangulation $\calT$ and $A$ is dequeued due to
a flip operation on $b_T$. Also, suppose at any later moment before the reversal
of $A$, $A$ lies on $\partial T'$ of a
pseudo-triangle $T'$ with $T'=Ltri(b_{T'})$ in another good pseudo-triangulation
$\calT'$. Then $A$ cannot be dequeued due to the flip of $b_{T'}$.
\vspace*{0.08in}

By combining Sub-lemmas \ref{lem:80}(a) and \ref{lem:80}(b), we conclude that for any dequeued atom, it
can be dequeued at most twice before its next reversal (if any). An explanation of
this is as follows.

Suppose at the moment $\xi_1$,
an atom $A$ on $\partial T_1$ of a pseudo-triangle $T_1$
in a good pseudo-triangulation $\calT_1$ is dequeued for the
first time due to the flip of $b_{T_1}$.
We assume $T_1=Rtri(b_{T_1})$ (the case for $T_1=Ltri(b_{T_1})$
can be analyzed similarly). If $A$ will not be dequeued again
in the algorithm, then we are done. Otherwise, suppose at
a later moment $\xi_2$ ($A$ has not been reversed),
$A$ is on $\partial T_2$ of a pseudo-triangle $T_2$
in a good pseudo-triangulation $\calT_2$ and $A$ is dequeued for the
second time due to the flip of $b_{T_2}$.
By Sub-lemma \ref{lem:80}(a), the case $T_2=Rtri(b_{T_2})$ cannot occur.
Thus, only $T_2=Ltri(b_{T_2})$ is possible.
For any moment $\xi_3$ after $\xi_2$, suppose $A$ is on $\partial T_3$
of a pseudo-triangle $T_3$ in a good pseudo-triangulation $\calT_3$
and $A$ has not been reversed.
Now, if $T_3=Rtri(b_{T_3})$,
then by Sub-lemma \ref{lem:80}(a), $A$ cannot be dequeued due
to the flip of $b_{T_3}$.
But if $T_3=Ltri(b_{T_3})$, then
by Sub-lemma \ref{lem:80}(b), $A$ cannot be dequeued either.

Thus, $A$ can be dequeued
at most twice in Phase I of the entire algorithm before its next
reversal (if any).
Lemma \ref{lem:80} then follows.

Note: It appears possible to show that there is at most {\em one}
dequeue per atom before its reversal.  But, proving this seems to
make the already long and complicated proof of Lemma \ref{lem:80}
even longer and more complicated, and this stronger statement,
although nicer, is not essential to our result.

In addition, we briefly discuss why $R'$ ($=Rtri(b^*)$) is also $Ltri(b'_R)$ in
Case 2 with the help of Lemma \ref{lem:50}. In Case 2.1 (i.e.,
$x_R=Head(b'_R)$), this is obviously true. We discuss Case 2.2 (i.e.,
$x_R=Head(b)$) below. It suffices to show that
$b'_R$ lies on $\widehat{x_Ry_R}$ of $\partial R$. Assume to the
contrary that $b'_R$ does not lie on $\widehat{x_Ry_R}$ of $\partial R$.
Then, $\widehat{x_Ry_R}$ must be an obstacle arc on an obstacle, say
$P$. Thus, it is easy to see that $b'_R$ must be tangent to $P$ and
the tangent point on $P$ is $y_R$, and $b'_R$ lies on
$\widehat{y_Rz_R}$. Further, since in Case 2, $p^*$ does not lie on the obstacle
arc between $b$ and $b'_R$, we have $p^*\not\in
\widehat{x_Ry_R}$, and thus $b'_R\in \widehat{y_Rp^*}$. Note that
$\widehat{y_Rp^*}$ is part of $\widehat{x_Rq_R}$. Therefore, $b'_R\in
\widehat{y_Rq_R}$. However, since $y_R\in \widehat{x_Rq_R}$, by Lemma
\ref{lem:50}, the portion $\widehat{y_Rq_R}$ of $\partial R$ is an
obstacle arc, contradicting with $b'_R\in \widehat{y_Rq_R}$ since
$b'_R$ is a free bitangent. Hence, $b'_R$ must lie on
$\widehat{x_Ry_R}$ and $R'$ is $Ltri(b'_R)$.

\subsection{Bounding the Number of Enqueue Operations in Phase II}
\label{sec:PhaseII}

In this section, we prove $n_E\leq n_Q+n_D+n_S+k$ and $n_S=O(n+k)$.
Consequently, due to $n_Q=O(n+k)$, $n_D=O(n+k)$, and $k=|\calB|$,
we obtain $n_E=O(n+k)$ and Lemma
\ref{lem:keylemma} thus follows.

Let $Q$ be the
set of all reversed atoms in the entire algorithm, $D$ be the set of all
dequeue operations in Phase I, and $S$ be the set of {\em
special enqueue} operations in Phase II that will be defined later.
Thus, $n_Q=|Q|$, $n_D=|D|$, and $n_S=|S|$.

To prove $n_E\leq n_Q+n_D+n_S+k$,
we will show that every enqueue operation in
Phase II corresponds to an element in $D$, $Q$, $S$, or
$\calB$, and we {\em charge} the enqueue operation to that element;
each such element will be charged only $O(1)$ times in
the entire algorithm. Further, we will prove that $n_S=|S|=O(n+k)$.
We discuss the three main cases individually.

For Case 1, refer to the pseudocode Algorithm \ref{algo:case1} in
Section \ref{subsubsec-Phase-II}.
There are three enqueue sequences, i.e., Lines \ref{line:case110},
\ref{line:case120}, and \ref{line:case130}, and their corresponding
charges are already shown in the pseudocode. We briefly explain why we
can charge them in those ways. For Line \ref{line:case110}, it is easy
to see that $\widehat{z_Lp_{R'}}$ (recall $p_{R'}=Tail(b'_R)$)
is reversed due to the flip of $b$. So we can charge the enqueue on
$\widehat{z_Lp_{R'}}$ to $Q$. For Line \ref{line:case120}, since
$b^*\in \calB$, we can charge the enqueue on $b^*$ to $\calB$.
For Line \ref{line:case130}, recall that when computing $b^*$ in
Phase I, all atoms in $\widehat{q'_Rp^*}$ have been dequeued (from
$AwakeMax[R]$). Note that the atoms in the enqueued portion of Line
\ref{line:case130} are all in $\widehat{q'_Rp^*}$, implying that they
have been dequeued in Phase I. So we can charge them to $D$.

For Case 2, again, we discuss the two subcases Case 2.1 (i.e.,
$x_R=Head(b'_R)$, see Fig.~\ref{fig:maincases}) and
Case 2.2 (i.e., $x_R=Head(b)$, see Fig.~\ref{fig:case21}).

For Case 2.1, refer to the pseudocode Algorithm \ref{algo:case21}.
There are three enqueue sequences, i.e., Lines \ref{line:case210},
\ref{line:case220}, and \ref{line:case230}, and their corresponding
charges are already shown in the pseudocode.
For Line \ref{line:case210}, note that $\widehat{z_Lx_R}$ is reversed
due to the flip of $b$, so we charge the enqueue operations on
$\widehat{z_Lx_R}$ to $Q$. For Line \ref{line:case220}, we charge the
enqueue on $b^*$ to $\calB$. For Line \ref{line:case230}, note that
the atoms in $\widehat{y_Rp^*}$ are dequeued when computing $b^*$ in
Phase I, so we charge the enqueue to $D$.

For Case 2.2, refer to the pseudocode Algorithm \ref{algo:case22}.
There are five enqueue sequences, i.e., Lines \ref{line:case2210},
\ref{line:case2220}, \ref{line:case2230}, \ref{line:case2240}, and
\ref{line:case2250}. For Line \ref{line:case2210}, which is for the
case when $q_{R'}\in\widehat{p_{R'}p^*}$, note that the enqueued
portion is $\widehat{x_{R'}q_{R'}}$. Since
$q_{R'}\in\widehat{p_{R'}p^*}$, $q_{R'}$ is before $p^*$, and thus
all the enqueued atoms in Line \ref{line:case2210} are dequeued for
computing $b^*$ in Phase I. Therefore, we can charge them to $D$.
Line \ref{line:case2220} is trivial. For Line \ref{line:case2230},
again, the atoms in the enqueued portion $\widehat{y_Rp^*}$ have
been dequeued for computing $b^*$, we charge those enqueue
operations to $D$. For Line \ref{line:case2240}, note that
$p_{R'}=Head(b'_R)$. Since $x_{R}=Head(b)$, the enqueued portion
$\widehat{x_Rp_{R'}}$ is reversed due to the flip of $b$, so we can
charge the enqueue to $Q$. The enqueue in Line \ref{line:case2250}
needs special treatment. Below we
discuss that the enqueued atoms there have some special properties
and we call those enqueue the {\em special enqueue operations} and
charge them to $S$. Later, we will prove $|S|=O(n+k)$.

We assume that the reader has read the
detailed algorithm discussion for Algorithm \ref{algo:case22} in
Section \ref{subsubsec-Phase-II}. Note that Line \ref{line:case2250}
is for the case $w_L=q_L$ (recall that $w_L$ is defined to be $q_L$ if $q_L\in \widehat{y_Lz_L}$ and $y_L$ otherwise). First, note that the enqueue sequence in
Line \ref{line:case2250} is on $\widehat{w_{R'}q_L}$ of $\partial
R'$, which is part of $\widehat{q_{R'}q_L}$ regardless of whether
$w_{R'}$ is $y_{R'}$ or $q_{R'}$. We claim that
$\widehat{w_{R'}q_L}$ is part of $\widehat{y_Lq_L}$ on $\partial L$.
To prove this claim, it suffices to show that $w_{R'}$ is after
$y_L$ on $\partial L$. Clearly, $w_{R'}$ is always after $y_{R'}$.
Recall that $y_{R'}$ is $q^*$ or $y_L$. If $y_{R'}$ is $q^*$, then
$q^*$ is after $y_L$ on $\partial L$; else, $y_{R'}$ is $y_L$. In
either case, $y_{R'}$ is always after $y_L$ on $\partial L$.
Consequently, $w_{R'}$ is after $y_L$ on $\partial L$. The claim
thus follows. Hence, the special enqueue sequence is on
$\widehat{w_{R'}q_L}$, which is part of both $\widehat{q_{R'}q_L}$
and $\widehat{y_Lq_L}$.

Due to $w_L=q_L$, $y_L$ is before $q_{L}$ on $\partial L$, and in
other words $y_L$ lies on $\widehat{x_Lq_L}$. By Lemma \ref{lem:50},
$\widehat{y_Lq_L}$ is an obstacle arc and thus $\widehat{w_{R'}q_L}$
also lies on that obstacle. For any point $p$ on an atom $A$ of
$\widehat{w_{R'}q_L}$ on $\partial R'$, let $l(p)$ be the directed
tangent line of the pseudo-triangle $R'$ at $p$. Suppose we move
from $p$ along $l(p)$ towards its inverse direction (resp., the
direction of $l(p)$), and let $a$ (resp., $a'$) be the first point
encountered on any {\em obstacle} in $\calP$; we call the point $a$
(resp., $a'$) the {\em backward $\calP$-view} (resp., {\em forward
$\calP$-view}) of $p$. Let $\calP_{back}(A)$ (resp.,
$\calP_{for}(A)$) denote the set of backward (resp., forward)
$\calP$-view points of the points of $A$. Since
$\widehat{x_Rp_{R'}}$ (recall $p_{R'}=Head(b'_R)$) on $\partial R$
is reversed due to the flip of $b$, by Observation \ref{obser:10},
$\widehat{x_Rp_{R'}}$ is an obstacle arc.  An easy but critical
observation is that for any point on $\widehat{q_{R'}q_L}$, its
backward $\calP$-view is on $\widehat{x_Rp_{R'}}$ (see
Fig.~\ref{fig:case21}). Since the enqueued portion
$\widehat{w_{R'}q_L}$ is part of $\widehat{q_{R'}q_L}$, for any
point on $\widehat{w_{R'}q_L}\setminus\{q_{R'}\}$, its backward
$\calP$-view is on $\widehat{x_Rp_{R'}}$, which is reversed due to
the flip of $b$. To summarize what have been deduced above, we have
(i) the portion of $\partial R'$ involved in the special enqueue
sequence is $\widehat{w_{R'}q_L}$, which is an obstacle arc, and
(ii) the backward $\calP$-view points of all points on
$\widehat{w_{R'}q_L}\setminus\{q_{R'}\}$ lie on
$\widehat{x_Rp_{R'}}$, which is an obstacle arc reversed due to the
flip of $b$. We then have the following observation.

\begin{observation}\label{obser:numberofS1}
Consider a flip operation on a minimal bitangent $b$. If an atom $A$
on $\partial R'$ with $R'=Rtri(\varphi(b))$ is involved in a special
enqueue operation (for processing the flip of $b$), then $A$ is an
obstacle arc and $\calP_{back}(A)$ is an obstacle arc reversed due
to the flip of $b$.
\end{observation}

In addition, if an atom $A$ is an obstacle arc, say, on the obstacle
$P$, then for any point $p\in A$, the tangent line of $P$ at $p$ is
also the tangent line of the pseudo-triangle (at $p$) on which $A$
lies, and vice versa. Thus, as long as the atom $A$ is not reversed
in the algorithm, both $\calP_{back}(A)$ and $\calP_{for}(A)$ will
not change. After $A$ is reversed (if this ever happens),
$\calP_{back}(A)$ and $\calP_{for}(A)$ still refer to the same two
obstacle arcs but switch names with each other.

The bound of $|S|$ will be discussed later in Lemma
\ref{lem:numberofS}. Some special enqueue
operations also appear in Case 3.

For Case 3 (see Fig.~\ref{fig:case3}), refer to the pseudocode
Algorithm \ref{algo:case3}. There are three enqueue sequences, i.e.,
Lines \ref{line:case310}, \ref{line:case320}, and
\ref{line:case330}. For Line \ref{line:case310}, since
$\widehat{x_Rp^*}$ is reversed due to the flip of $b$, we charge the
enqueue to $Q$. Line \ref{line:case320} is trivial. For Line
\ref{line:case330}, we explain below that the enqueue can also be
viewed as special enqueue and charged to $S$.

Note that Line \ref{line:case330} is on the case $w_L=q_L$. Thus,
$y_L\in \widehat{x_Lq_L}$. By Lemma \ref{lem:50}, $\widehat{y_Lq_L}$
is an obstacle arc. Recall that in Case 3 $x_{R'}$ is either $y_L$
or $q^*$. If $x_{R'}=q^*$, $q^*$ is on $\widehat{y_Lz_L}$;
otherwise, $x_{R'}=y_L$. In either case, the enqueued portion in Line
\ref{line:case330}, i.e.,
$\widehat{x_{R'}q_L}$, lies on $\widehat{y_Lq_L}$, which is an
obstacle arc. Note that the backward $\calP$-view of any point on
$\widehat{x_{R'}q_L}$ is on $\widehat{x_Rp^*}$ ($x_R=Head(b)$,
see Fig.~\ref{fig:case3}),
which is reversed due to the flip of $b$ and is an obstacle arc by
Observation \ref{obser:10}.

In summary, we have: (i) the enqueued portion $\widehat{x_{R'}q_L}$
in Line \ref{line:case330} is an obstacle arc, and (ii) the backward
$\calP$-view points of all points on $\widehat{x_{R'}q_L}$ lie on
$\widehat{x_Rp^*}$, which is an obstacle arc reversed due to the
flip of $b$. Thus, Observation \ref{obser:numberofS1} also applies
to the enqueue on $\widehat{x_{R'}q_L}$ in Line \ref{line:case330}.
Therefore, we also treat the enqueue as special enqueue and charge
them to $S$.

We have finished the discussion on charging the enqueue operations
in Phase II for all cases. It remains to show $|S|=O(n+k)$. For
this, we first give a similar observation on the (possible) special
enqueue operations on $\partial L'$ with $L'=Ltri(\varphi(b))$ for
the flip operation on $b$.

\begin{observation}\label{obser:numberofS2}
Consider a flip operation on a minimal bitangent $b$. If an atom $A$
on $\partial L'$ with $L'=Ltri(\varphi(b))$ is involved in a special
enqueue operation (for processing the flip of $b$), then $A$ is an
obstacle arc and $\calP_{for}(A)$ is an obstacle arc reversed due to
the flip of $b$.
\end{observation}

Based on Lemma \ref{lem:70} and Observations \ref{obser:numberofS1}
and \ref{obser:numberofS2}, we prove the following lemma.

\begin{lemma}\label{lem:numberofS}
The number $|S|$ of all special enqueue operations in Phase II of
the entire algorithm is $O(n+k)$.
\end{lemma}

\begin{proof}
Consider an arbitrary atom $A$ in a good pseudo-triangulation at a
moment $\xi_0$ during the algorithm ($A$ may have been reversed).
First, it is important to note that $A$ can be involved in a special
enqueue operation only if $A$ is an obstacle arc and the special
enqueue operation is for processing a flip operation on a minimal
bitangent $b$ such that $A$ is on $\partial T$ with $T\in
\{Rtri(\varphi(b)), Ltri(\varphi(b))\}$. In the following, we show
that the atom $A$ can be involved in at most two special enqueue
operations before its next reversal (if any). If $A$ is not an
obstacle arc, the above statement simply holds. We assume $A$ is an
obstacle arc. We assume that the discussion below is on the time
period after the moment $\xi_0$ and before the next reversal of $A$
(if any), unless otherwise stated.

If the atom $A$ is not involved in any special enqueue operation in
the algorithm later on, then we are done. Otherwise, suppose at a
later moment $\xi_1>\xi_0$, $A$ is involved in a special enqueue
operation for processing the flip of a
 minimal bitangent $t_1$. Without loss of generality, we assume that $A$ is on
$\partial Rtri(\varphi(t_1))$. Then, by Observation
\ref{obser:numberofS1}, the obstacle arc $\calP_{back}(A)$ is
reversed due to the flip of $t_1$. By Lemma \ref{lem:70},
$\calP_{back}(A)$ cannot be reversed again later in the algorithm.

If $A$ is not involved in any special enqueue operation in the
algorithm after the moment $\xi_1$, then we are done. Otherwise,
suppose at a later moment $\xi_2>\xi_1$, $A$ is involved in a
special enqueue operation for processing the flip of a minimal
bitangent $t_2$. Because $A$ has not been reversed since the moment
$\xi_0$, as discussed earlier, both $\calP_{back}(A)$ and
$\calP_{for}(A)$ do not change. Then, $A$ cannot be on $\partial
Rtri(\varphi(t_2))$ since otherwise, by Observation
\ref{obser:numberofS1}, the obstacle arc $\calP_{back}(A)$ would be
reversed again due to the flip of $t_2$. Hence, $A$ can only be on
$\partial Ltri(\varphi(t_2))$. By Observation
\ref{obser:numberofS2}, the obstacle arc $\calP_{for}(A)$ is
reversed due to the flip of $t_2$. By Lemma \ref{lem:70},
$\calP_{for}(A)$ cannot be reversed again later.

Consider any flip operation on a minimal bitangent $t_3$ in the
algorithm at any later moment $\xi_3>\xi_2$. Because $A$ has not
been reversed since the moment $\xi_0$, both $\calP_{back}(A)$ and
$\calP_{for}(A)$ do not change. No matter whether $A$ is on
$\partial Rtri(\varphi(t_3))$ or $\partial Ltri(\varphi(t_3))$, $A$
cannot be involved in any special enqueue operation for processing
the flip of $t_3$, since otherwise, by Observations
\ref{obser:numberofS1} and \ref{obser:numberofS2}, the obstacle arc
$\calP_{back}(A)$ or $\calP_{for}(A)$ would be reversed again due to
the flip of $t_3$.

Therefore, we obtain that $A$ can be involved in at most two special
enqueue operations before its next reversal (if any). By Lemma
$\ref{lem:70}$, $A$ can be reversed at most once. We now claim that
$A$ can be involved in at most two special enqueue operations in the
entire algorithm. Indeed, if $A$ is involved in two special enqueue
operations before its reversal (if any), then both $\calP_{back}(A)$
and $\calP_{for}(A)$ are reversed before the reversal of $A$. Note
that after $A$ is reversed, $\calP_{back}(A)$ and $\calP_{for}(A)$
refer to the same two obstacle arcs but switch names with each
other. Since neither of $\calP_{back}(A)$ and $\calP_{for}(A)$ can
be reversed again, the reversed atom $A$ cannot be involved in any
special enqueue operation in the rest of the algorithm. If $A$ is
involved in one special enqueue operation before its reversal, then
exactly one of $\calP_{back}(A)$ and $\calP_{for}(A)$ is reversed
before the reversal of $A$. After $A$ is reversed, only one of
$\calP_{back}(A)$ and $\calP_{for}(A)$ (with their names switched
with each other) can possibly be reversed, and thus in this
situation the reversed $A$ can be involved in at most one special
enqueue operation in the rest of the algorithm.  Finally, if $A$ has
not been involved in any special enqueue operation before its
reversal, then the claim simply holds. Therefore, the claim is true
and the atom $A$ can be involved in at most two special enqueue
operations in the entire algorithm.

Since the number of all atoms in the algorithm is $O(n+k)$, the
number of all special enqueue operations in Phase II of the entire
algorithm is $O(n+k)$, i.e., $|S|=O(n+k)$.
\end{proof}

\section{Computing a Shortest Path in the Relevant Visibility Graph}
\label{sec:shortestpath}

In this section, we compute a shortest path from $s$ to $t$ in $G$,
which is the relevant visibility graph of the $O(h)$ pairwise
disjoint convex splinegons in $\calS'$ with a total of $O(n)$ vertices.
Recall that $k$ is the number of the free common tangents of all
splinegons in $\calS'$. For convenience, we assume that the number
of convex splinegons in $\calS'$ is $h$ and the total number of
splinegon vertices is $n$. Let $\calS'=\{S_1, S_2, \ldots, S_h\}$.

To find a shortest path from $s$ to $t$ in the graph $G$, since $G$
has $O(k)$ nodes and $O(k)$ edges, simply running Dijkstra's
algorithm on $G$ would take $O(k\log k)$ time. To avoid the $\log k$
factor, we transform $G$ to a {\em coalesced graph} $G^c$ such that:
(1) $G^c$ has only $O(h)$ nodes and $O(k)$ edges; (2) a shortest
$s$-$t$ path in $G$ corresponds to a shortest $s$-$t$ path in $G^c$,
which can be found in $O(h\log h+k)$ time. This approach is quite
similar to that in \cite{ref:ChenCo11} for computing a shortest
$s$-$t$ path among $n$ convex pseudodisks of $O(1)$ complexity each.
In general, the approach in \cite{ref:ChenCo11} relies only on the
convexity of the objects involved and thus is applicable to our
problem setting. Note that the idea of using a coalesced graph was
first proposed by Hershberger and Guibas \cite{ref:HershbergerAn88},
but the definition of the coalesced graph and its construction in
\cite{ref:ChenCo11} are both different from those in
\cite{ref:HershbergerAn88}. We extend the method in
\cite{ref:ChenCo11} to solving our problem in the splinegon setting.

As the approach in \cite{ref:ChenCo11}, a key to our algorithm is to
compute a set of $O(h)$ ``distinguished points" on the boundaries of
the splinegons in $\calS'$, which are then used to construct $G^c$.
By a proof similar to that in \cite{ref:ChenCo11}, a set of $O(h)$
distinguished points can be obtained easily once the Voronoi diagram
of the convex splinegons in $\calS'$ is available. Denote by
$\Vor(\calS')$ the Voronoi diagram of the $h$ convex splinegons in
$\calS'$. The next lemma follows from the results in \cite{ref:ChenCo11}.

\begin{lemma}{\em \cite{ref:ChenCo11}}\label{lem:coalconstruct}
After the Voronoi diagram $\Vor(\calS')$ is built, the
coalesced graph $G^c$ with $O(h)$ nodes and $O(k)$ edges can be
constructed in $O(n+k+h\log h)$ time.
\end{lemma}

It remains to describe how to compute $\Vor(\calS')$. In Section
\ref{sec:voronoidiagram}, we will show that $\Vor(\calS')$ can be
computed in $O(n+h\log h)$ time and $O(n)$ space, which is optimal. Thus, we have the
following result.

\begin{theorem}\label{theo:201}
A shortest $s$-$t$ path for the convex \spp\ can be found in
$O(n+h\log h+k)$ time, where $k=O(h^2)$ is the number of free common
tangents among the convex splinegons of $\calS'$.
\end{theorem}

Recently, Chen {\em et al.} \cite{ref:ChenCo13} have shown the following result:
Suppose we can prove that a distinguished point set $P$ with $|P|=O(h)$
exists; then another distinguished point set $P'$ with $|P'|=O(h)$ can
be found by a simple greedy algorithm in $O(n+k)$ time. Since it is proved that
such a set $P$ exists \cite{ref:ChenCo11},
we can use the algorithm in \cite{ref:ChenCo13} to compute
another set $P'$ without computing the Voronoi diagram
$\Vor(\calS')$. In this
way, the coalesced graph $G^c$ can still be constructed
in $O(n+k+h\log h)$
time by using the distinguished points in $P'$.
However, since computing $\Vor(\calS')$ itself is an interesting
problem, we choose to present our optimal solution for it in Section
\ref{sec:voronoidiagram}.

\subsection{The Voronoi Diagram of Convex Splinegons}
\label{sec:voronoidiagram}

In this section, we compute the Voronoi diagram $\Vor(\calS')$ for a
set $\calS'$ of $h$ pairwise disjoint convex splinegons of totally
$n$ vertices. To our best knowledge, no efficient algorithm was
given previously for it. By extending Fortune's sweeping algorithm
\cite{ref:FortuneA87}, one may obtain an $O(n+h\log h\log n)$ time
solution for it. For the convex polygon case (i.e., all splinegons in $\calS'$ are
convex polygons), $\Vor(\calS')$ can be computed in $O(n+h\log h)$
time \cite{ref:McAllisterA96}.
We show that by generalizing the algorithm in
\cite{ref:McAllisterA96}, $\VD(\calS')$ for the convex splinegon
case can also be computed in $O(n+h\log h)$ time (as in \cite{ref:McAllisterA96},
we assume that the edges of each input splinegon are represented
as a cyclically ordered list). Note that since the combinatorial complexity of each splinegon edge is $O(1)$, we assume the bisector of any two splinegon edges can be computed in constant time.

In fact, as in \cite{ref:McAllisterA96}, we achieve a stronger result:
The {\em compact diagram} (to be
defined below) of the convex splinegons in $\calS'$ can be computed
in $O(h\log n)$ time, from which $\VD(\calS')$ can be derived in
additional $O(n)$ time. Note that $h\log n=O(n+h\log h)$. As in
\cite{ref:McAllisterA96} and to be discussed later, the compact
diagram has several advantages over the ``normal" Voronoi diagram.

\begin{figure}[t]
\begin{minipage}[t]{\linewidth}
\begin{center}
\includegraphics[totalheight=1.7in]{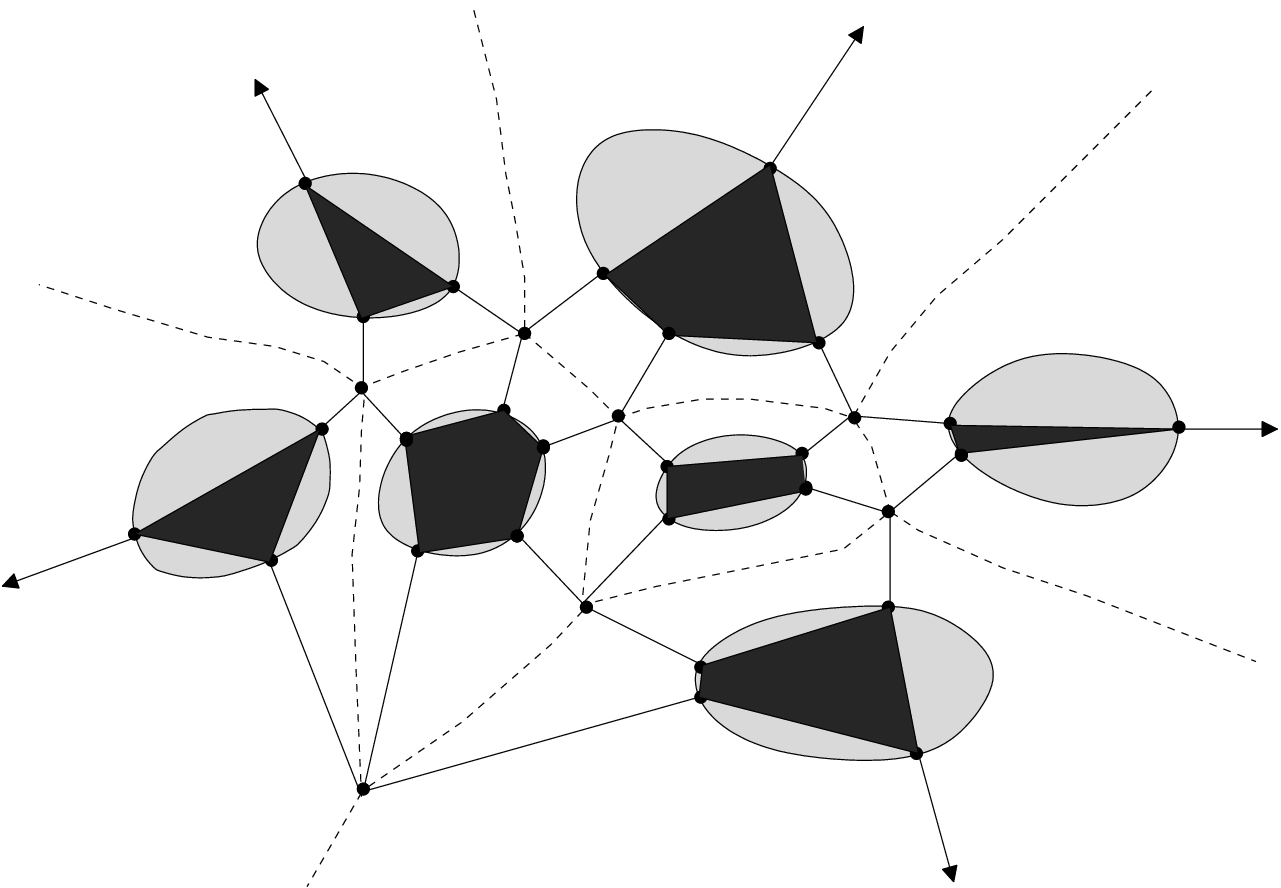}
\caption{\footnotesize Illustrating the compact diagram of a set of
convex splinegons: The black polygons are the cores of the
splinegons. The Voronoi diagram of the splinegons is shown with
dashed curves.} \label{fig:compactdiagram}
\end{center}
\end{minipage}
\vspace*{-0.15in}
\end{figure}

We first formally define the compact diagram of $\calS'$, denoted by
$\CD(\calS')$. We follow the terminology in
\cite{ref:McAllisterA96}. Consider a convex splinegon $S\in\calS'$, which
is contained in a Voronoi cell of $\VD(\calS')$, say $C_S$. For each
Voronoi vertex $v$ on the boundary of $C_S$, we draw a line segment
from $v$ to its closest point, say $p_v$, on $S$. The segment
$\overline{vp_v}$ is called the {\em spoke} from $v$ to $S$ and the
point $p_v$ is called the {\em spoke attachment point}. If the cell
$C_S$ is unbounded, then there is a point on $\partial S$ whose
normal does not intersect the cell boundary; we view this normal as
a spoke from an infinite Voronoi vertex to $S$. The {\em core} of
$S$ is the convex hull of all spoke attachment points on $S$. The
{\em compact diagram} $\CD(\calS')$ is the union of all spokes and
cores of the splinegons in $\calS'$ (see
Fig.~\ref{fig:compactdiagram}). Theorem \ref{theo:compact} gives the
algorithm for computing $\CD(\calS')$.

Besides its efficient construction, the compact diagram has several
advantages over the Voronoi diagram. Note that while the Voronoi
diagram $\VD(\calS')$ may have $O(n)$ high degree (but still constant)
algebraic curves (whose shapes depend on the boundaries of the convex splinegons of $\calS'$), the compact diagram $\CD(\calS')$ consists of only
$O(h)$ line segments. This feature makes the compact diagram easier
and more efficient to display and represent. In addition, for
applications in which knowing only two candidates for the closest
splinegons is sufficient, the original splinegons can be discarded
and only the $O(h)$ segments of $\CD(\calS')$ need to be stored,
using $O(h)$ instead of $O(n)$ space. In \cite{ref:McAllisterA96},
two applications of the compact diagram were discussed, i.e., the
post-office problem and the retraction motion planning problem, in
which the sites were modeled as convex polygons. With our results,
if the sites are modeled as convex splinegons, then the
corresponding post-office problem and the retraction motion planning
problem can be handled similarly with the same performance as in
\cite{ref:McAllisterA96}. Our results may also find other
applications.

\begin{theorem}\label{theo:compact}
The compact diagram $\CD(\calS')$ of the convex splinegons in
$\calS'$ can be computed in $O(h\log n)$ time, from which the
Voronoi diagram $\VD(\calS')$ can be derived in additional $O(n)$
time.
\end{theorem}

\begin{proof}
We first focus on computing $\CD(\calS')$. For this, we generalize
the corresponding algorithm in \cite{ref:McAllisterA96}. Given a set
$\calP$ of $h$ pairwise disjoint convex polygons of totally $n$
vertices (each polygon is represented in a standard fashion),
McAllister, Kirkpatrick, and Snoeyink gave an algorithm
\cite{ref:McAllisterA96} for computing the compact diagram of the
convex polygons of $\calP$ in $O(h\log n)$ time.
We refer to their algorithm as the MKS algorithm.  We first sketch
the MKS algorithm and then discuss our generalization of it on the
convex splinegon set $\calS'$.

Like Fortune's approach \cite{ref:FortuneA87}, the MKS algorithm is
a sweeping algorithm, sweeping the convex polygons of $\calP$ from
(say) left to right. As in Fortune's algorithm, the Voronoi cell
boundary maintained by the sweepline is called the {\em sweep front}
or {\em beach line}, which consists of Voronoi edges between the
sweepline and some polygons. A maximal connected portion of the
sweep front between the sweepline and a single polygon is called a
{\em front arc}. There are two types of events in the sweeping
process. A {\em site event} occurs when the sweepline reaches the
leftmost point of a polygon. A {\em circle event} occurs when the
sweepline reaches the rightmost point of a circle that is tangent to
three polygons of some consecutive front arcs on the sweep front.
Clearly, there are $O(h)$ site events and $O(h)$ circle events. The
MKS algorithm focuses on computing the vertices of the Voronoi
diagram as well as identifying the polygons that generate those
Voronoi vertices. Two data structures are maintained by the sweeping
algorithm: A balanced binary search tree that stores the sweep front
and a priority queue that schedules the events in the order that the
sweepline will encounter them.

Generalizing Fortune's algorithm in a straightforward manner would
take $O(h\log h\log n)$ time since every site event is processed in
$O(\log h\log n)$ time. Specifically, at each site event, the
sweepline is at the leftmost point, say $p$, of a polygon, and it
needs to determine the front arc on the sweep front that is closest
to $p$. A straightforward processing of this task takes $O(\log
h\log n)$ time. Based on a critical observation
\cite{ref:McAllisterA96}, the MKS algorithm handles each site event
in $O(\log n)$ time.
This observation
states that the sweepline can be partitioned into disjoint intervals
such that finding the nearest front arc to the point $p$ is
equivalent to locating in which interval the point $p$ lies, which
can be carried out in $O(\log n)$ time. Two subroutines heavily used
in the MKS algorithm are $spoke(p,A)$ and $vertex(A,B,C)$. Given a
convex polygon $A$ and a point $p$ outside $A$, $spoke(p,A)$ returns
the closest point on $A$ to $p$, which can be implemented in $O(\log
n)$ time by binary search. The subroutine $vertex(A,B,C)$ takes
three convex polygons $A,B$, and $C$ as input and computes a finite
or an infinite Voronoi vertex $v$ such that the Voronoi cells for
the polygons $A$, $B$, and $C$ occur in a counterclockwise order
around $v$. An advanced technique developed in
\cite{ref:KirkpatrickTe95}, called {\em tentative prune-and-search},
is used to implement $vertex(A,B,C)$ in $O(\log n)$ time.

A generalization of the MKS algorithm to computing $\CD(\calS')$ for
our convex splinegon set $\calS'$ turns out to be quite natural.
First, for each convex splinegon $S$ of $\calS'$, we consider its
leftmost and rightmost points as two new vertices of $S$, which may
partition at most two splinegon edges of $S$ into two new edges
each. This step can be done in $O(h\log n)$ time by binary search on
each splinegon in $\calS'$. After this step, for each splinegon edge
of the splinegons in $\calS'$, any vertical line can intersect it at
most once. Next, we sweep the splinegons of $\calS'$ from left to
right. We define the {\em site events} and {\em circle events}
similarly as in the MKS algorithm. Clearly, there are still $O(h)$
site events and $O(h)$ circle events. We also maintain the two data
structures for storing the sweep front and for scheduling the
events. Essentially, the MKS algorithm relies on the convexity of
the objects involved. Since the splinegons in $\calS'$ are convex,
the scheme of the MKS algorithm is still applicable. For example,
the critical observation used by the MKS algorithm for handling site
events is based on the convexity of the polygons. In our problem, at
each site event, the sweepline is at the leftmost point, say $p$, of
a convex splinegon. To determine the nearest front arc on the sweep
front that is closest to $p$, by following the same approach as for
the MKS algorithm, we can also partition the sweepline into disjoint
intervals such that finding the nearest front arc to the point $p$
is equivalent to locating in which interval the point $p$ lies.

For the implementation details and the running time of our compact
diagram algorithm, in general, since each splinegon edge is of
$O(1)$ complexity, the operations on splinegons edges needed in the
algorithm can be performed in the same order of time asymptotically
as those on polygon edges in the MKS algorithm. Obviously, the
subroutine $spoke(p,A)$ can be implemented in $O(\log n)$ time by a
binary search. For the subroutine $vertex(A,B,C)$, the tentative
prune-and-search technique \cite{ref:KirkpatrickTe95} is also
applicable to our problem. Specifically, this technique defines
three continuous, monotone-decreasing functions that rely only on
the convexity of the objects involved. In our problem, since all
splinegons are convex, we can define three such functions in exactly
the same way as those for convex polygons in the MKS algorithm. One
basic operation needed in our algorithm is: For any point $p$ on the
boundary of a convex splinegon $S_i\in\calS'$, compute the normal of
$S_i$ at $p$. Since each splinegon edge is of $O(1)$ complexity,
this operation takes $O(1)$ time, as in the convex polygon case.
Other operations can also be performed in the same order of time as
their counterparts in the convex polygon case. We omit the details.
Therefore, the subroutine $vertex(A,B,C)$ can be implemented in
$O(\log n)$ time.

We conclude that the compact diagram $\CD(\calS')$ can be computed
in $O(h\log n)$ time.

Since each splinegon edge is of $O(1)$ complexity, as in
\cite{ref:McAllisterA96}, the Voronoi diagram $\VD(\calS')$ can be
derived from $\CD(\calS')$ in an additional $O(n)$ time. The theorem
thus follows.
\end{proof}

\section{Wrapping Things Up}
\label{sec:polygoncase}

We now show how to find a shortest $s$-$t$ path for our original
\spp\ problem on the splinegon set $\calS$. In Section
\ref{sec:corridors}, we build a corridor structure to obtain $O(h)$
corridor paths and a set $\calS'$ of $O(h)$ pairwise disjoint convex splinegons with a total of
$O(n)$ vertices such that a
shortest $s$-$t$ path for \spp\ is also a shortest $s$-$t$ path
avoiding
the convex splinegons in $\calS'$ and possibly utilizing some corridor
paths. In Section \ref{sec:shortestpath}, based on $\calS'$, we
construct a coalesced graph $G^c$ such that: (1) $G^c$ has only
$O(h)$ nodes and $O(k)$ edges; (2) a shortest $s$-$t$ path avoiding
the convex splinegons in $\calS'$ corresponds to a shortest $s$-$t$
path in $G^c$.

To compute a shortest $s$-$t$ path for our original \spp\ problem, our
final step is to incorporate the $O(h)$ corridor paths into the
graph $G^c$ to obtain a new graph $G_a^c$ such that a shortest
$s$-$t$ path for \spp\ corresponds to a shortest $s$-$t$ path in
$G_a^c$, as follows.

Recall that a corridor path connects the two apices of two funnels.
When building $G^c$, in addition to other distinguished points, we
also treat all the $O(h)$ funnel apices as distinguished points. In
this way, every funnel apex defines
two vertices in $G^c$ since every distinguished point defines two
vertices in $G^c$ (refer to \cite{ref:ChenCo11} on this).
Consider a corridor path connecting two funnel
apices $u$ and $v$. Suppose the two vertices in $G^c$ defined by $u$
(resp., $v$) are $u_1$ and $u_2$ (resp., $v_1$ and $v_2$). After
obtaining $G^c$, we add to $G^c$ eight directed edges $e(u_1,v_1),
e(u_1,v_2), e(u_2,v_1), e(u_2,v_2), e(v_1,u_1), e(v_1,u_2),
e(v_2,u_1)$, and $e(v_2,u_2)$, whose weights are the length of the
corresponding corridor path. We do this for each corridor path, and
then obtain the graph $G_a^c$. Since there are $O(h)$ corridor paths, the
graph $G_a^c$, which still has $O(h)$ nodes and $O(k)$ edges, can be
constructed in $O(n+h\log h+k)$ time. A shortest $s$-$t$ path for
\spp\ can then be computed by running Dijkstra's algorithm on
$G_a^c$, in $O(h\log h+k)$ time.

In summary, we have the following result.

\begin{theorem}\label{theo:30}
Given a set $\calS$ of $h$ pairwise disjoint splinegons with a total of $n$
vertices in the plane, a shortest
$s$-$t$ path in the free space can be computed in $O(n\log n+k)$ time or
$O(n+h\log^{1+\epsilon} h+k)$ time for any arbitrarily small constant $\epsilon>0$,
where $k$ is the size of the
relevant visibility graph and $k=O(h^2)$.
\end{theorem}

\section{Conclusions}

In this paper, we present an efficient algorithm for computing
shortest paths among
curved obstacles in the plane. For curved obstacles, previous
solutions were known only for convex curved obstacles while our approach
works for non-convex curved obstacles. Even if applied to polygonal
obstacles, our algorithm is faster than the previous best
$O(n\log n)$ time solution \cite{ref:HershbergerAn99}
when the number of obstacles is small
(e.g., $h=o(\sqrt{n\log n})$).
As a subproblem that is interesting in its own right,
we give an output sensitive optimal $O(n+h\log h+k)$ time
algorithm for computing
the relevant visibility graph of convex curved obstacles;
even if applied to convex polygonal obstacles, our algorithm is better
than the previous best $O(n+h^2\log n)$ time solution
\cite{ref:KapoorAn97,ref:RohnertSh86}.

It would be interesting to see whether the techniques developed in this paper can be used to solve other related problems on curved objects. 


\bibliographystyle{plain}
\bibliography{reference}

\end{document}